\documentclass[aps, prd, letterpaper, showkeys, showpacs, twocolumn, nofootinbib, preprintnumbers, longbibliography, unsortedaddress, 10pt]{revtex4-1}

\usepackage[top=2.8cm, bottom=2.8cm, left=2cm, right=2cm]{geometry}
\usepackage[utf8]{inputenc}  
\usepackage{amsmath,amssymb,lmodern,amsfonts} 
\usepackage{graphicx}
\usepackage{natbib}
\usepackage[caption=false]{subfig}
\usepackage{hyperref} 
\usepackage{color}
\usepackage[table]{xcolor}
\usepackage{booktabs}
\usepackage{tabulary}
\usepackage{tabularray}
\usepackage{natbib}
\usepackage{placeins}

\usepackage{comment}
\usepackage{soul}
\setstcolor{red}
\setul{}{0.7pt}

\definecolor{darkgreen}{rgb}{0,0.6,0}

\newcommand{\dd}{{\text{d}}}
\newcommand{\mc}[1]{{\mathcal{#1}}}
\newcommand{\mpl}{m_{\text{P}}}
\newcommand{\AQ}{\text{AQ}}
\newcommand{\FQ}{\text{FQ}}
\newcommand{\VF}{\mathrm{VF}}
\newcommand{\twoF}{\mathrm{2F}}
\newcommand{\eff}{\text{eff}}

\begin{document}

\preprint{}

\title{Anisotropic Universes in Light of Background Cosmological Observations}

\author{Jose L. Palacios-C\'ordoba}
\email{palacios.jose@correounivalle.edu.co}
\author{J. Bayron Orjuela-Quintana}
\email{john.orjuela@correounivalle.edu.co}
\author{Gabriela A. Valencia-Zu\~niga}
\email{gabriela.zuniga@correounivalle.edu.co}
\author{C\'esar A. Valenzuela-Toledo}
\email{cesar.valenzuela@correounivalle.edu.co}
\affiliation{Departamento  de  F\'isica,  Universidad  del Valle, \\ Ciudad  Universitaria Mel\'endez,  Santiago de Cali  760032,  Colombia}

%%%%%%%%%%%%%%%%%%%%%%%%
\begin{abstract}
%%%%%%%%%%%%%%%%%%%%%%%%

The cosmological principle is a cornerstone of the standard cosmological model. However, recent observations suggest potential deviations from this assumption, hinting at a small anisotropic expansion. Such an expansion can arise from sources that break rotational invariance. A minimal realization of this scenario is described by a Bianchi I geometry, where the degree of anisotropy is quantified by the shear parameter $\Sigma$. In this work, we constrain the present-day value of the shear, $\Sigma_0$, by confronting theoretical predictions with recent cosmological data. We implement various anisotropic models within the Boltzmann code \texttt{CLASS} and explore their parameter space using the sampler \texttt{MontePython}. Although our results show that $\Sigma_0$ is model-dependent, notably, in one specific scenario considering a homogeneous scalar field coupled to a 2-form field, $\Sigma_0 = 0$ is excluded at the $2\sigma$ confidence level, with mean value around $|\Sigma_0| \sim 10^{-4}$ while remaining consistent with observations. These findings challenge the conventional assumption that cosmic shear is negligible in the present universe. Moreover, the anisotropic expansion in this model is driven by a steep scalar field potential, a feature often found in supergravity-inspired scenarios. While anisotropic models offer interesting alternatives and could help explain some cosmological anomalies, they generally introduce additional parameters, making the standard $\Lambda$CDM model statistically favored in most cases. Still, they remain compatible with current observations and provide new perspectives on features not fully explained within the standard framework. These results highlight the importance of further exploring anisotropic cosmologies to better understand their implications.

%%%%%%%%%%%%%%%%%%%%%%%%
\end{abstract}
%%%%%%%%%%%%%%%%%%%%%%%%

%%%%%%%%%%%%%%%%%%%%%%%%%%%%%%
\pacs{98.80.Cq, 95.36.+x}
%%%%%%%%%%%%%%%%%%%%%%%%%%%%%%

%%%%%%%%%%%%%%%%%%%%%%%%%%%%%%%%%%%%%%%%%%%%%%%%%%%%%%%%%%%%%%%%%%%%%%%%%%%%%%%%%%%%%%%%%%%
\keywords{Anisotropic dark energy, observable anisotropy, cosmological constraints.} 
%%%%%%%%%%%%%%%%%%%%%%%%%%%%%%%%%%%%%%%%%%%%%%%%%%%%%%%%%%%%%%%%%%%%%%%%%%%%%%%%%%%%%%%%%%%

%%%%%%%%%%%%%
\maketitle
%%%%%%%%%%%%%

%%%%%%%%%%%%%%%%%%%%%%%%%%
\section{Introduction} 
\label{Sec: Intro}
%%%%%%%%%%%%%%%%%%%%%%%%%%

The discovery of the accelerated expansion of the universe through Type Ia supernovae (SnIa) observations marked a turning point in modern cosmology~\cite{SupernovaSearchTeam:1998fmf, SupernovaCosmologyProject:1998vns}. This phenomenon, attributed to an unknown component termed dark energy, established the $\Lambda$CDM model as the standard cosmological paradigm, where the cosmological constant $\Lambda$ drives the acceleration while cold dark matter (CDM) supports structure formation~\cite{Peebles:2022bya}. 

This model relies on two key pillars: General Relativity (GR), which governs gravitational interactions, and the cosmological principle, which postulates that the universe is homogeneous and isotropic on large scales~\cite{Scott:2018adl}.

Despite its success in describing diverse observations—including late-time acceleration~\cite{SupernovaSearchTeam:2004lze, Haridasu:2017lma}, structure formation~\cite{SDSS:2003eyi}, the cosmic microwave background (CMB)~\cite{WMAP:2012nax, Planck:2018vyg}, and baryon acoustic oscillations (BAO)~\cite{SDSS:2009ocz, Blake:2011en, Aubourg:2014yra}—growing evidence challenges some of its assumptions~\cite{Perivolaropoulos:2021jda, Abdalla:2022yfr}. Observational discrepancies such as the Hubble tension, where local SnIa measurements of the Hubble constant $H_0$ differ from CMB-based values~\cite{Wong:2019kwg, Riess:2020fzl, Riess:2021jrx}, and the $S_8$ tension, reflecting inconsistencies in large-scale structure clustering between Planck constraints and weak lensing surveys~\cite{Huang:2021tvo, Anchordoqui:2021gji, Sabogal:2024yha}, suggest deviations from $\Lambda$CDM. Additionally, anomalies including the CMB lowest multipoles alignment~\cite{Rakic:2007ve}, large-scale bulk flows~\cite{Kashlinsky:2008ut, Watkins:2008hf, Kashlinsky:2009dw, Atrio-Barandela:2014nda, Watkins:2023rll, Lopes:2024vfz}, and unexpectedly massive structures~\cite{Gott:2003pf,Clowes:2012pn,Horvath:2013kwa,Lopez:2022kbz} question the validity of the cosmological principle~\cite{Aluri:2022hzs}. 

Theoretical challenges such as the cosmological constant problem, which stems from the vast discrepancy between theoretical estimates of vacuum energy and its observed value~\cite{Weinberg:1988cp, Martin:2012bt}, and the coincidence problem, which questions why dark energy and matter densities are of the same order today despite evolving independently~\cite{Amendola_Tsujikawa_2010,Steinhardt:1999nw,Zlatev:1998tr,Amendola:1999er,Copeland:2006wr}, further motivate alternative scenarios. These issues suggest that $\Lambda$CDM may be an effective rather than a fundamental description of cosmic acceleration~\cite{Padmanabhan:2002ji,Peebles:2002gy}.

Anisotropic expansion has emerged as a promising avenue to address these discrepancies~\cite{Land:2005jq, Perivolaropoulos:2021jda, Akarsu:2021max, Adam:2024kgs}. Unlike isotropic models, anisotropic cosmologies introduce directional dependencies in the universe’s expansion history, potentially explaining anomalies such as the CMB quadrupole suppression~\cite{Battye:2009ze, Perivolaropoulos:2014lua, Schwarz:2015cma}, large-scale bulk flows~\cite{Koivisto:2008ig,Palle:2010waf,BeltranJimenez:2008rei,Cembranos:2019jlp}, and misalignments between cosmic dipoles, such as the quasar and CMB dipoles~\cite{Secrest:2020has, Dalang:2021ruy,vonHausegger:2024jan,Secrest:2025wyu, Dam:2022wwh,Wagenveld:2025ewl,Wagenveld:2023kvi,Secrest:2022uvx,Oayda:2024hnu}.

Previous attempts to constrain anisotropic expansion from observational data have been successfully carried out in the literature. For example, Refs.~\cite{Campanelli:2010zx, Amirhashchi:2018nxl} used supernovae and observational Hubble data to place upper bounds on the present-day shear of $\sim 10^{-2}$ and $\sim 10^{-3}$, respectively. Similarly, Ref.~\cite{Saadeh:2016sak} analyzed temperature and polarization measurements from the CMB and reported a much tighter constraint of less than $10^{-11}$ at the 95\% confidence level. However, these bounds were derived under specific model assumptions,\footnote{Model-independent approaches have been explored in Refs.~\cite{Wang:2014vqa, Wang:2017ezt, Salehi:2016sta}.} such as an anisotropic fluid or cosmological constant, which may not be capable of sustaining prolonged periods of anisotropic expansion. It remains unclear whether alternative anisotropic sources could lead to different observational signatures or less stringent bounds on the shear.

However, sustaining anisotropy on cosmological timescales poses theoretical challenges. Wald’s no-hair theorem states that an initially anisotropic universe with a positive cosmological constant evolves toward an isotropic de Sitter state, implying that anisotropy cannot persist indefinitely in a $\Lambda$-dominated universe~\cite{Wald:1983ky}. Consequently, viable anisotropic models require additional mechanisms—such as modified gravity~\cite{Loo:2023anl,Sahu:2016ccd,Pawde:2024jtr,Kibaroglu:2024mov,Deliduman:2023caa}, multiple interacting fields~\cite{Sharma:2021ayk,Amirhashchi:2014sia,Katore:2018cxb,Hossienkhani:2016kba}, or unconventional energy-momentum components~\cite{Sharif:2010vp,Verma:2024lex,Singh:2014ica}—to counteract isotropization.

On the other hand, scalar field dynamics alone cannot maintain anisotropic expansion. Therefore, one has to consider additional degrees of freedom—such as vector fields~\cite{Orjuela-Quintana:2021zoe}, $p$-forms~\cite{BeltranAlmeida:2019fou}, or gauge fields~\cite{Orjuela-Quintana:2020klr, Guarnizo:2020pkj}-introducing anisotropic stresses that influence cosmic evolution. While various theoretical models propose anisotropic mechanisms, their observational viability remains largely unexplored.

Most studies in this context focus on dynamical properties, assessing stability and phase-space behavior, yet systematic comparisons with cosmological data are scarce. Although some works analyze $w$CDM-like parameterizations where the equation of state is split into parallel and orthogonal components~\cite{Campanelli:2010zx, Amirhashchi:2018nxl}, and $w$CDM models with anisotropic perturbations~\cite{PhysRevD.47.3151}, a broader investigation of anisotropic dark energy scenarios within a more general framework is still needed.

In this work, we take a step toward bridging this gap by exploring a range of anisotropic dark energy models and confronting them with observational constraints. Our analysis seeks to answer key questions: How much anisotropy can we expect from these models? Which physical mechanisms are most effective in producing observable anisotropic signatures? To address these, we implement various anisotropic cosmological models in the Boltzmann code \texttt{CLASS}~\cite{Blas:2011rf} and explore their parameter space using the Monte Carlo sampler \texttt{MontePython}~\cite{Audren:2012wb, Brinckmann:2018cvx}. By incorporating multiple datasets—including SnIa~\cite{Scolnic:2021amr,Melia:2018kyb,DES:2022zrg}, Hubble parameter measurements~\cite{Jiao:2022aep,Guo:2015gpa, Moresco:2016mzx,refId0,Moresco2012}, BAO~\cite{DESI:2025zgx,eBOSS:2020gbb,2011MNRAS.416.3017B,Ross:2014qpa}, and CMB shift parameters~\cite{Wang:2013mha,WMAP:2003tof,Wang:2007mza}—we derive constraints on the present-day shear parameter $\Sigma_0$.

Our results show that $\Sigma_0$ is model-dependent, with one scenario—a homogeneous scalar field coupled to a 2-form field—excluding isotropy at the $2\sigma$ confidence level. This model exhibits additional intriguing features, which we discuss throughout the work. On the other hand, although model selection criteria generally favor $\Lambda$CDM due to its minimal parameter set, anisotropic models remain competitive within observational uncertainties. Our findings highlight the viability of anisotropic cosmologies and the need for further exploration of their role in modern cosmology.

All modifications to the \texttt{CLASS} code were performed at the background level. Specifically, the following modules were altered: \texttt{background.h} and \texttt{input.h}, where the model parameters and dynamical variables are defined; \texttt{input.c}, which reads the parameter values from the \texttt{.ini} file; \texttt{background.c}, where the modified system of differential equations is solved and the Hubble parameter is computed. The modified version of the code is publicly available at: \href{https://github.com/navidadxxJose/class_anisotropic_DE_models}{\texttt{GitHub}}.\footnote{\href{https://github.com/navidadxxJose/class_anisotropic_DE_models}{https://github.com/navidadxxJose/class\_anisotropic\_DE\_models}.}

This paper is organized as follows: Section~\ref{Sec: Framework} provides a theoretical overview of anisotropic cosmological models in a Bianchi I background. Section~\ref{Sec: Sources} details the models analyzed and their anisotropy-sustaining mechanisms. Section~\ref{Sec: Results} describes our methodology, including model implementation in \texttt{CLASS}, parameter estimation with \texttt{MontePython}, and the datasets used. We then present constraints on model parameters and $\Sigma_0$. Section~\ref{Sec: Analysis} discusses the implications of our results and evaluates model viability using Bayesian criteria. Finally, Section~\ref{Sec: Conclusions} summarizes our conclusions and outlines future research directions.

%%%%%%%%%%%%%%%%%%%%%%%%%%%%%%%%%%%%%%%%%%%%%%%%%%%%%%%%%%%%%%%
\section{Theoretical Framework: Cosmology from Bianchi I}
\label{Sec: Framework}
%%%%%%%%%%%%%%%%%%%%%%%%%%%%%%%%%%%%%%%%%%%%%%%%%%%%%%%%%%%%%%%

We will begin our study by formulating the general evolution equations for a cosmological model within the framework of the most general Bianchi I background. We will closely follow the formalism detailed in Ref.~\cite{Pereira:2007yy}. The Bianchi I metric is given by:
\begin{equation}
    \dd s^2 = - \dd t^2 + a^2(t)\gamma_{i j}(t) \dd x^i \dd x^j,  
 \label{eq: Bianchi I}   
\end{equation}
where $a(t)$ represents an average scale factor that depends solely on cosmic time $t$, and $x^i$ denotes the spatial coordinates. The spatial metric, $\gamma_{ij}(t)$, satisfies the volume-preserving condition, $\det (\gamma_{ij}) = 1$. Since a constant $\gamma_{ij}$ results in an isotropic metric (as its components can be absorbed through a redefinition of spatial coordinates), the impact of anisotropy on the evolution of the geometry is characterized by the geometric shear tensor, defined as: 
\begin{equation}
    \tilde{\sigma}_{ij} = \frac{1}{2} \dot{\gamma}_{ij}, \quad \tilde{\sigma}^2 = \tilde{\sigma}_{ij} \tilde{\sigma}^{ij},
\end{equation}
where an overdot denotes derivative with respect to time. In the Misner representation, the spatial metric components and the volume-preserving condition are expressed as:
\begin{equation}
    \gamma_{ij}(t) = \left[ e^{2 \beta_i(t)} \right] \delta_{i j}, \quad \sum_i \beta_i = 0,
\label{eq: Bianchi I gamma}    
\end{equation}
where $\beta_i(t)$ are time-dependent functions, and no summation is implied here. Throughout this work, we employ square brackets to explicitly indicate such quantities avoiding summation conventions. This representation elucidates that, despite initially considering three distinct scale factors, only two degrees of freedom remain due to the volume-preserving constraint. Under this formalism, the geometric shear tensor and geometric shear scalar take the form:
\begin{equation}
    \tilde{\sigma}_{ij} = \left[ \dot{\beta}_i \right] g_{ij}, \quad \tilde{\sigma}^2 = \sum_i \dot{\beta}_i^2.
\end{equation}
A cosmological model is generally described by the action:
\begin{equation}
    \mc{S} = \int \dd x^4 \, \sqrt{-g} \left\{ \frac{\mpl^2}{2}R + \mc{L} \right\},
\end{equation}
where $g$ denotes the determinant of the metric $g_{\mu\nu}$, $\mpl$ is the reduced Planck mass, $R$ is the Ricci scalar, and $\mc{L}$ encapsulates the Lagrangian densities of all matter components in the universe, including terms that explicitly break spatial rotational invariance.

By varying the action with respect to the metric $g^{\mu\nu}$, we obtain the gravitational field equations, which are given by:
\begin{equation}
    \mpl^2 G_{\mu\nu} = T_{\mu\nu},
\end{equation}
where $G_{\mu\nu}$ is the Einstein tensor, and the energy-momentum tensor $T_{\mu\nu}$ takes the general form:
\begin{equation}
    T_{\mu\nu} = (\rho + p)u_\mu u_\nu + p g_{\mu\nu} + \Pi_{\mu\nu},
\end{equation}
where $\rho$ is the total energy density of the cosmic fluids, $p$ denotes the total isotropic pressure, and $u_\mu$ is the total 4-velocity, normalized as $u_\mu u^\mu = -1$. The total anisotropic stress, $\Pi_{\mu\nu}$, accounts for deviations from isotropy and satisfies the traceless and transverse conditions, $\Pi^\mu_{~\mu} = 0$ and $u_\mu \Pi^{\mu\nu} = 0$, respectively.

For the Bianchi I metric, the components of the Einstein tensor are:
\begin{align}
    G^0_{~0} &= -3H^2 + \frac{1}{2}\tilde{\sigma}^2, \\
    G^i_{~j} &= - \left(2\dot{H} + 3H^2 + \frac{1}{2}\tilde{\sigma}^2 \right)\delta^i_{~j} + \dot{\tilde{\sigma}}^i_{~j} + 3H \tilde{\sigma}^i_{~j},
\end{align}
where $H \equiv \dot{a}/a$ is the average expansion rate. These expressions allow us to derive the first Friedman equation ($\mpl^2 G^0_{~0} = T^0_{~0}$) and the second Friedman equation ($\mpl^2 G^i_{~i} = T^i_{~i}$) as:
\begin{align}
    3 \mpl^2 H^2 &= \rho + \frac{1}{2} \mpl^2 \tilde{\sigma}^2, \\
    -2 \mpl^2 \dot{H} &= \rho + p + \mpl^2 \tilde{\sigma}^2,
\end{align}
while the evolution equation for the geometric shear tensor components follows from the traceless part of the Einstein equations:
\begin{equation}
    \mpl^2 \left( \dot{\tilde{\sigma}}^i_{~j} + 3H\tilde{\sigma}^i_{~j} \right) = \Pi^i_{~j}.
\end{equation}
Energy-momentum conservation, $\nabla_\mu T^{\mu\nu} = 0$, leads to the continuity equation:
\begin{equation}
    \dot{\rho} + 3H(\rho + p) = - \tilde{\sigma}_{ij} \Pi^{ij},
\end{equation}
where the right-hand term explicitly accounts for anisotropic contributions, absent in isotropic expansion or a perfect fluid scenario.

In comparison with an isotropic model, the Friedman equations are modified only by the geometric shear scalar. Given this, it is natural to consider the simplest anisotropic extension: a universe that expands preferentially along a single direction while preserving residual symmetry in the transverse plane. This assumption is well-motivated both physically and mathematically. Physically, it corresponds to the most straightforward deviation from isotropy. Mathematically, axial symmetry significantly reduces the number of independent shear components, simplifying the dynamics without sacrificing generality for our purposes. 

To impose the residual symmetry in a plane, we consider an axisymmetric case where the anisotropy parameters satisfy:
\begin{equation}
    \beta_1 = -2\sigma, \quad \beta_2 = \beta_3 = \sigma, \quad \Rightarrow \quad \tilde{\sigma}^2 = 6\dot{\sigma}^2.
\end{equation}

Thus, by adopting the axially symmetric decomposition of the shear introduced in the previous equation, the spatial part $\gamma_{ij}$, given in Eq.~\eqref{eq: Bianchi I gamma}, leads to the following simplified form of the metric in Eq.~\eqref{eq: Bianchi I}:
\begin{equation}
\label{Eq: Bianchi I}
    \dd s^{2} = -\dd t^{2} + a(t)^{2}\left[e^{-4\sigma} \dd x^{2} + e^{2\sigma}\left(\dd y^{2} + \dd z^{2} \right)\right]. 
\end{equation}
The Friedman equations simplify to:
\begin{align}
    3 \mpl^2 H^2 &= \rho + 3\mpl^2 \dot{\sigma}^2, \\
    -2 \mpl^2 \dot{H} &= \rho + p + 6 \mpl^2 \dot{\sigma}^2,
\end{align}
while the evolution equation for $\sigma$ reads:
\begin{equation}
\label{Eq: Evo Geometrical Shear}
    \mpl^2(\ddot{\sigma} + 3H\dot{\sigma}) = T^1_{~1} - T^2_{~2}.
\end{equation}
In the following analysis, we investigate the cosmological dynamics arising from various sources of anisotropy. To this end, we decompose the total Lagrangian as:
\begin{equation}
    \mc{L} = \mc{L}_r + \mc{L}_b + \mc{L}_c + \mc{L}_\text{DE},
\end{equation}
where $\mc{L}_{i = r, b, c}$ corresponds to the Lagrangians for radiation, baryons, and CDM, respectively. These components are modeled as perfect fluids, implying that they do not contribute to the anisotropic stress tensor, $\Pi_{\mu\nu}$. In contrast, $\mc{L}_\text{DE}$ represents the Lagrangian for dark energy, which will serve as the primary source of anisotropy in our framework.

%%%%%%%%%%%%%%%%%%%%%%%%%%%%%%%%%%%%%%
\section{Sources of Anisotropy}
\label{Sec: Sources}
%%%%%%%%%%%%%%%%%%%%%%%%%%%%%%%%%%%%%%

In general, any field that breaks spatial rotational invariance can, in principle, sustain an anisotropic expansion of the Universe. Homogeneous scalar fields, which are among the most commonly employed fields in cosmology, cannot generate such anisotropy, as they do not contribute to the anisotropic stress~\cite{Fadragas:2013ina, Mimoso:1995ge}. However, it has been demonstrated that coupling a homogeneous scalar field to other types of fields, such as vector fields or $p$-forms, can give rise to a stable anisotropic expansion~\cite{Kaneta:2022kjj, Socorro:2022iqt, Lee:2023azx}. In this work, we investigate the present-day anisotropy induced by specific realizations of anisotropic dark energy, firstly focusing on scalar fields coupled to additional degrees of freedom. Specifically, we consider a homogeneous scalar field coupled to: $(i)$ a vector field~\cite{Orjuela-Quintana:2021zoe}, and $(ii)$ a 2-form field~\cite{BeltranAlmeida:2019fou}.

Furthermore, it has been shown that a configuration of inhomogeneous scalar fields~\cite{Motoa-Manzano:2020mwe}, and a scalar field with internal symmetries~\cite{Orjuela-Quintana:2020klr}, can also sustain a homogeneous yet anisotropic expansion. These models will be examined as well.

In some of these scenarios, the scalar field drives cosmic acceleration, while its coupling to an additional field introduces anisotropy into the expansion. However, certain fields that explicitly break isotropy can, by themselves, sustain an anisotropic accelerated expansion. Among these are vector fields~\cite{Koivisto:2007bp} and $p$-forms~\cite{Orjuela-Quintana:2022jrg}, which we will also analyze in this work.

More details about the models under investigation are outlined below. While this section does not provide an exhaustive theoretical derivation for each case—relevant references are cited throughout the text—we present the essential steps required to obtain the equations of motion used in our numerical analysis. However, we clarify that a proper dynamical analysis of each model has been carried out in previous works, which are appropriately cited throughout the text.

%%%%%%%%%%%%%%%%%%%%%%%%%%%%%%%%%%%%%%%%%%%%%%%%%%%%%%%%
\subsection{Scalar field coupled to a vector field}
%%%%%%%%%%%%%%%%%%%%%%%%%%%%%%%%%%%%%%%%%%%%%%%%%%%%%%%%

The first model under consideration involves the interaction between a homogeneous scalar field, $\phi$, and a vector field, $A_\mu$, referred to as \textit{scalar field coupled to a vector field} (AQ) model. For a dynamical analysis of this model, we refer the reader to Refs.~\cite{Orjuela-Quintana:2021zoe, Thorsrud:2012mu}, and for its application to inflation, to Ref.~\cite{Ohashi:2013pca}. The Lagrangian describing the dark energy component is given by:
\begin{equation}
\label{Eq: action_AQ}
    \mc{L}_\text{DE} = - \frac{1}{2} \nabla_\mu \phi \nabla^\mu \phi - \frac{1}{4}f_\text{AQ}^{2}(\phi) F_{\mu\nu} F^{\mu\nu} - V_\text{AQ}(\phi),
\end{equation} 
where $F_{\mu\nu} \equiv \nabla_{\mu}A_{\nu} - \nabla_{\nu}A_{\mu}$ is the field strength tensor of the vector field, $f_\text{AQ}(\phi)$  is a coupling function governing the interaction between the scalar and vector fields, and $V_\text{AQ}(\phi)$ is the scalar field potential defined as:
\begin{equation}
    V_{\text{AQ}} \equiv V_a e^{-\lambda_{\text{AQ}}\phi}/m_{\text{p}},
\end{equation}
where $V_a$ and $\lambda_{\text{AQ}}$ are constants.

Regarding the Bianchi I metric in Eq.~\eqref{Eq: Bianchi I}, the compatible vector field profile is $A_{\mu} = \left(0, A(t),0,0\right)$, with $A(t)$ being a function of time. We consider an exponential coupling function of the form:
\begin{equation}
    f_\text{AQ}(\phi) = f_{a}e^{-\mu_{\mathrm{AQ}} \left(\frac{\phi}{\mpl}\right)}, 
\end{equation} 
where $\mu_{\mathrm{AQ}}$ and $f_{a}$ are constants. 

To compute the Hubble parameter $H$, we introduce the following dimensionful dynamical variables:
\begin{equation}
    x \equiv \frac{\dot{\phi}}{\sqrt{6} \mpl}, \quad z^{2} \equiv \frac{\rho_{A}}{3\mpl^2}, \quad y^2 \equiv \frac{V_\text{AQ}(\phi)}{3\mpl^2},
\end{equation} 
where $\rho_A$ is the vector field density defined in Eq.~\eqref{Eq: Densities fields AQ}.
From the equations of motion~\eqref{Eq: field_equation_A_AQ}, \eqref{Eq: field_equation_phi_AQ}, and \eqref{Eq: dd sigma_AQ} in Appendix~\ref{App: AQ Model}, we construct the following first-order differential equations governing the evolution of the dynamical variables $x, y, z$, and the geometric shear $\dot{\sigma}$:
\begin{align}
    \frac{\dd x}{\dd N} &= -3 \left(x - \lambda_\AQ \frac{x^2 + y^2}{\sqrt{6}H} + \mu_\AQ \frac{2z^2}{\sqrt{6}H} \right), \label{Eq: x_AQ} \\
    \frac{\dd y}{\dd N} &= -\lambda_\AQ\frac{\sqrt{6}}{2} \frac{x y}{H}, \label{Eq: y_AQ} \\
    \frac{\dd z}{\dd N} &= z\left(\sqrt{6} \, \mu_\AQ \frac{x}{H} - 2\frac{\dot{\sigma}}{H} - 2 \right), \label{Eq: z_AQ} \\
    \frac{d\dot{\sigma}}{dN} &= -3\dot{\sigma}  + \frac{2 z^{2}}{H}, \label{Eq: sigma_AQ}
\end{align}
where the derivatives are computed with respect to the number of $e$-folds defined as $N \equiv \log a$. Note that, unlike the standard approach in dynamical system analyses of cosmological models where variables are normalized by $H$~\cite{Bahamonde:2017ize}, we retain their explicit dependence on $H$. The conventional normalization by $H$ effectively removes it from the system, requiring its determination via integration of the e.g. deceleration parameter $q \equiv - a\ddot{a}/\dot{a}^2$. Instead, our approach allows for a direct computation of $H$ within the dynamical system, albeit at the cost of dealing with a non-autonomous set of differential equations. This trade-off is justified as it provides an explicit evolution for $H$, facilitating the implementation in the Boltzmann solver \texttt{CLASS}, and allowing a direct calculation of background observables. Then, $H$ can be written  from the first Friedman equation in Eq.~\eqref{Eq: Friedman_1_AQ} as:
\begin{equation}
    H^2 = \frac{1}{3\mpl^2}\left( \rho_m + \rho_r \right) + x^2 +y^2 + z^2 + \dot{\sigma}^2, 
\label{Eq: Friedman_1_1_AQ}
\end{equation} 
where we have defined the matter density $\rho_m \equiv \rho_b + \rho_c$. By substituting Eq.~\eqref{Eq: Friedman_1_1_AQ} into Eqs.~\eqref{Eq: x_AQ}-\eqref{Eq: sigma_AQ}, we obtain a closed system that can be solved numerically. From the numerical solution for $x$, $y$, $z$, and $\dot{\sigma}$, $H$ is determined as a function of $N$ according to Eq.~\eqref{Eq: Friedman_1_1_AQ}. 

Note that the $\Lambda$CDM limit of this model is recovered when $\lambda_\AQ = f_\AQ = 0$. In this case, the potential becomes flat, and there is no interaction between the scalar field and the vector field. Recall that, in the absence of coupling, the scalar field $\phi$ evolves as $\phi \propto a^{-6}$.

%%%%%%%%%%%%%%%%%%%%%%%%%%%%%%%%%%%%%%%%%%%%%%%%%%%%%%
\subsection{Scalar Field Coupled to a 2-form Field}
%%%%%%%%%%%%%%%%%%%%%%%%%%%%%%%%%%%%%%%%%%%%%%%%%%%%%%

The second model consists of a canonical scalar field, $\phi$, coupled to a 2-form field, $B_{\mu\nu}$, which we refer to as the \textit{anisotropic quintessence coupled to a 2-form} (FQ) model. The Lagrangian describing the dark energy component is given by:
\begin{align}
    \mc{L}_\text{DE} = &- \frac{1}{2}\nabla_\mu \phi \nabla^\mu \phi - \frac{1}{12} f_\FQ(\phi) H_{\mu\nu\alpha}H^{\mu\nu\alpha} \nonumber \\
    &- V_\FQ(\phi)
\label{Eq: action_2F} 
\end{align}
where $H_{\mu\nu\alpha} = 3\nabla_{[\mu}B_{\nu\alpha]}$ represents the field strength tensor of the 2-form field. The scalar potential, $V_\FQ(\phi)$, and the coupling function, $f_\FQ(\phi)$, take the exponential forms:
\begin{align}
    V_\FQ(\phi) &= V_{f}e^{-\lambda_\FQ\left(\frac{\phi}{\mpl}\right)}, \\
    f_\FQ(\phi) &= f_{f}e^{-\mu_\FQ \left(\frac{\phi}{\mpl}\right)},
\end{align}
where $V_{f}$, $\lambda_\FQ$, $f_f$, and $\mu_\FQ$ characterize the model. For a detailed dynamical analysis of this model we refer the reader to Ref.~\cite{BeltranAlmeida:2019fou}.

Considering the metric given in Eq.~\eqref{Eq: Bianchi I}, we impose a configuration for $B_{\mu\nu}$ that preserves residual symmetry in the $(y,z)$ plane:
\begin{equation}
\label{Eq: 2form Profile}
    B_{\mu\nu} \, \dd x^{\mu} \wedge \dd x^{\nu} = 2 v_B(t) \, \dd y \wedge \dd z,
\end{equation} 
where $v_{B}$ is a scalar function. 

For this model, we introduce the following dimensionful dynamical variables:
\begin{equation}
    x_1 = \frac{\dot{\phi}}{\sqrt{6} \mpl^2}, \quad x_2^2 = \frac{V_\FQ}{3 \mpl^2}, \quad \tilde{\rho}_{B} = \frac{\rho_{B}}{3 \mpl^2}.
\end{equation}
Then, using the equations of motion for the scalar field and the 2-form field, and the evolution equation for the geometric shear in Appendix~\ref{App: 2F Model}, we obtain the following system of differential equations:
\begin{align}
    \frac{\dd x_1}{\dd N} &= -3x_1 + \frac{\sqrt{6}}{2} \left( \lambda_\FQ \frac{x_2^2}{H} - \mu_\FQ\frac{\tilde{\rho}_{B}}{H} \right), \\
    \frac{\dd x_2}{\dd N} &= -\lambda_\FQ\frac{\sqrt{6}}{2} \frac{x_1 x_2}{H}, \\
    \frac{\dd \dot{\sigma}}{\dd N} &= -3 \dot{\sigma} - \frac{2 \tilde{\rho}_{B}}{H}, \\ 
    \frac{\dd \tilde{\rho}_{B}}{\dd N} &= -\tilde{\rho}_{B} \left(2 - \frac{\dot{\sigma}}{H} - \sqrt{6} \, \mu_\FQ\frac{\tilde{\rho}_{B} x_1}{H} \right).
\end{align}
Finally, the first Friedman equation, rewritten in terms of these variables, is given by:
\begin{equation} 
    H^{2} = \frac{1}{3\mpl^{2}}\left(\rho_{m} + \rho_{r} \right) + x_{1}^{2} + x_{2}^{2} +\tilde{\rho}_{B} + \dot{\sigma}^{2}.
\end{equation}  

The $\Lambda$CDM limit of this model is recovered when $\lambda_\FQ = f_\FQ = 0$. In this case, the potential reduces to a cosmological constant, and there is no interaction between the scalar field and the 2-form field.

%%%%%%%%%%%%%%%%%%%%%%%%%%%%%%%%%%%%%%%%%%%%%
\subsection{Inhomogeneous scalar fields}
%%%%%%%%%%%%%%%%%%%%%%%%%%%%%%%%%%%%%%%%%%%%%

An inhomogeneous scalar field naturally breaks spatial rotational invariance but also violates homogeneity. However, it is possible to construct a configuration of inhomogeneous scalar fields that preserves homogeneity while supporting spatial anisotropy. Such is the case of the \textit{anisotropic solid dark energy} (SDE) model. In this framework, a ``solid'' is represented by a triad of inhomogeneous scalar fields $\phi^{I} \equiv x^{I}$, where $I = 1, 2, 3$ and $x^{I}$ denotes the comoving Cartesian coordinates~\cite{Armendariz-Picon:2007umg}. A specific realization of this model was introduced in Ref.~\cite{Motoa-Manzano:2020mwe}, where the cosmological dynamics of the following dark energy Lagrangian was analyzed:
\begin{equation}
    \mc{L}_\text{DE} = - \sum_I F^I \left( X^I \right), \quad X^I \equiv g^{\mu\nu} \nabla_\mu \phi^I \nabla_\nu \phi^I.
\label{Eq: action_SDE}
\end{equation}
Here, $F^I$ represents the Lagrangian contribution of each scalar field comprising the solid, while $X^I$ denotes its kinetic term.

In a Bianchi I universe described by the metric in Eq.~\eqref{Eq: Bianchi I}, the kinetic terms $X^I$ take the form: 
\begin{equation}
    X^1(t) = \frac{e^{4\sigma(t)}}{a^2(t)}, \qquad X^2(t) = X^3(t) = \frac{e^{-2\sigma(t)}}{a^2(t)}.
\end{equation}
To maintain compatibility with the axial symmetry of the metric, the functions $F^I$ must satisfy $F^2\left( X^2 \right) = F^3 \left( X^3 \right)$. For this analysis, we assume power-law dependencies for $F^1$ and $F^2$, given by $F^1 \propto \left( X^1 \right)^n$ and $F^2 \propto \left( X^2 \right)^m$, where $n$ and $m$ are constant parameters of the model.

To describe the dynamics of this system in terms of a closed set of first-order differential equations, we introduce the dimensionful variables:
\begin{equation}
    f_1^2 \equiv \frac{F^1}{3 \mpl^2}, \qquad f_2^2 \equiv \frac{F^2}{3 \mpl^2}.
\end{equation} 
Differentiating these variables with respect to the number of $e$-folds,
$N$, and incorporating the shear contribution in the first Friedman equation, along with the evolution equations for the scalar fields in Eq.~\eqref{Eq: field_equation_SDE} and the geometrical shear in Eq.~\eqref{Eq: field_equation_sigma_SDE} in Appendix~\ref{App: Sources}, we obtain the system:
\begin{align}
    \frac{\dd f_1}{\dd N} &= - n f_1 \left(1 - 2\frac{\dot{\sigma}}{H} \right), \\
    \frac{\dd f_2}{\dd N} &= - m f_2 \left(1 + \frac{\dot{\sigma}}{H} \right), \\
    \frac{\dd \dot{\sigma}}{\dd N} &= - 3 \dot{\sigma} - \frac{2 \left( n f_1^2 - m f_2^2 \right)}{H}.
\end{align}
This system will be solved to determine $H$ as a function of $N$, which can be expressed as:
\begin{equation}
    H^2 = \frac{1}{3\mpl^2} \left(\rho_{m} + \rho_{r} \right) + f_1^2 + 2f_2^2 + \dot{\sigma}^2.
\end{equation}
It is important to note that the scalar field configuration above is not the only possible way for the solid to break spatial rotational invariance. In fact, Ref.~\cite{Endlich:2012pz} considers a different approach, modeling the solid as a system of three scalar fields, each possessing a shift symmetry and an internal SO(3) symmetry. As discussed in Ref.~\cite{Bartolo:2013msa}, relaxing this SO(3) symmetry allows the solid to sustain a prolonged phase of anisotropic inflation. 

Likewise, as with the other models, the $\Lambda$CDM limit for the solid model is reached when $m = n = 0$. In this case, the Lagrangian contributions $F^{I}$ reduce to constants.
%%%%%%%%%%%%%%%%%%%%%%%%%%%%%%%%%%%%%%%%%%%%%%%%%%%%%
\subsection{Scalar field with internal symmetry}
%%%%%%%%%%%%%%%%%%%%%%%%%%%%%%%%%%%%%%%%%%%%%%%%%%%%%

From the preceding discussion, it is evident that scalar fields endowed with internal symmetries can give rise to anisotropic accelerated expansion. This property was explored in Ref.~\cite{Orjuela-Quintana:2020klr}, where a gauge field with a spontaneously broken internal SO(3) symmetry, akin to the Higgs mechanism, was proposed as a driver of cosmic acceleration. In this framework, the Higgs field is represented by a triplet of scalar fields, $\mc{H}^{a}$, which interacts with a gauge vector field, $A^{a}_{~\nu}$, due to the underlying SO(3) symmetry.\footnote{Latin indices correspond to the SO(3) gauge components or spatial coordinates.} Since the gauge field $A^a_{~\nu}$ evolves according to its Yang-Mills action, this framework is referred to as the \textit{anisotropic Einstein-Yang-Mills-Higgs} (EYMH) model. For isotropic studies related to this model see Refs.~\cite{Rinaldi:2015iza, Alvarez:2019ues, Frasca:2025bcx} The corresponding Lagrangian density is given by: 
\begin{equation}
\label{Eq: action_EYMH}
    \mc{L}_\text{DE} = - \frac{1}{4}F^a_{~\mu\nu}F_a^{~\mu\nu} - \frac{1}{2} D_\mu \mc{H}^a D^\mu \mc{H}_a - V_\mc{H}(\mc{H}^2), 
\end{equation}
where the field strength tensor associated with the gauge field is defined as: 
\begin{equation}
    F^{a}_{~\mu\nu} \equiv \nabla_\mu A^a_{~\nu} - \nabla_\nu A^a_{~\mu} + \tilde{g} \varepsilon^a_{~bc} A^b_{~\mu} A^c_{~\nu}, 
\end{equation} 
with $\tilde{g}$ denoting the SO(3) structure constant and $\varepsilon^a_{~bc}$ the Levi-Civita antisymmetric tensor. The Higgs triplet is expressed as $\mc{H} \equiv \left(\mc{H}_1, \mc{H}_2, \mc{H}_3 \right)^\mathrm{T}$, and its gauge-covariant derivative takes the form:
\begin{equation}
D_\mu \mc{H}^a \equiv \nabla_\mu \mc{H}^a + \tilde{g} \varepsilon^{a}_{~bc} A^b_{~\mu} \mc{H}^c. 
\end{equation}
The symmetry-breaking potential, $V_\mc{H}$, is given by
\begin{equation}
    V_\mc{H}(\mc{H}^{2}) \equiv \frac{\lambda_\mc{H}}{4} \left(\mc{H}^{2} -\mc{H}_\text{v}^2 \right)^2,
\end{equation}
where $\lambda_\mc{H}$ and $\mc{H}_\text{v}$ correspond to the quartic coupling constant and the vacuum expectation value of the Higgs field, respectively.

Following the formulation and conventions of Ref.~\cite{Orjuela-Quintana:2020klr}, the gauge and Higgs fields are assumed to take the ansatz:
\begin{equation} 
\mc{H} = \left( \Phi(t), 0, 0\right), \ \ A^1_{~1} = I(t), \ \ A^2_{~2} = A^3_{~3} = J(t), 
\end{equation} 
where $\Phi$, $I$, and $J$ are time-dependent scalar functions.

To establish a closed system governing the model's dynamics, we introduce the following dimensionful dynamical variables:
\begin{align} 
    x &\equiv \frac{G_{1}\dot{I}}{\sqrt{3}\mpl}, \quad y \equiv \frac{G_{2}\dot{J}}{\sqrt{3}\mpl}, \quad z \equiv \frac{\dot{\Phi}}{\sqrt{6}}, \nonumber \\
    w &\equiv \frac{\tilde{g}G_{2}J\Phi}{\sqrt{3}\mpl}, \quad v^2 \equiv \frac{V_\mc{H}}{3\mpl^2}, \quad \xi \equiv \frac{\sqrt{3}\mpl}{G_{2}J}, \nonumber \\
    p^2 &\equiv \frac{\tilde{g}}{\sqrt{3}\mpl} (G_{1}I)^2, \quad s^2 \equiv \frac{\tilde{g}}{\sqrt{3}\mpl} (G_{2}J)^2.
\end{align} 
Here, we have introduced the shorthand notation \mbox{$G_1 \equiv \sqrt{g^{11}}$} and \mbox{$G_2 \equiv \sqrt{g^{22}} = \sqrt{g^{33}}$} to represent metric components. Employing the field equations~\eqref{Eq: field_equation_A_EYMH}, \eqref{Eq: field_equation_PHI_EYMH}, and \eqref{Eq: sigma_EYMH} derived in Appendix~\ref{App: Sources}, the resulting system of first-order differential equations in terms of these dynamical variables is:
\begin{align}
    \frac{\dd x}{\dd N} &= -2 x \left(1 + \frac{\dot{\sigma}}{H}\right) - \frac{ps^{2}\xi}{H}, \\
    \frac{\dd y}{\dd N} &= y \left( - 2 + \frac{\dot{\sigma}}{H} \right) - \frac{\xi}{H} \left(p^2 s^2 + s^4 w^2 \right), \\
    \frac{\dd z}{\dd N} &= - 3z - \frac{w \xi}{H} \left(\sqrt{2} s^2 + \alpha v^2 \right), \\
    \frac{\dd w}{\dd N} &= w \left( -1 + \frac{\xi y}{H} - \frac{\dot{\sigma}}{H} \right) + \sqrt{2}\xi \frac{z s^2}{H}, \\
    \frac{\dd v}{\dd N} &= \alpha \frac{w z \xi}{H}, \\
    \frac{\dd \xi}{\dd N} &= \xi \left( 1 - \frac{y \xi}{H} + \frac{\dot{\sigma}}{H} \right), \\
    \frac{\dd p}{\dd N} &= p \left( -1 + \frac{\dot{\sigma}}{H}\right) + \frac{\xi s x}{H}, \\
    \frac{\dd s}{\dd N} &= s \left( -1 + \frac{y \xi}{H} - \frac{\dot{\sigma}}{H}\right), \\
    \frac{\dd \dot{\sigma}}{\dd N} &= - 3\dot{\sigma} + \frac{s^2}{H} \left(s^2 - p^2 \right) + \frac{w^2}{H} - \frac{y^2}{H} + \frac{x^2}{H},
\end{align}
where we have defined $\alpha \equiv \sqrt{2 \lambda_\mc{H}/\tilde{g}^2}$. Finally, the Hubble equation in terms of the dynamical variables is expressed as:
\begin{eqnarray} 
H^{2} &=& \frac{1}{3\mpl^{2}}\left(\rho_{m} + \rho_{r} \right) + \frac{1}{2}x^{2}+ y^{2} + p^{2}s^{2} \nonumber  \\
&+& \frac{1}{2}s^{4} + v^{2} + w^{2} + z^{2} + \dot{\sigma}^{2}.
\end{eqnarray} 

It is worth noting that this model does not admit $\Lambda$CDM as a limiting case.

%%%%%%%%%%%%%%%%%%%%%%%%%%%%%
\subsection{Vector field}
%%%%%%%%%%%%%%%%%%%%%%%%%%%%%

Up to this point, we have described several models based on the dynamics of scalar fields that can sustain anisotropic accelerated expansion. In some of these models, a homogeneous scalar field drives the accelerated expansion of the universe, while anisotropy arises from its interaction with another field capable of breaking spatial rotational invariance, such as a vector field or a 2-form field. However, these anisotropy-inducing fields can, in principle, also drive accelerated expansion without the need for a scalar field~\cite{Rodriguez-Benites:2023otm}. Here, we explore this possibility.  

We begin by considering the dynamics of a vector field within the Bianchi I spacetime given in Eq.~\eqref{Eq: Bianchi I}, which we refer to as the \textit{anisotropic vector field} (VF) model. The action governing this model is given by:
\begin{equation}
\label{Eq: action_avf}
    \mc{L}_\text{DE} = - \frac{1}{4}F_{\mu\nu} F^{\mu\nu} - V_\VF \left(X_\VF \right),
\end{equation}
where $A_\mu$ is the vector field, $F_{\mu\nu}$ is its field strength tensor, $X_\VF \equiv - A_\mu A^\mu/2$ is a quadratic term, and $V_\VF \left( X_\VF \right)$ is its potential, defined as:
\begin{equation}
\label{Eq: potential_vf}
    V_\VF \equiv V_v e^{-\lambda_\VF \left( \frac{X_\VF}{\mpl^{2}} \right)}, 
\end{equation} 
where $\lambda_\VF$ and $V_v$ are constants. Similar models in an isotropic context have been studied in e.g. Refs~\cite{Gomez:2022okq, Rodriguez-Benites:2023otm, Coelho:2025vmo}.

Given the metric in Eq.~\eqref{Eq: Bianchi I}, a compatible configuration for the vector field is  $A_{\mu} = \left(0, \psi(t), 0, 0 \right)$, where $\psi(t)$ is a time-dependent function. Then, for the VF model, we define the following dimensionful dynamical variables:
\begin{equation}
   z \equiv \frac{\dot{\psi}e^{2\sigma}}{\sqrt{6} \mpl a}, \quad v^2 \equiv \frac{V_\VF}{3 \mpl^2}, \quad x \equiv \frac{\psi  e^{2\sigma}}{\mpl a}.
\end{equation}
Using the evolution equations for the vector field and the geometric shear, given in Eqs.~\eqref{Eq: field_equation_vf} and \eqref{Eq: Evo sigma_vf}, respectively, we obtain a closed system of coupled ordinary differential equations:
\begin{align}
    \frac{\dd z}{\dd N} &= - z \left(1 + \frac{\dot{\sigma}}{H} \right) - \lambda_\VF \sqrt{\frac{3}{2}}\frac{x v^2}{H}, \\
    \frac{\dd v}{\dd N} &= \lambda_\VF\frac{\sqrt{6}}{2}\frac{x z v}{H} + \lambda_\VF \left( \frac{\dot{\sigma}}{H} - \frac{1}{2} \right) x^2 v, \\
    \frac{\dd x}{\dd N} &= \sqrt{6}\frac{z}{H} + x\left( 2\frac{\dot{\sigma}}{H} - 1 \right), \\ 
    \frac{\dd \dot{\sigma}}{\dd N} &= -3 \dot{\sigma} + 3\frac{z^2}{H} - \lambda_\VF \frac{x^2 v^2}{H}. 
\end{align}
The Hubble parameter is then expressed as: 
\begin{equation}
    H^{2} = \frac{1}{3\mpl^{2}}\left(\rho_{m}+ \rho_{r} \right) + z^{2} + v^{2} + \dot{\sigma}^{2}.
\end{equation}

For $\lambda_\VF = 0$, the potential of the vector field becomes flat, and the model reaches the $\Lambda$CDM limit.
%%%%%%%%%%%%%%%%%%%%%%%%%%%%%
\subsection{2-form field} 
%%%%%%%%%%%%%%%%%%%%%%%%%%%%%

As analyzed in Ref.~\cite{Orjuela-Quintana:2022jrg}, a 2-form field alone can sustain an anisotropic accelerated expansion phase. That study demonstrated that when the 2-form field is coupled to cold dark matter, the strength of this coupling influences the amount of observable anisotropy today. We refer to this model as the \textit{anisotropic 2-form coupled to CDM} (2F) model. In this paper, we provide a brief review of the key results obtained in that work, aiming to compare them with other models, particularly the FQ model described earlier.

The model is described by the following action:
\begin{equation}
\mc{L}_\text{DE} = - \frac{1}{12} H_{\mu\nu\alpha}H^{\mu\nu\alpha} - V_\twoF \left( X_\twoF \right) + f_\twoF(X_\twoF)\mc{L}_c.
\end{equation}
Here, $V_\twoF(X_\twoF)$ is the potential associated with the 2-form $B_{\mu\nu}$, where $X_\twoF \equiv B_{\mu\nu}B^{\mu\nu}$. The function $f_\twoF(X_\twoF)$ represents the coupling between cold dark matter and the 2-form field. The 2-form field is assumed to have the configuration given in Eq.~\eqref{Eq: 2form Profile}, and both the potential and the coupling function are taken as exponential functions: 
\begin{align}
    V_\twoF(X_\twoF) &\equiv V_b e^{-\lambda_\twoF \left( \frac{X_\twoF}{\mpl^2}\ \right)}, \\
    f_\twoF(X_\twoF) &\equiv f_b e^{\mu_\twoF \left( \frac{X_\twoF}{\mpl^2}\ \right)},    
\end{align}
with $\lambda_\twoF$, $\mu_\twoF$, $V_b$ , and $f_b$ constants.

Following the procedure outlined earlier, the profile of the 2-form field compatible with the symmetries of the metric in Eq.~\eqref{Eq: Bianchi I} is given in Eq.~\eqref{Eq: 2form Profile}. However, we replace $v_B(t)$ by $\varphi(t)$ in order to keep notations apart. Then, we introduce the following dimensionful dynamical variables: 
\begin{equation}
    x \equiv \frac{\varphi e^{-2\sigma}}{\mpl a^2}, \quad z \equiv \frac{1}{\sqrt{6}\mpl}\frac{\dot{\varphi}e^{-2\sigma}}{a^2}, \quad v^2 \equiv \frac{V_\twoF}{3\mpl^2}.
\end{equation}
The system of differential equations governing these variables, along with the anisotropic stress $\dot{\sigma}$, is derived using Eqs.~\ref{Eq: field_equation_B} and \eqref{Eq: field_equation_sigma_B} in Appendix~\ref{App: Sources}. We obtain:
\begin{align}
    \frac{\dd z}{\dd N} &= \left(-1 + 2\frac{\dot{\sigma}}{H} \right)z + 2\sqrt{6}\left(\lambda_\twoF v^2 -\mu_\twoF \tilde{\rho}_c \right)\frac{x}{H}, \\
    \frac{\dd v}{\dd N} &= - 2\lambda_\twoF \left\{\sqrt{6}\frac{z}{H} - 2 x\left(1 + \frac{\dot{\sigma}}{H} \right)\right\}v x , \\
    \frac{\dd x}{\dd N} &= \sqrt{6}\frac{z}{H} - 2 x \left(1 + \frac{\dot{\sigma}}{H} \right), \\
    \frac{\dd \dot{\sigma}}{\dd N} &= -3\dot{\sigma} -2\frac{z^2}{H} - 4\left(\lambda_\twoF v^2 - \mu_\twoF \tilde{\rho}_c \right)\frac{x^2}{H}.
\end{align}
The coupled CDM density is given by:
\begin{equation}
    \tilde{\rho}_c \equiv \rho_\mathrm{c, 0} e^{2\mu_\twoF x^2}e^{-3N},
\end{equation}
where $\rho_\mathrm{c, 0}$ is the present-day value of $\rho_c$. Note that the constant $f_b$ from the coupling function $f_\twoF(X_\twoF)$ can be absorbed into $\rho_\mathrm{c, 0}$. 

Thus, the first Friedmann equation in terms of the new variables is expressed as:
\begin{equation}
    H^{2} = \frac{1}{3\mpl^2}\left(\rho_b + \rho_r \right)+ \tilde{\rho}_c + z^2 + v^2 + \dot{\sigma}^2.
\end{equation}

Note that, as in the 2-form coupled to scalar field model, the $\Lambda$CDM limit is reached when $\lambda_\twoF = f_\twoF = 0$, in which case the potential reduces to a cosmological constant and there is no interaction between CDM and the 2-form field.

%%%%%%%%%%%%%%%%%%%%%%%%
\subsection{Summary}
%%%%%%%%%%%%%%%%%%%%%%%%

In this section, we have introduced various sources of anisotropy in the late-time accelerated expansion of the Universe at background level. Since scalar fields dominate theoretical studies, we have considered scenarios where a homogeneous and isotropic scalar field is coupled to a field that breaks spatial rotational invariance. Specifically, we have examined two models: one in which a scalar field is coupled to a vector field and another where it is coupled to a 2-form field.

Additionally, scalar fields alone can sustain anisotropic expansion under certain conditions. We have introduced two homogeneous but anisotropic models: the first involves a specific configuration of three inhomogeneous scalar fields, known as a ``solid'', while the second considers a scalar field with an internal SO(3) symmetry, allowing it to interact with a gauge vector field.

Furthermore, fields that explicitly break isotropy can independently drive an anisotropic accelerated expansion. Therefore, we have also examined two models in which either a vector field or a 2-form field alone is responsible for this expansion.

It is generally believed that any anisotropy in the Universe should be small, as suggested by cosmological observations. Our goal is to critically assess this assumption by constraining the parameter space of various models that introduce late-time sources of anisotropy. In the following section, we explore this issue in greater detail.

%%%%%%%%%%%%%%%%%%%%%%%%%%%%%%%%%%%%%%%%%%
\section{Cosmological Constraints}
\label{Sec: Results}
%%%%%%%%%%%%%%%%%%%%%%%%%%%%%%%%%%%%%%%%%%

As anticipated from the discussion in the previous section, the evolution of anisotropy generally depends on the nature of the source that breaks spatial rotational invariance. Consequently, the predicted shear at present varies across different models. The primary objective of this work is to constrain the level of present-day anisotropy predicted by various theoretical frameworks, using recent observational data.

To achieve this, we adopt a two-step approach. First, at the background level, the Hubble parameter, $H$, serves as the fundamental quantity characterizing cosmological observations. Thus, we begin by determining the evolution of $H$ by solving the system of differential equations governing the dynamics of each model. Second, we explore the parameter space of each model to identify the optimal parameter values that best fit the observational datasets. Using these best-fit parameters and their associated confidence intervals, we derive constraints on the allowed level of anisotropy at present.

In the following section, we provide a detailed account of this methodology.

%%%%%%%%%%%%%%%%%%%%%%%%%%%%%%%%%%%%%%%%%%%%%%%%%%%
\subsection{Boltzmann Solver Implementation}
%%%%%%%%%%%%%%%%%%%%%%%%%%%%%%%%%%%%%%%%%%%%%%%%%%%

The key quantity describing cosmological observables at the background level is the Hubble rate, $H$. To obtain $H$ as a function of time for a given model, one must solve its dynamics, which are encoded in a system of first-order differential equations. In general, this task is performed numerically. Consequently, Boltzmann solvers play a crucial role, as they are optimized for handling the types of differential equation systems typically encountered in theoretical cosmology. The most widely used solvers are \texttt{CLASS}~\cite{Lesgourgues:2011re,Blas:2011rf} and \texttt{CAMB}~\cite{Lewis:1999bs}. Here, leveraging the user-friendly structure of \texttt{CLASS}, we implement the equations of motion for our models within this framework.

Several challenges arise when attempting to numerically compute $H$ in \texttt{CLASS}. One of the most significant is determining suitable initial conditions for the fields driving the dark energy mechanism, as well as identifying parameter values that yield a viable cosmology—i.e., one that aligns with current observational constraints on cosmic fluid abundances at present, such as $\Omega_{c, 0} \sim 0.25$ for CDM, $\Omega_{b, 0} \sim 0.05$ for baryons, $\Omega_{r, 0} \sim 10^{-4}$ for radiation, and $H_0 \sim 70 \ \text{km}/(s \ \text{Mpc})$ for the Hubble constant. Ensuring consistency with these observational benchmarks requires a careful selection of initial conditions for the dark energy fields. To address this, we utilize a \textit{shooting} algorithm already implemented in \texttt{CLASS}, which systematically determines appropriate initial conditions as functions of the present-day cosmological parameters. This approach effectively reinterprets the initial conditions as derived parameters in terms of their present-day values.

However, this method is not without its challenges. A key limitation is that suitable initial conditions may not exist for arbitrarily large regions of a model’s parameter space. For instance, in the case of the SDE model, if the shooting routine is configured to match $H_{0} = 70 \ \text{km} / (s \ \text{Mpc})$, the algorithm may fail to find valid initial conditions for $f_1$ and $f_2$ for certain values of the parameters $n$ and $m$. Consequently, the numerically obtained $H$ at present may deviate from the target value. Given that each model presents its own numerical challenges, we will provide specific details regarding their implementation in \texttt{CLASS} in the following sections. 

%%%%%%%%%%%%%%%%%%%%%%%%%%%%%%%%%%%%%%%%%%%%%%%%%%%
\subsection{Parameter Space Sampler}
%%%%%%%%%%%%%%%%%%%%%%%%%%%%%%%%%%%%%%%%%%%%%%%%%%%

After incorporating our model into \texttt{CLASS}, we proceed to explore its parameter space using the Metropolis-Hastings algorithm, a widely employed Markov Chain Monte Carlo (MCMC) method. For this purpose, we utilize the publicly available \texttt{MontePython} code \cite{Brinckmann:2018cvx}, which is specifically designed for cosmological parameter inference.

Once the time evolution of $H$ is obtained from a given model, we can compare its predictions against observational data, such as the luminosity distance derived from Sn Ia measurements. However, since the expansion rate depends on the model parameters, different parameter sets lead to different cosmological evolutions. To determine the best-fit parameters, we must efficiently explore a dense region of the parameter space. The MCMC approach provides a robust statistical framework for this task, allowing us to estimate both the best-fit parameters and their associated confidence intervals.

\texttt{MontePython} is particularly useful for this purpose, as it operates in conjunction with \texttt{CLASS}. The process follows an iterative scheme: \texttt{MontePython} supplies an initial set of model parameters to \texttt{CLASS}, such as $\{\Omega_{c, 0}, \Omega_{b,0}, \Omega_{r, 0}, H_0\}$. \texttt{CLASS} then solves the cosmological dynamics and computes the corresponding observables. These theoretical predictions are subsequently compared with the observational dataset, and \texttt{MontePython} uses this information to propose a new set of parameters. This process continues until the predefined maximum number of steps is reached. The resulting MCMC chains encapsulate the posterior probability distributions of the model parameters, allowing for an efficient exploration of the multidimensional parameter space.

Our goal is to leverage the combination of \texttt{MontePython} and \texttt{CLASS} to impose constraints on the parameters of each model, ensuring consistency with recent and precise cosmological observations. In the next section, we will detail the observational datasets employed in our analysis.

%%%%%%%%%%%%%%%%%%%%%%%%%%%%%%%%%%%
\subsection{Observational Data}
%%%%%%%%%%%%%%%%%%%%%%%%%%%%%%%%%%%

In this work, we will use cosmological datasets related to observable quantities at the background level only. Before discussing each dataset in detail, we first clarify an important aspect regarding the use of cosmological observations in the presence of anisotropic expansion. In general, anisotropic expansion leads to direction-dependent observations. However, at the background level, all observational quantities, such as the luminosity distance, depend on the Hubble rate $H$. In our case, $H$ is modified only by the inclusion of a homogeneous term proportional to $\dot{\sigma}^2$. This allows us to use standard cosmological datasets to constrain the present-day shear by treating anisotropy as a ``correction term'' to $H$. 

A similar approach is taken when analyzing non-flat cosmologies. In such cases, observational datasets are not explicitly modified to account for curvature, as it enters the Friedman equations as an additional homogeneous term. This allows constraints on curvature to be derived using standard datasets~\cite{Denissenya:2018zcv,Handley:2019tkm,Ryan:2019uor}. Despite this analogy, we acknowledge that a thorough assessment of the validity of this approach is necessary. Specifically, the potential impact of anisotropic redshift effects on observations remains an open question~\cite{Appleby:2014kea,Sorrenti:2022zat,Bengaly:2015dza}. While we provide an estimation of how a small anisotropy affects the luminosity distance in Appendix~\ref{App: Luminosity Distance}, a more comprehensive investigation of this aspect is left for future research, in line as the analysis in Ref.~\cite{Verma:2024lex} or in Ref.~\cite{BeltranAlmeida:2021ywl}.

With this clarification in place, we now proceed to describe the datasets used in our cosmological parameter inference approach. \\

%%%%%%%%%%%%%%%%%%%%%%%%%%%%%%%%%%%%%%%%%%%%
$\bullet$ \textbf{CMB:} 
%%%%%%%%%%%%%%%%%%%%%%%%%%%%%%%%%%%%%%%%%%%%
At the background level, the CMB provides crucial insights into the universe’s evolution. In particular, the CMB constrains cosmological models through the position of the first peak in its temperature power spectrum and its corresponding angular scale, typically analyzed within the framework of the standard flat $\Lambda$CDM model. This information is encapsulated in the CMB shift parameters, $R$ and $l_a$, defined as \cite{Wang:2013mha}:
\begin{equation}
    R \equiv \frac{r(z_{\star})}{c}\sqrt{\Omega_{m, 0} H_{0}^{2}}, \quad 
    l_{a} \equiv \pi \frac{r(z_\star)}{r_{s}(z_\star)}.
\end{equation}
Here, $c$ is the speed of light in vacuum, and $r_s$ is the comoving sound horizon at decoupling, given by:
\begin{equation}
    r_s(z_\star) \equiv \int_{z_\star}^\infty \dd z \left(\frac{c_s}{H(z)}\right), 
\end{equation}
where $z_{\star}$ is the redshift at photon decoupling, i.e., the epoch when the CMB was emitted. The sound speed in the baryon-photon fluid, $c_s$, is given by:
\begin{equation}
    c_s = \frac{c}{\sqrt{3}}\left(1 + \frac{3\rho_b}{4\rho_\gamma} \right)^{-1/2},
\end{equation}
where $\rho_\gamma$ is the photon energy density. The comoving distance to decoupling is:
\begin{equation}
    r(z_\star) \equiv \int_0^{z_\star} \dd z \left( \frac{c}{H(z)} \right).
\end{equation}
The values of these parameters are estimated using the Planck satellite measurements\cite{Planck:2018vyg}, where the effective number of relativistic species, $N_\eff$, is included in the computation of $\rho_\gamma$.

It is worth noting that the CMB shift parameters are derived under the assumption of the $\Lambda$CDM model, which may lead to systematic biases when applied to non-$\Lambda$CDM scenarios. To mitigate this potential source of bias, we complement our study with a constraint analysis that does not rely on CMB data. Further details regarding this consideration are provided at the end of this subsection.  \\ 

%%%%%%%%%%%%%%%%%%%%%%%%%%%%%%%%%%%%%%%%%%%%
$\bullet$ \textbf{BAO:}
%%%%%%%%%%%%%%%%%%%%%%%%%%%%%%%%%%%%%%%%%%%%
The interaction between photons and baryons in the early universe generates periodic fluctuations in the density of matter, known as baryon acoustic oscillations (BAO). These oscillations imprint a characteristic scale on the large-scale structure of the universe, serving as a cosmic standard ruler. Specifically, BAO measurements constrain the ratio $D_{V}/r_{d}$, where $D_{V}$ is the volume-averaged distance, and $r_{d} \equiv r(z_d)$ is the comoving sound horizon at the drag epoch, $z_d$, which marks the moment when baryons decoupled from photons and began to gravitationally collapse, driving structure formation.

The volume-averaged distance, $D_{V}(z)$, is defined as:
\begin{equation}
    D_{V}(z) = \left[(1+z)^{2} \, d_{A}^{2}(z) \, \frac{cz}{H(z)}\right]^{1/3},
\end{equation}
where the angular diameter distance, $d_{A}(z)$, is related to the comoving distance $r(z)$ by:
\begin{equation}
    d_A(z) \equiv \frac{r(z)}{1 + z}.
\end{equation}
BAO measurements play a crucial role in constraining cosmological models, particularly those involving dark energy and deviations from the standard $\Lambda$CDM paradigm. Since BAO data provides independent geometric constraints on cosmic expansion, it is especially valuable when combined with other probes, such as Sn Ia and CMB data, to break degeneracies between different dark energy models. In this work, we employ BAO data from the 6dF Galaxy Survey \cite{2011MNRAS.416.3017B} and SDSS DR7 MGS \cite{Ross:2014qpa} to test anisotropic dark energy scenarios, assessing their viability against observational constraints. \\

%%%%%%%%%%%%%%%%%%%%%%%%%%%%%%%%%%%%%%
$\bullet$ \textbf{H(z):} 
%%%%%%%%%%%%%%%%%%%%%%%%%%%%%%%%%%%%%%
The Hubble rate can be computed in terms of the redshift as:
\begin{equation}
    H(z) = -(1 +z)\frac{\dd z}{\dd t}.
\end{equation}
This relation implies that by measuring the redshift of an astronomical object and tracking its time evolution, we can infer the expansion rate of the universe at different epochs. This principle underlies the \textit{cosmic chronometers} method, which estimates the \textit{differential age} of passively evolving galaxies to determine $H(z)$ independently of any cosmological model.

Additionally, direct measurements of $H(z)$ can be obtained through baryon acoustic oscillation (BAO) observations. In particular, the BAO peak in the radial direction, traced by galaxy or quasar clustering, provides a geometric probe of the Hubble parameter by effectively measuring the expansion rate along the line of sight. The combination of these two techniques offers a robust dataset of $H(z)$ values at various redshifts, enabling precise constraints on cosmic expansion history. In this work, we employ $H(z)$ measurements from both cosmic chronometers and BAO studies \cite{Guo:2015gpa, Moresco:2016mzx} to test the consistency of anisotropic dark energy models with observational data. \\

%%%%%%%%%%%%%%%%%%%%%%%%%%%%%%%%%%%%%%%%%%%%
$\bullet$ \textbf{SnIa:} 
%%%%%%%%%%%%%%%%%%%%%%%%%%%%%%%%%%%%%%%%%%%%
Type Ia supernovae serve as standardizable candles, providing key data on the luminosity distance through measurements of the distance modulus, $\mu_0$, defined as:
\begin{equation}        
    \mu_{0} \equiv m - M = 5 \log_{10}\left(\frac{d_{L}}{\text{Mpc}}\right) + 25,
\end{equation}
where $m$ is the apparent magnitude of a Sn Ia, and $M$ is its absolute magnitude. The luminosity distance, $d_L$, is given by: 
\begin{equation}
    d_{L}(z) = (1 + z) \int_{0}^{z} \frac{\dd \tilde{z}}{H(\tilde{z})}.
\end{equation}
This quantity provides a direct observational probe of cosmic expansion, making Sn Ia one of the most powerful tools for constraining dark energy models.

In this work, we employ the Pantheon+ dataset~\cite{Brout:2022vxf}, which consists of 1701 light curves from 1550 distinct Sn Ia spanning redshifts $z \in [0.001, 2.26]$. This dataset benefits from an improved calibration through the SH0ES Cepheid host distance anchors, refining constraints on the expansion history of the universe. Given the extensive redshift coverage and systematic corrections incorporated in Pantheon+, it provides a robust basis for testing the effects of anisotropic dark energy on the cosmic distance ladder.

To conclude this subsection, we emphasize that the presence of dark energy influences key properties of the CMB power spectrum, particularly the position of the first acoustic peak and its angular scale. These effects are encapsulated in the shift parameters $R$ and $l_a$. However, the ``observational'' values of these parameters are derived from Planck's measurements, which assume the $\Lambda$CDM model as the underlying cosmology. Consequently, incorporating this dataset into our analysis may introduce an unintended bias due to the model-dependent nature of these estimations.

To mitigate this potential bias and assess the robustness of our results, we will perform the parameter inference analysis using the full set of cosmological datasets considered in this work—CMB, BAO, H(z), and Sn Ia—and an analysis that excludes the CMB data as well. This complementary approach, entirely free from any assumptions derived from CMB-based constraints, aims to provide a more unbiased perspective on the viability of anisotropic dark energy models.

%%%%%%%%%%%%%%%%%%%%%%%%%%%%%%%%%%%%%%%%%%%%%%%%%
\subsection{Constraints on the Parameter Space}
%%%%%%%%%%%%%%%%%%%%%%%%%%%%%%%%%%%%%%%%%%%%%%%%%

After implementing our models in \texttt{CLASS}, we explore their parameter space using the Metropolis-Hastings algorithm, a widely employed MCMC method, facilitated by the \texttt{MontePython} code~\cite{Brinckmann:2018cvx}. The resulting MCMC chains encapsulate the posterior probability distributions of the model parameters, which we analyze using \texttt{MontePython}'s built-in statistical tools. This approach enables efficient sampling of the multidimensional parameter spaces, which, for our analysis, are encoded in parameter vectors of the form:
\begin{equation}
    \mathcal{P} \equiv \{\omega_b, \Omega_{c,0}, h, N_\eff\} + \{ \text{DE~parameters} \},
\end{equation}
where $h$ is the reduced Hubble parameter, defined via $H_0 = 100 h \, \text{km} \, s^{-1} \, \text{Mpc}^{-1}$, and $\omega_{b} \equiv \Omega_{b, 0} h^{2}$ is the reduced baryon density parameter. 

The MCMC analysis yields 2D posterior probability contours displaying the 68.3\% ($1\sigma$) and 95.4\% ($2\sigma$) confidence levels for all parameters across the six different models under consideration. The constrained parameters include the ``standard cosmological parameters''—$\omega_{b}$, $\Omega_{c,0}$, $h$, and $N_\mathrm{eff}$—alongside model-specific parameters for each dark energy scenario. For example, the scalar field coupled to a vector field model is characterized by the parameters $\lambda_\AQ$ and $\mu_\AQ$, while the anisotropic solid dark energy model includes $n$ and $m$. Additionally, we include the present-day shear parameter, $\Sigma_0$, and the dark energy density parameter, $\Omega_{\mathrm{DE}}$, as derived quantities—meaning they are indirectly constrained through the primary model parameters. We also include the absolute magnitude of Type Ia supernovae, $M$, since it is directly fitted in the analysis using the Pantheon+ dataset.

The standard cosmological parameters are sampled within the following flat priors: $100\omega_b \in [2.0, 2.5]$, $\Omega_c \in [0.1, 1.0]$, $h \in [0.1, 1.0]$, and $N_\text{eff} \in [1.0, 5.0]$. The priors for each model-specific parameter are chosen individually. In the following sections, we present the results of our MCMC analysis for each model, discussing the inferred parameter constraints and their implications. Further details about the MCMC results are given in Appendix~\ref{App: Full Posteriors}. \\

%%%%%%%%%%%%%%%%%%%%%%%%%%%%%%%%%%%%%%%%%%%%%%%%%
\subsubsection{$\mathbf{CMB+BAO+H(z)+SnIa}$}
%%%%%%%%%%%%%%%%%%%%%%%%%%%%%%%%%%%%%%%%%%%%%%%%%

%%%%%%%%%%%%%%%% Anisotropic Quintessence %%%%%%%%%%%%%%%%%%%%%%%%%%
$\bullet$ \textbf{Scalar field coupled to a vector field (AQ):} 
%%%%%%%%%%%%%%%% Anisotropic Quintessence %%%%%%%%%%%%%%%%%%%%%%%%%%

The AQ model introduces two additional parameters, $\lambda_\AQ$ and $\mu_\AQ$. To perform the MCMC analysis, we impose prior constraints on these parameters based on the dynamical system analysis presented in Refs.~\cite{Thorsrud:2012mu, Orjuela-Quintana:2021zoe}. Specifically, we restrict $\lambda_\AQ^2 < 2$ to ensure the existence of an anisotropic accelerated attractor. For $\mu_\AQ$, we adopt the range $0 < \mu_\AQ < 100$, as no strong constraints arise from the dynamical system itself—only that $\mu_\AQ$ remains positive. This choice accommodates both weak and strong coupling scenarios between the scalar and vector fields.

\begin{figure}[t!]
\centering
\includegraphics[width=\linewidth]{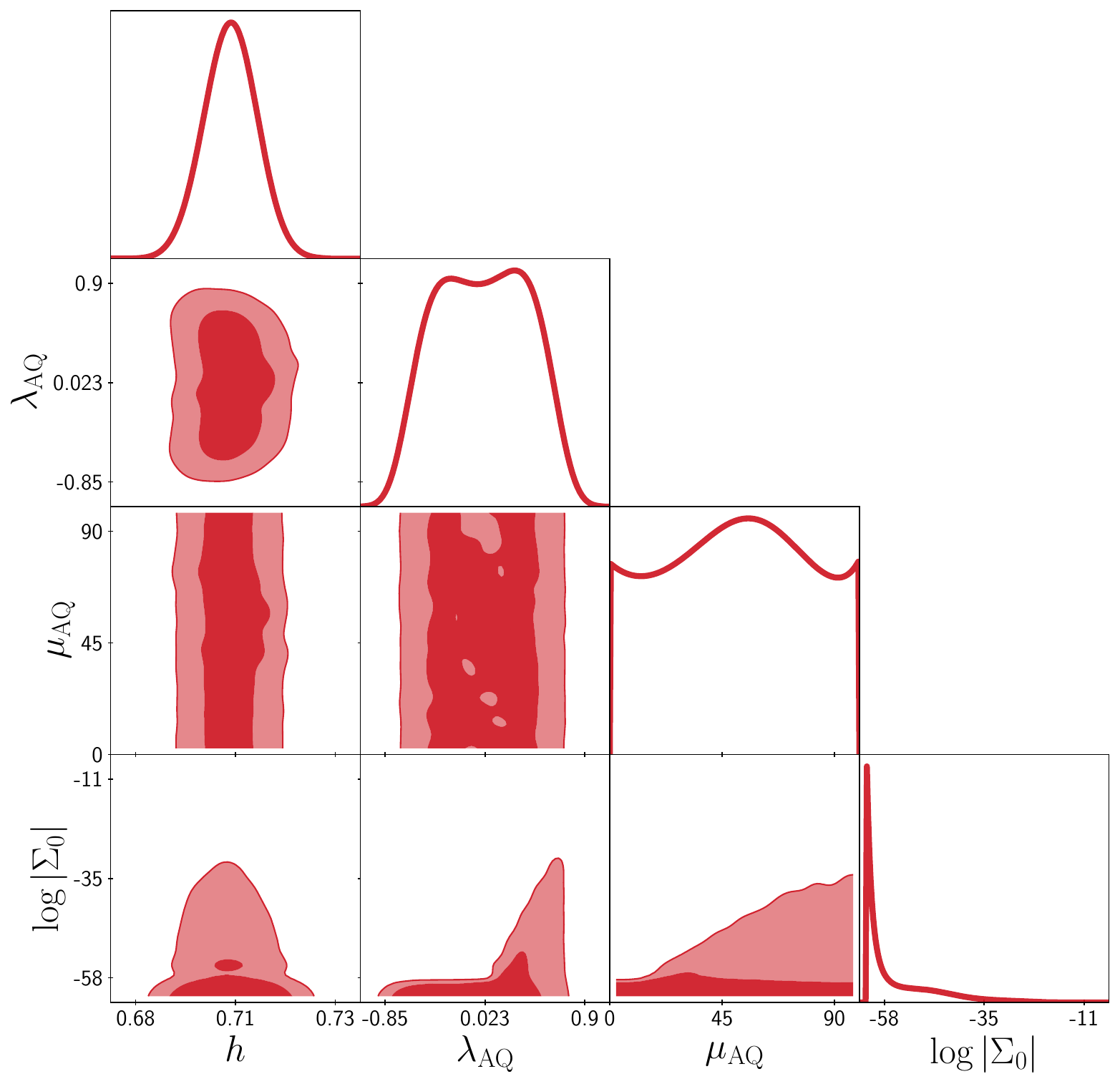}
\caption{Confidence contours for the parameters $\{h, \lambda_{\AQ}, \mu_{\AQ}, \Sigma_0\}$ of the scalar field coupled to a vector field model at the $1\sigma$ and $2\sigma$ confidence levels. The scalar field potential slope, $\lambda_\AQ$, is preferred to be small, while the coupling strength $\mu_\AQ$ between the scalar and vector fields remains unconstrained by the datasets. The mean value of the present-day shear is $|\Sigma_0| \sim 10^{-27}$, and thus negligible; the anisotropic is displayed in logarithmic scale given its large dispersion.}
\label{Fig: Contours_anq}
\end{figure}

Figure~\ref{Fig: Contours_anq} presents the 2D posterior distributions of the AQ parameters. Our MCMC analysis indicates that $\mu_\AQ$ remains unconstrained given the available datasets, while $\lambda_\AQ$ exhibits a preference for small values ($|\lambda_\AQ| < 1$). Additionally, we report the mean values of the present-day anisotropic stress and the reduced Hubble parameter as $|\Sigma_{0}| = \left(9.67^{+1.09}_{-4.19}\right)\times10^{-27}$ and $h = 0.7041_{-0.0073}^{+0.0077}$, respectively. The complete set of parameter constraints are listed in Table~\ref{Table: means}. 

%%%%%%%%%%%%%%%% TABLE %%%%%%%%%%%%%%%%
\begin{center}
\begin{table*}
\begin{centering}
\begin{tblr}{|X[1.2, c] X[1.2, c] X[1.2, c] X[1.2, c] X[1.2,c] |}
\hline \hline
Param & $\Lambda$CDM & AQ & SDE & FQ  \\
\hline \hline
$\Omega_{c,0}$ & $0.2436_{-0.007}^{+0.0068}$ & $0.2432_{-0.0070}^{+0.0069}$ & $0.2437_{-0.0071}^{+0.0067}$ &  $0.2444_{-0.0071}^{+0.0068}$ \\
%%%%%%%%
$100\omega_{b}$ & $2.276_{-0.015}^{+0.015}$ & $2.279_{-0.016}^{+0.016}$ &
$2.28_{-0.015}^{+0.015}$ & $2.281_{-0.016}^{+0.016}$ \\
%%%%%%%%
$h$ & 
$0.7049_{-0.0075}^{+0.0074}$ & $0.7041_{-0.0073}^{+0.00742}$ & $0.7037_{-0.00753}^{+0.00741}$ & $0.7034_{-0.0075}^{+0.0075}$  \\
%%%%%%%%
$N_\text{eff}$ &
$3.369_{-0.12}^{+0.12}$ & 
$3.375_{-0.120}^{+0.130}$ &
$3.384_{-0.12}^{+0.12}$ & 
$3.38_{-0.12}^{+0.12}$ \\
%%%%%%%%
$\Omega_{\mathrm{DE},0}$ & $0.7105_{-0.0074}^{+0.0076}$ & $0.7107_{-0.0075}^{+0.0076}$ &  $0.7102_{-0.0073}^{+0.0077}$ & $0.7094_{-0.0074}^{+0.0077}$  \\
%%%%%%%%
$M$ &
$-19.35_{-0.021}^{+0.021}$ & $-19.35_{-0.021}^{+0.021}$&  $-19.35_{-0.021}^{+0.021}$ &
$-19.35_{-0.021}^{+0.021}$ \\
%%%%%%%%
$10^{4}|\Sigma_{0}|$ &
- & $ 9.67^{+1.09}_{-4.19}\times10^{-23}$ &  $|-6.269_{-23.9}^{+49.1}|$ &
$0.530_{-0.285}^{+2.37}$ \\
%%%%%%%%
$\mathrm{Param}_{1}$ & 
- & 
$0.0026_{-0.44}^{+0.44}$ & 
$0.039_{-0.039}^{+0.014}$ & 
$1.072_{-1.1}^{+0.14}$ \\
%%%%%%%%
$\mathrm{Param}_{2}$ & 
- &
$50.37$ & 
$0.0386_{-0.0386}^{+0.0139}$ & $32.45_{-6.0}^{+18.0}$ \\
\hline \hline
%%%%%%%%
\end{tblr}
\par\end{centering}
\caption{Mean values and corresponding $68.3\%$ confidence limits for the parameters of the $\Lambda$CDM model, the scalar field coupled to a vector field (AQ) model, the anisotropic solid dark energy (SDE) model, and the scalar field coupled to a 2-form (FQ) model, constrained by the combined dataset CMB + BAO + H(z) + Sn Ia. Here, $\text{Param}1$ and $\text{Param}2$ correspond to $\lambda_\AQ$ and $\mu_\AQ$ for the AQ model, $n$ and $m$ for the SDE model, and $\lambda_\FQ$ and $\mu_\FQ$ for the FQ model. Since no constraint is found for $\mu_\AQ$, as shown in Fig. \ref{Fig: Contours_anq}, only its mean value is reported without a standard deviation. Note that, in all cases, $N_\eff$ is significantly higher than the value 3.046 determined by Planck~\cite{Planck:2018vyg}. This is due to the degeneracy between $N_\eff$ and $H_0$, where large values of $H_0$ tend to push $N_\eff$ higher (see Ref.~\cite{Escudero:2018mvt}).}
\label{Table: means}
\end{table*}
\par\end{center}

%%%%%%%%%%%%%%%% TABLE %%%%%%%%%%%%%%%%
\begin{center}
\begin{table*}
\begin{centering}
\begin{tblr}{|X[1.2, c] X[1.2, c] X[1.2, c] X[1.2, c] X[1.2,c]|}
\hline \hline
Param & EYMH & VF & 2F & IQ  \\
\hline \hline
$\Omega_{c,0}$ & $0.2436_{-0.0071}^{+0.0068}$ & $0.2436_{-0.0071}^{+0.0068}$ & $0.244_{-0.007}^{+0.007}$ &  
$0.2436_{-0.0071}^{+0.0068}$\\
%%%%%%%%
$100\omega_{b}$ & $2.277_{-0.015}^{+0.015}$ & $2.276_{-0.015}^{+0.015}$ &
$2.28_{-0.01}^{+0.01}$ & 
$2.28_{-0.016}^{+0.015}$\\
%%%%%%%%
$h$ & 
$0.7047_{-0.0075}^{+0.0074}$ & $0.7049_{-0.0075}^{+0.0074}$ & $0.705_{-0.007}^{+0.007}$ & 
$0.7036_{-0.0076}^{+0.0074}$  \\
%%%%%%%%
$N_\text{eff}$ &
$3.372_{-0.12}^{+0.12}$ & 
$3.37_{-0.12}^{+0.12}$ &
$3.369_{-0.12}^{+0.12}$ &
$3.377_{-0.12}^{+0.12}$ \\
%%%%%%%%
$\Omega_{\mathrm{DE},0}$ & $0.7104_{-0.0074}^{+0.0077}$ & $0.7105_{-0.0074}^{+0.0077}$ &   $0.7105_{-0.0073}^{+0.0077}$ & $0.7103_{-0.0074}^{+0.0077}$ \\
%%%%%%%%
$M$ &
$-19.35_{-0.021}^{+0.021}$  &
$-19.35_{-0.021}^{+0.021}$ &  $-19.35_{-0.021}^{+0.021}$ &
$-19.35_{-0.021}^{+0.021}$ \\
%%%%%%%%
$\Sigma_{0}$ &
$8.744_{-0.041}^{+1851}\times 10^{-8}$  &
$8.53_{-2.08}^{+4.45}\times10^{-17}$ &  $-3.501_{-0.53}^{+3.5}\times 10^{-4}$ &
- \\
%%%%%%%%
\hline \hline
%%%%%%%%
\end{tblr}
\par\end{centering}
\caption{Mean values and corresponding $68.3\%$ confidence limits for the anisotropic Einstein-Yang-Mills-Higgs (EYMH) model, the anisotropic vector field (VF) model, and the anisotropic 2-form model coupled to CDM (2F), and the isotropic quintessence model constrained by the combined dataset CMB + BAO + H(z) + Sn Ia. The specific parameters introduced by these dark energy models are not included, as they remain unconstrained. It is worth noting that in the EYMH model, large difference between the upper and lower bounds are due merely to the large parameter range spanned by $\log |\Sigma_0|$ (see Fig.~\ref{Fig: Contours_eymh}).}
\label{Table: means_2}
\end{table*}
\par\end{center}

%%%%%%%%%%%%%%%% Anisotropic 2-form %%%%%%%%%%%%%%%%
$\bullet$ \textbf{Scalar field coupled to a 2-form (FQ):}
%%%%%%%%%%%%%%%% Anisotropic 2-form %%%%%%%%%%%%%%%%

This model introduces two additional parameters: $\lambda_\FQ$, which characterizes the slope of the scalar field potential, and $\mu_\FQ$, which quantifies the coupling strength between the scalar field and the 2-form field. The prior ranges are chosen as $0 < \lambda_\FQ < 20$ and $0 < \mu_\FQ < 50$, following the dynamical system analysis in Ref.~\cite{BeltranAlmeida:2019fou}, to ensure that the attractor solution corresponds to an anisotropic dark energy-dominated regime.

\begin{figure}[t!]
\centering
\includegraphics[width=\linewidth]{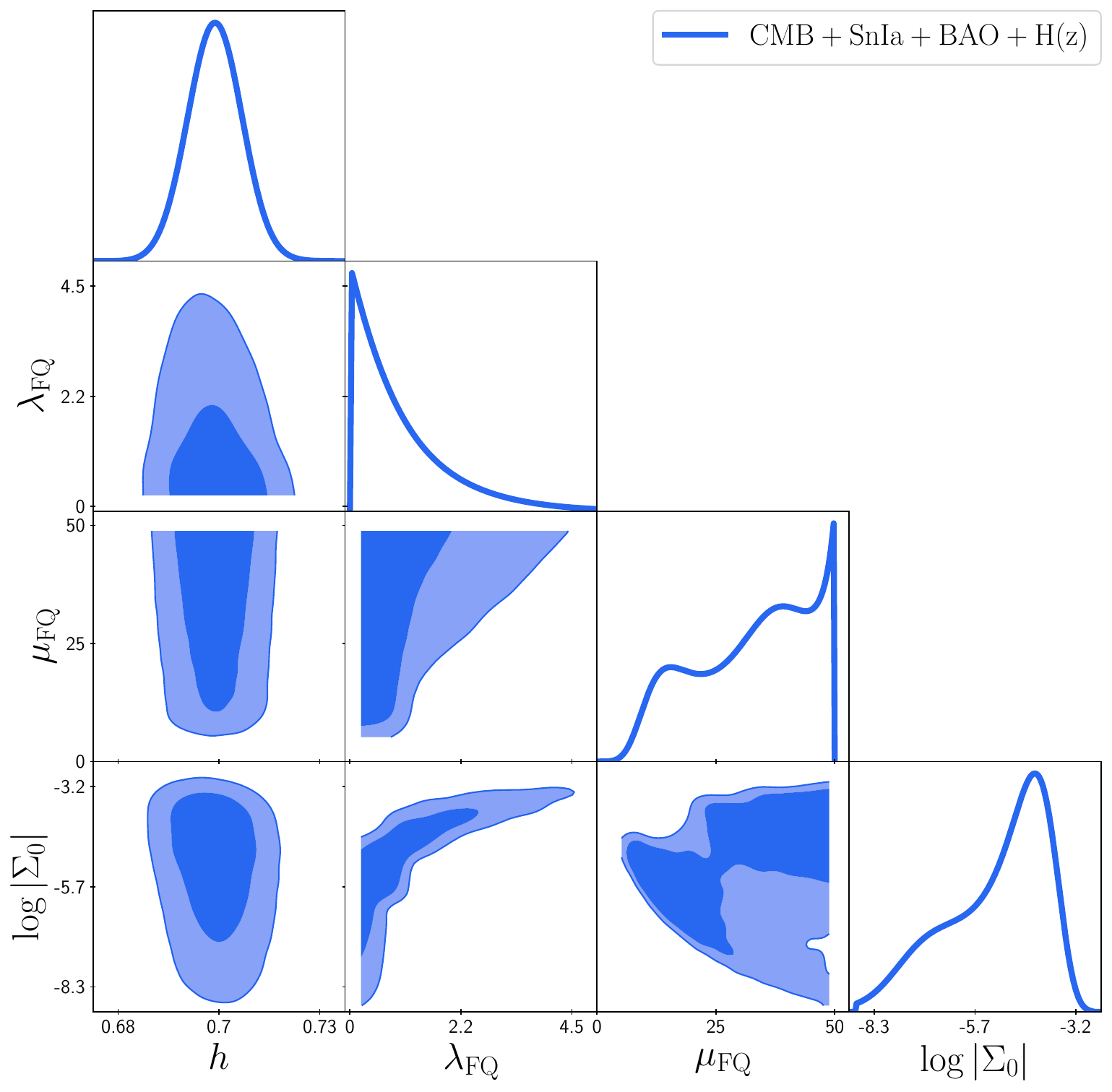}
\caption{Confidence contours for the parameters $\{h, \lambda_\FQ, \mu_\FQ, \Sigma_0\}$ of the anisotropic 2-form model at the $1\sigma$ and $2\sigma$ confidence levels. While both parameters, $\lambda_\FQ$ and $\mu_\FQ$, are constrained by the datasets, the obtained limits are not sufficiently stringent to precisely determine their values. The mean value of the present-day shear is $\Sigma_0 \sim 10^{-3}$, with the notable result that $\Sigma_0 = 0$ is excluded.}
\label{Fig: Contours_a2f}
\end{figure}

Figure~\ref{Fig: Contours_a2f} and Table~\ref{Table: means} present the results of the MCMC analysis for the FQ model. The inferred mean values for the model parameters are $\lambda_\FQ = 1.072_{-1.1}^{+0.14}$ and $\mu_\FQ = 32.45_{-6.0}^{+18.0}$. Additionally, the Hubble parameter is constrained to $h = 0.7034_{-0.0075}^{+0.0075}$. Notably, the mean value of the anisotropic stress is found to be $10^{4}|\Sigma_{0}| = 0.530_{-0.285}^{+2.37}$, indicating that $\Sigma_0 = 0$ is excluded at the $1\sigma$ confidence level. This result suggests that while the level of anisotropy remains small, it is non-negligible. \\

%%%%%%%%%%%%%%%% Anisotropic Solid Dark Energy %%%%%%%%%%%%%%%%
$\bullet$ \textbf{Inhomogeneous scalar fields (SDE):} 
%%%%%%%%%%%%%%%% Anisotropic Solid Dark Energy %%%%%%%%%%%%%%%%
The anisotropic solid model introduces two additional parameters, $n$ and $m$, which correspond to the exponents of the Lagragian functions $F^1$ and $F^2$. As shown in Ref.~\cite{Motoa-Manzano:2020mwe}, this model admits three anisotropic accelerated attractors, with their selection depending on the values of $n$ and $m$, which can range from small to large values. However, achieving both a small level of anisotropy and a dark energy equation of state close to $w_\text{DE} \approx -1$ requires $n$ and $m$ to remain small. Furthermore, we observed that the shooting method implemented in \texttt{CLASS} frequently fails for values $n, m \gtrsim 1$. To circumvent this issue, we empirically determined a viable parameter region where the shooting routine remains stable, leading to the chosen priors: $0 < \{n, m\} < 0.087$.

\begin{figure}[t!]
\centering
\includegraphics[width=\linewidth]{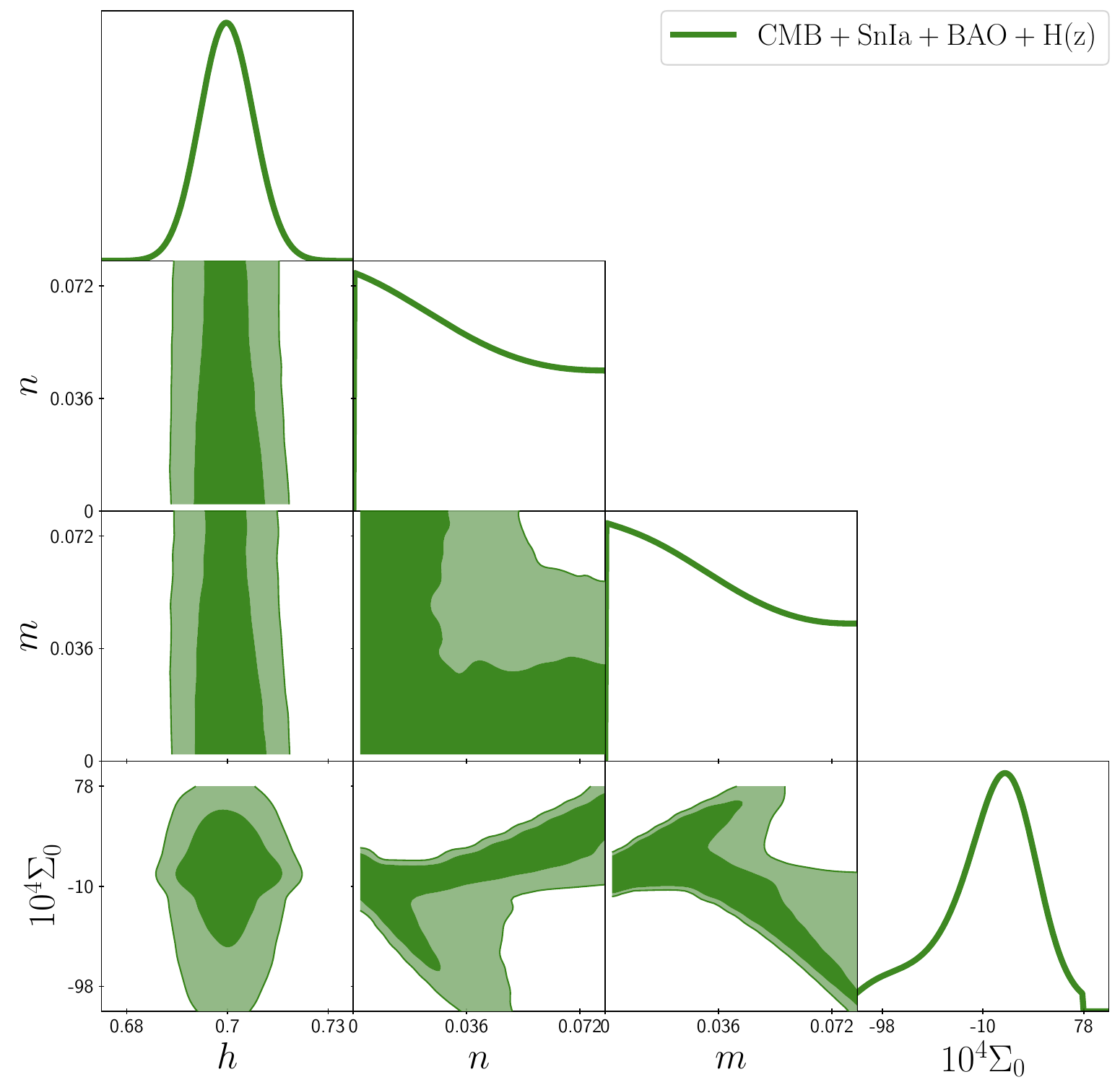}
\caption{Confidence contours for the parameters $\{h, n, m, \Sigma_0\}$ of the anisotropic solid model at $1\sigma$ and $2\sigma$ confidence levels. Since $\Sigma_{0}$ does not have a large dispersion, a logarithmic scale is not used. Although the mean value of the present-day shear is $\Sigma_0 \sim 10^{-4}$, a negligible shear is favored by the observations.}
\label{Fig: Contours_sde}
\end{figure}

Table~\ref{Table: means} presents the mean values of the parameters along with their associated uncertainties, while Fig.~\ref{Fig: Contours_sde} displays the corresponding 2D posterior distributions. The mean values obtained for the model parameters are $n = 0.03898_{-0.039}^{+0.014}$ and $m = 0.03864_{-0.039}^{+0.014}$. Additionally, the mean value of the Hubble parameter is found to be $h = 0.7037_{-0.00753}^{+0.00741}$. The present-day shear is constrained to $10^{4}\Sigma_{0} = -6.27_{-23.9}^{+49.1}$, indicating that a negligible level of anisotropy is consistent with the confidence contours. This result is expected, as discussed in Ref.~\cite{Motoa-Manzano:2020mwe}, since the condition $n = m$ leads to an isotropic accelerated expansion. From Table~\ref{Table: means}, it is evident that the data favor $n \approx m$ thereby supporting a scenario with negligible shear. \\

%%%%%%  Anisotropic Einstein Yang Mill Higgs %%%%%%%%%%%%%%%%%%%%
$\bullet$ \textbf{Scalar field with internal symmetry (EYMH):} 
%%%%%%  Anisotropic Einstein Yang Mill Higgs %%%%%%%%%%%%%%%%%%%%
The EYMH model introduces a single additional free parameter, $w_{i}$, which is related to the initial condition of the variable $w$ through $w(a_{i}) = H(a_{i})w_{i}$, where $a_{i}$ denotes the initial scale factor. The parameter $w_{i}$ is also linked to the gauge field charge $\tilde{g}$ and serves as an indicator of the interaction strength between the Higgs field and the gauge field. The second parameter of the model, $\alpha$, is fixed at $\alpha = 1$, in accordance with the results derived from the Standard Model of particle physics~\cite{kane2017modern}.

Similar to the anisotropic solid model, the shooting method fails for arbitrary values of $w_i$, though it is known that $w_i$ should remain small. To ensure the numerical stability of the shooting routine, the prior range for this parameter is chosen as $-40 < \log(w_{i}) < -28.92$ through an iterative trial-and-error approach. This selection guarantees that within the specified interval, suitable initial conditions can always be found to satisfy the target value required for the shooting method, as discussed at the beginning of this section.

\begin{figure}[t!]
\centering
\includegraphics[width=\linewidth]{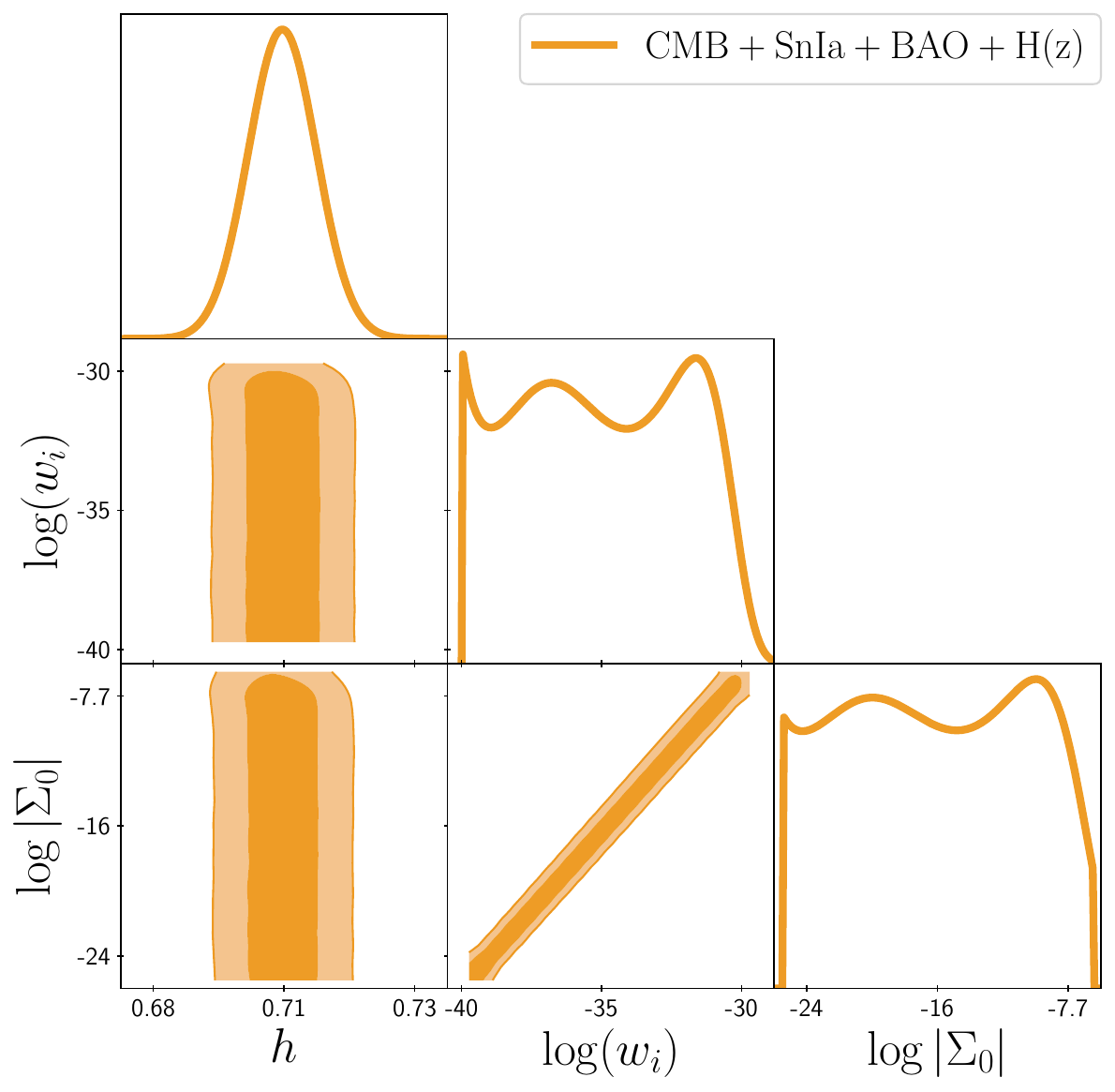}
\caption{Confidence contours for the parameters $\{h, \log (w_i), \log |\Sigma_0|\}$ of the anisotropic EYMH model at the $1\sigma$ and $2\sigma$ confidence levels. The parameter $w_i$ which characterizes the coupling strength between the Higgs field and the gauge vector field, remains unconstrained by the observational datasets. Additionally, the mean value of the present-day shear is $\Sigma_0 \sim 10^{-8}$.}
\label{Fig: Contours_eymh}
\end{figure}

Despite this setup, the parameter $w_{i}$, remains unconstrained by the observational data employed in this analysis, as illustrated in Fig.~\ref{Fig: Contours_eymh}. The constraints on all model parameters are summarized in Table~\ref{Table: means_2}. The mean value of the Hubble parameter is found to be $h = 0.7047_{-0.00749}^{+0.00739}$. Additionally, the present-day shear remains practically unconstrained, exhibiting a large dispersion around its mean value. This behavior arises because the parameter $w_{i}$ , which governs the anisotropy, is itself unconstrained. \\ %Consequently, the anisotropy, as a derived quantity, can take any value within the prior range of $w_{i}$

\begin{figure}[t!]
\centering
\includegraphics[width=\linewidth]{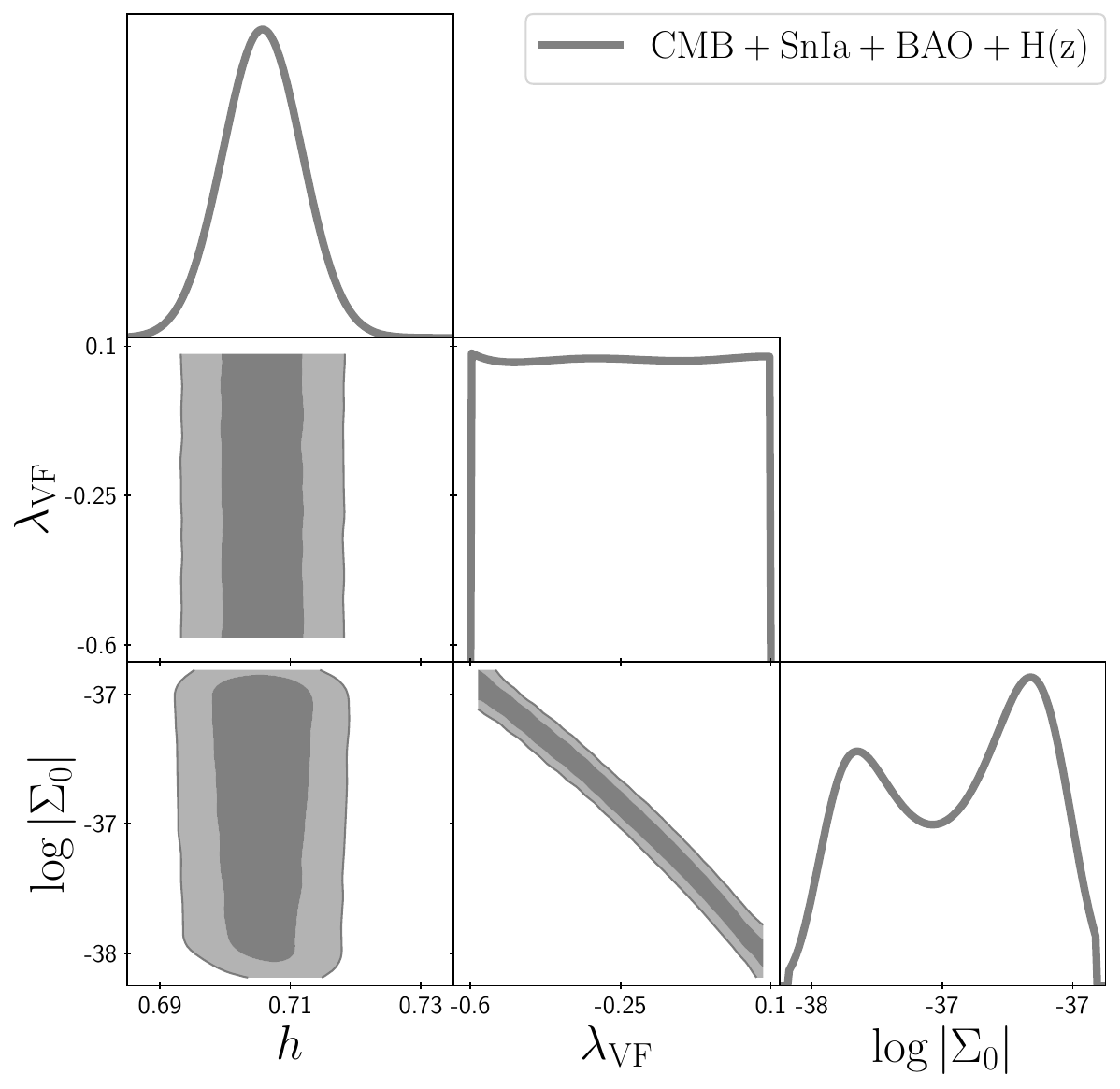}
\caption{Confidence contours for the parameters $\{h, \lambda_{\mathrm{VF}}, \Sigma_0\}$ of the anisotropic VF model at the $1\sigma$ and $2\sigma$ confidence levels. The parameter $\lambda_{\mathrm{VF}}$ remains unconstrained by the observational datasets. The mean value of the present-day shear is $10^{17}|\Sigma_{0}| = 8.53_{-2.08}^{+4.45}$.}
\label{Fig: Contours_avf}
\end{figure}

%%%%%%%%%%%%%%%%%%%%%%%%%%%%%%%%%%%%%%%%%
$\bullet$ \textbf{Vector field (VF):} 
%%%%%%%%%%%%%%%%%%%%%%%%%%%%%%%%%%%%%%%%%
Table~\ref{Table: means_2} also presents the results for the vector field (VF) model, in which a homogeneous vector field alone drives anisotropic accelerated expansion and Fig. \ref{Fig: Contours_avf} shows the 2D posterior distribution. In this scenario, the model parameter $\lambda_\VF$ remains unconstrained, while the mean value of the anisotropic stress is found to be $10^{17}|\Sigma_{0}| = 8.53^{+4.45}_{-2.08}$. This result suggests that the observational data favor $\Sigma_{0} \approx 0$ for the VF model, indicating that a homogeneous vector field alone is insufficient to generate an appreciable shear. This finding challenges the conventional assumption that vector field dynamics necessarily induce significant anisotropy in the cosmic expansion. In contrast, as observed in the coupled quintessence model—where the scalar field interacts with a vector field—anisotropy remains small. In the VF model, however, anisotropy is entirely negligible. \\

%%%%%%%%%%%%%%%%%%%%%%%%%%%%%%%%%%%%%%%%%%%%%%%%%%%%%%%%%%%%%
$\bullet$ \textbf{2-Form Field Coupled to CDM (2F):} 
%%%%%%%%%%%%%%%%%%%%%%%%%%%%%%%%%%%%%%%%%%%%%%%%%%%%%%%%%%%%%
This model has two additional parameters $\lambda_{\mathrm{2F}}$ and $\mu_{\mathrm{2F}}$ associated to the slope of the potential of the 2-form and the coupling with CDM respectively. The priors chosen for them are $\lambda_{\mathrm{2F}} \in [0,0.5]$ and $\mu_{\mathrm{2F}} \in [-0.05, 0.05]$ based on the results presented in Refs.~\cite{Orjuela-Quintana:2022jrg}.

\begin{figure}[t!]
\centering
\includegraphics[width=\linewidth]{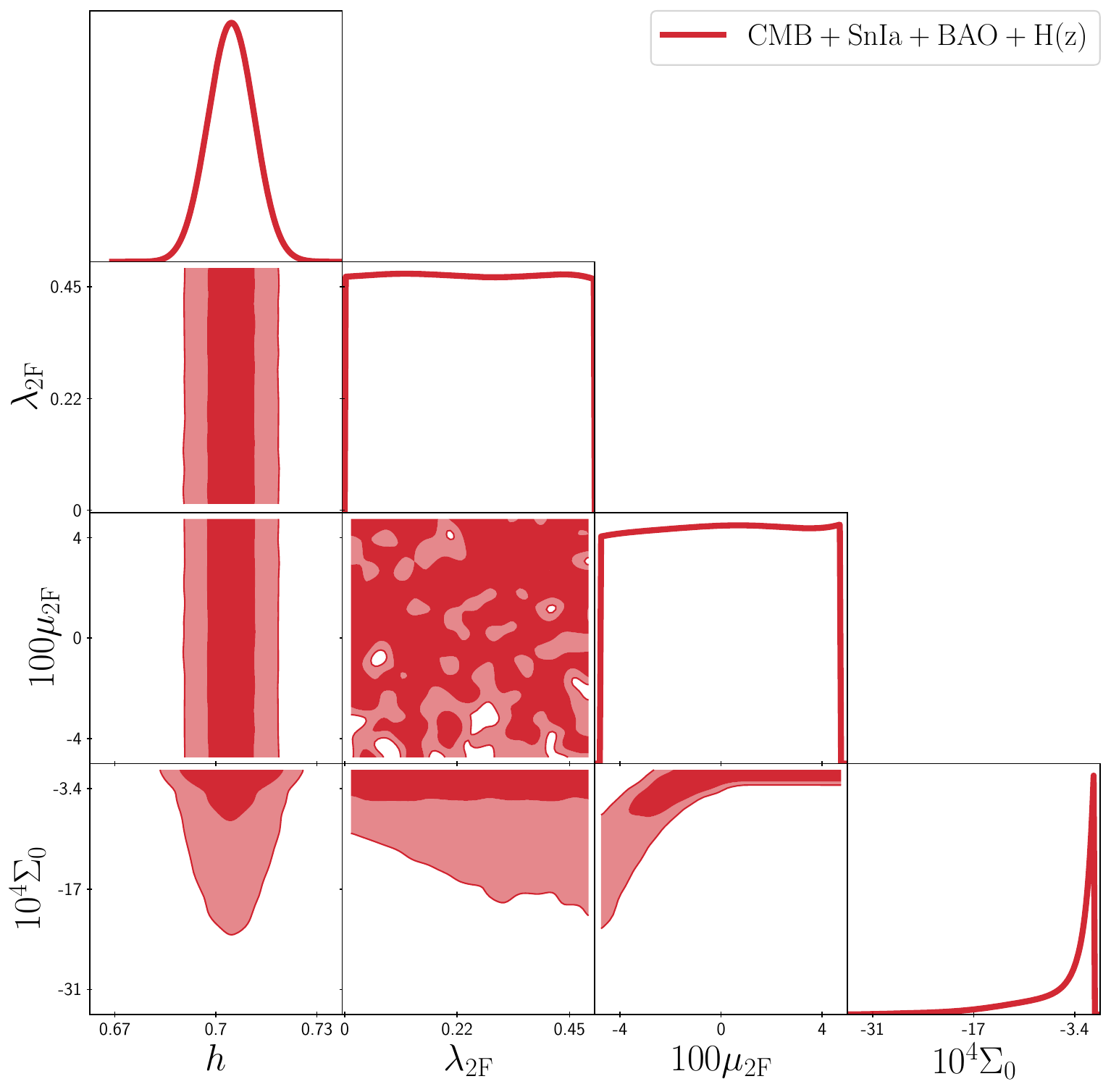}
\caption{Confidence contours for the parameters $\{h, \lambda_{\mathrm{2F}}, \mu_\mathrm{2F}, \Sigma_0\}$ of the 2-form model coupled to CDM at the $1\sigma$ and $2\sigma$ confidence levels. Both parameters, $\lambda_{\mathrm{2F}}$ and $\mu_{\mathrm{2F}}$, remain unconstrained by the datasets. The mean value and the $1\sigma$ deviation of the present-day shear is $10^{4}\Sigma_0 =  -3.501_{-0.53}^{3.5}$ which does not include the solution $\Sigma_0 = 0$.}
\label{Fig: Contours_TF}
\end{figure}

Table \ref{Table: means_2} and Fig. \ref{Fig: Contours_TF} present the mean values for the cosmological parameters and the confidence contours respectively. The model parameters $\lambda_{\mathrm{2F}}$ and $\mu_{\mathrm{2F}}$ remain unconstrained for the data used, while the mean value for the present-day shear is $10^{4}\Sigma_{0} = -3.50_{-0.53}^{3.48}$, which excludes the case $\Sigma_{0} = 0$ as the FQ model.

As part of our results, we present the values of the absolute magnitude of Sn Ia, $M$, for the different dark energy models considered. These values are consistent with the $\Lambda$CDM result shown in Table~\ref{Table: means}. Although a detailed analysis of the implications of $M$ lies beyond the scope of this work, we wish to briefly comment on the debate regarding its role in the Hubble tension. Several studies have suggested that the tension may originate from the physical correlation used in the SH0ES methodology to infer $H_0$ from the absolute peak magnitude $M$~\cite{Nunes:2021zzi}. In this context, the $M$ value derived by SH0ES is in a $3.4\sigma$ tension with the Planck-inferred value—obtained by constraining the sound horizon and the Sn Ia absolute magnitude—which is higher than the SH0ES estimate~\cite{Nunes:2021zzi,Huang:2024erq}.

Our analysis yields a value of $M$ that is slightly higher than the expected value in the $\Lambda$CDM framework using the Pantheon+ dataset, namely $M = -19.25 \pm 0.023$ \cite{Perivolaropoulos:2023iqj}. This result suggests a mild reduction in the tension with the Planck estimate. Notably, other works have found that the inclusion of additional datasets alongside Pantheon+ tends to shift $M$ toward higher values \cite{Nunes:2021zzi,Dinda:2022jih}, which is consistent with our findings.

Now we present the results for the case when the CMB dataset is not used in the cosmological analysis.

%%%%%%%%%%%%%%%%%%%%%%%%%%%%%%%%%%%%%%%%%%%%%%%%%%%%%%%%%
\subsubsection{$\mathbf{CMB~Independent~Constraints}$}
%%%%%%%%%%%%%%%%%%%%%%%%%%%%%%%%%%%%%%%%%%%%%%%%%%%%%%%%%

As previously discussed, the CMB shift parameters are derived from Planck's $\Lambda$CDM results, potentially introducing bias into the inference analysis. To mitigate this issue and provide a more robust assessment, we exclude CMB data from our joint dataset. The importance of CMB-independent constraints in cosmological parameter estimation has been highlighted in previous studies~\cite{Yadav:2023yyb, Schoneberg:2019wmt, Schoneberg:2022ggi}. In the following, we present the results of the corresponding MCMC analysis for the derived shear parameter, $\Sigma_0$, predicted by the anisotropic models under consideration.

As shown in Table~\ref{Table: means no cmb}, while the exclusion of CMB data leads to slight modifications in the inferred values of $\Sigma_0$, the overall conclusions remain unchanged. In particular, the predicted level of anisotropy remains generally small, with $|\Sigma_0| \lesssim 10^{-4}$ for the majority of the models, and a negligible shear is not excluded. However, a notable exception arises in the case of the FQ model, which allows for a maximum anisotropy of $|\Sigma_0| \sim 0.1$, with a clearer exclusion of null shear. Meanwhile, the remaining cosmological parameters remain consistent with observational expectations, with values such as $\Omega_{c,0} \sim 0.2$, $\omega_b \sim 0.02$, $h \sim 0.7$, and $\Omega_{\text{DE},0} \sim 0.75$. This result challenges the prevailing assumption that observations universally favor an isotropic cosmic expansion. Furthermore, even a small shear, of the order of $\Sigma_0 \sim 10^{-4}$, has been proposed as a potential explanation for the observed quadrupole anomaly in the CMB~\cite{Appleby:2010PhR,Campanelli:2006vb}. These findings underscore the need for further investigation into the implications of this model for other cosmological observables, such as CMB temperature anisotropies and the matter power spectrum.

%%%%%%%%%%%%%%%% TABLE %%%%%%%%%%%%%%%%
\begin{center}
\begin{table*}
\begin{centering}
\begin{tblr}{|X[1.2, c] X[1.5, c] X[1.2, c] X[1.2, c] X[1.5,c] X[1.5,c] X[1.2,c] |}
\hline \hline
Param & AQ & FQ & SDE  & EYMH & VF & 2F \\
\hline \hline
$10^4 \Sigma_0$ & $5.06^{+1.28}_{-3.06}\times10^{-24}$ & $611^{+469}_{-140}$ & $-24.5^{+123}_{-7.41}$ & $0.7_{-0.031}^{+187.8}$& $7.87_{-1.68}^{+4.85}\times10^{-13}$ & $-3.44^{+3.42}_{-0.448}$ \\
\hline \hline
%%%%%%%%
\end{tblr}
\par\end{centering}
\caption{Mean values and corresponding $68.3\%$ confidence intervals for the present-day shear parameter, $\Sigma_0$, predicted by the anisotropic models analyzed in this study. The exclusion of the CMB dataset from the inference analysis does not significantly alter the conclusions drawn in the previous section. Notably, only the FQ model allows for a detectable shear consistent with other observational constraints, with $\Sigma_0 = 0$ being excluded at this confidence level.}
\label{Table: means no cmb}
\end{table*}
\par\end{center}

%%%%%%%%%%%%%%%%%%%%%%%%%%%%%%%%%%%%%%%%%%%%%%%%%
\section{Analysis}
\label{Sec: Analysis}
%%%%%%%%%%%%%%%%%%%%%%%%%%%%%%%%%%%%%%%%%%%%%%%%%

As discussed in the previous section, parameter inference for each model is conducted through an MCMC analysis, utilizing cosmological observations at the background level. The primary observable driving these analyses is the Hubble parameter, $H(z)$. It is well established that the $\Lambda$CDM model provides an excellent fit to observational data at the background level. Consequently, when alternative DE models are examined, observational constraints tend to favor parameter values that closely mimic the predictions of $\Lambda$CDM model. Therefore, by analyzing the deviations in $H(z)$ between $\Lambda$CDM and the considered DE models, we can better understand the restrictions placed on the free parameters and the extent to which these models differ from the standard cosmological paradigm. \\

%%%%%%%%%%%%%%%%%%%%%%%%%%%%%%%%%%%%%%%%%%%
\subsection{Analysis of the DE Models}
%%%%%%%%%%%%%%%%%%%%%%%%%%%%%%%%%%%%%%%%%%%

\begin{figure}[t!]
\centering
\includegraphics[width=8cm, height=8cm]{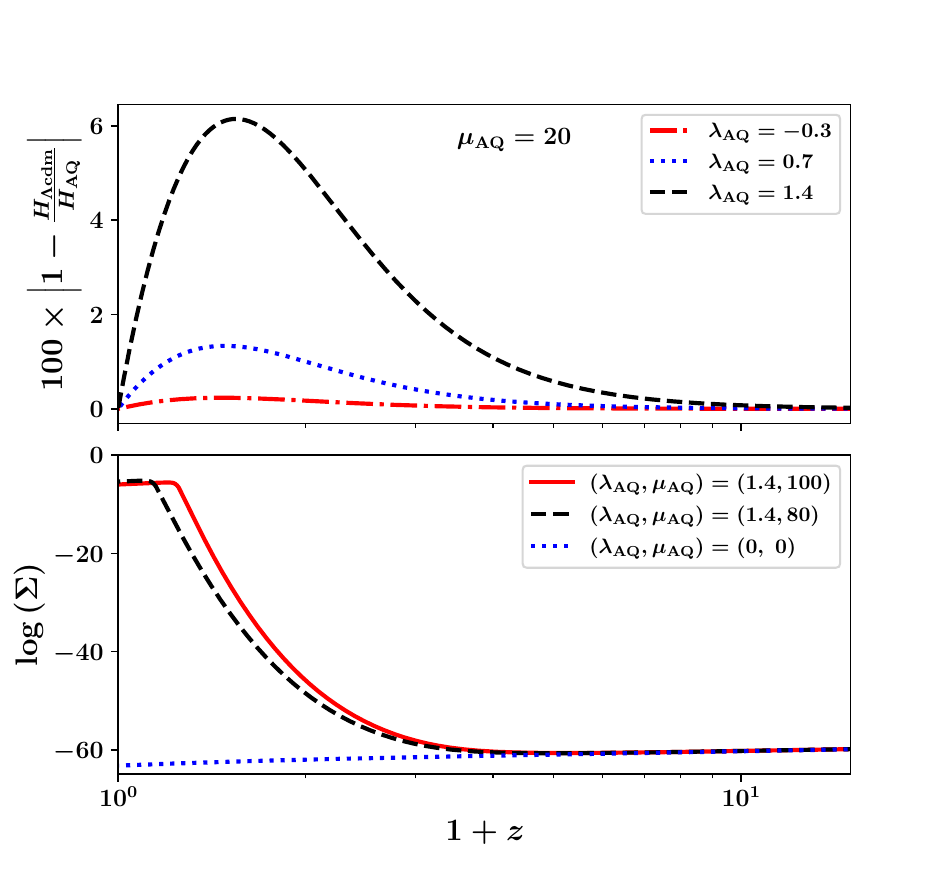}
\caption{\textbf{(Upper)} Percentage difference of \( H(z) \) between the AQ model and the \(\Lambda\)CDM model. For small values of \( \lambda_\AQ \), this difference remains minimal, aligning with cosmological observations. \textbf{(Lower)} Evolution of the shear \( \Sigma \) for varying \( \mu_\AQ \). The shear remains small throughout most of the expansion history, reaching approximately, in some cases, \( 10^{-3} \) only at very late times. Consequently, its impact on the overall expansion history is negligible.}
\label{Fig: H_aq}
\end{figure}

%%%%%%%%%%%%%%%% Anisotropic Quintessence %%%%%%%%%%%%%%%%
$\bullet$ \textbf{Scalar field coupled to a vector field (AQ):} 
%%%%%%%%%%%%%%%% Anisotropic Quintessence %%%%%%%%%%%%%%%%
For the AQ model, the data exhibit a preference for values of $\lambda_\AQ$ close to zero. The one-dimensional posterior distribution, shown in Fig.~\ref{Fig: Contours_anq}, deviates from a Gaussian profile, implying that the 68.3\% confidence region lacks a well-defined boundary. Since the quintessence model asymptotically approaches the $\Lambda$CDM model as $\lambda_\AQ \to 0$, the MCMC analysis naturally favors values that make the exponential potential nearly indistinguishable from a cosmological constant. The upper panel of Fig.~\ref{Fig: H_aq} illustrates the relative percentage difference in the Hubble parameter with respect to $\Lambda$CDM for fixed $\mu_\AQ$ and varying $\lambda_\AQ$. The results indicate that smaller absolute values of $\lambda_\AQ$ lead to minimal deviations from the standard cosmological model, explaining why the $1\sigma$ region in Fig.~\ref{Fig: Contours_anq} encompasses values of this parameter around zero.

Conversely, the MCMC analysis does not impose constraints on the coupling parameter $\mu_\AQ$. This is because the Hubble parameter, $H(z)$, remains largely insensitive to variations in the coupling function $f_\AQ(\phi) = f_ae^{-\mu_\AQ( \phi / \mpl)}$, which modulates the kinetic term $F_{\mu\nu}F^{\mu\nu}$. Consequently, the evolution of $H(z)$ is not displayed for different values of $\mu_\AQ$, as its impact on the expansion rate is negligible. The lower panel of Fig.~\ref{Fig: H_aq} illustrates the evolution of the shear for some values of $\lambda_\AQ$ and $\mu_\AQ$. The results indicate that $\Sigma_0$ ranges from approximately $10^{-28}$ to $10^{-3}$, consistent with the constraints derived from the MCMC analysis. The lower limit allowed by observations for the present-day shear is of the order $\Sigma_0 \sim 10^{-27}$, while the upper bound at the $1\sigma$ confidence level is approximately $\Sigma_0 \sim 10^{-26}$. Therefore, the data strongly favor small values of anisotropy. Note that $\Sigma$ is plotted for combinations of $\lambda_\AQ$ and $\mu_\AQ$ at the lower and upper limits of their respective priors. Consequently, the anisotropy at $z=0$ is expected to lie within the region shown in Fig.~\ref{Fig: H_aq}. \\

%%%%%%%%%%%%%%%% Anisotropic 2-form %%%%%%%%%%%%%%%%%%%%%%%%
$\bullet$ \textbf{Scalar field coupled to a 2-form (FQ):}
%%%%%%%%%%%%%%%% Anisotropic 2-form %%%%%%%%%%%%%%%%%%%%%%%%
Remarkably, both parameters introduced in the FQ model, $\lambda_\FQ$ and $\mu_\FQ$, are constrained by observations. However, the lower limit of $\lambda_\FQ$ is set by the prior, as is the upper limit of $\mu_\FQ$. As shown in Ref.~\cite{BeltranAlmeida:2019fou}, this model possesses two dark energy fixed points: one corresponding to an isotropic expansion and another to an anisotropic expansion. The isotropic solution is an attractor if  
\begin{equation}
  \lambda_\FQ^{2} + \lambda_\FQ \mu_\FQ -2 < 0,
\end{equation} 
while the anisotropic solution is an attractor if  
\begin{equation}
    \lambda_\FQ^{2} + \lambda_\FQ \mu_\FQ -2 > 0.
\end{equation}
Since the isotropic solution is a saddle point in this case, the parameter inference results indicate that observations favor an anisotropic attractor over an isotropic one.  

Additionally, the CMB constraint~\cite{Planck:2015bue} $\Omega_{\mathrm{DE}} < 0.02$ at redshift $z \sim 50$ imposes a lower bound $\mu_{\mathrm{2F}} > 5.5$. The inferred 95.8\% confidence region gives $\mu_\FQ = 11.18$, confirming agreement between theoretical predictions and observational constraints. Moreover, unlike in the AQ model, the mean value of $\lambda_\FQ$ is not close to zero, reinforcing the viability of accelerated expansion with a steep exponential potential in the 2-form model, as previously discussed in Ref.~\cite{BeltranAlmeida:2019fou}. Such solutions have gained interest due to their connections with string theory~\cite{Alestas:2024gxe, Bhattacharya:2024hep, Andriot:2024jsh, Gallego:2024gay, Akrami:2025zlb}.  

\begin{figure}[]
\centering
\includegraphics[width=8cm, height=8cm]{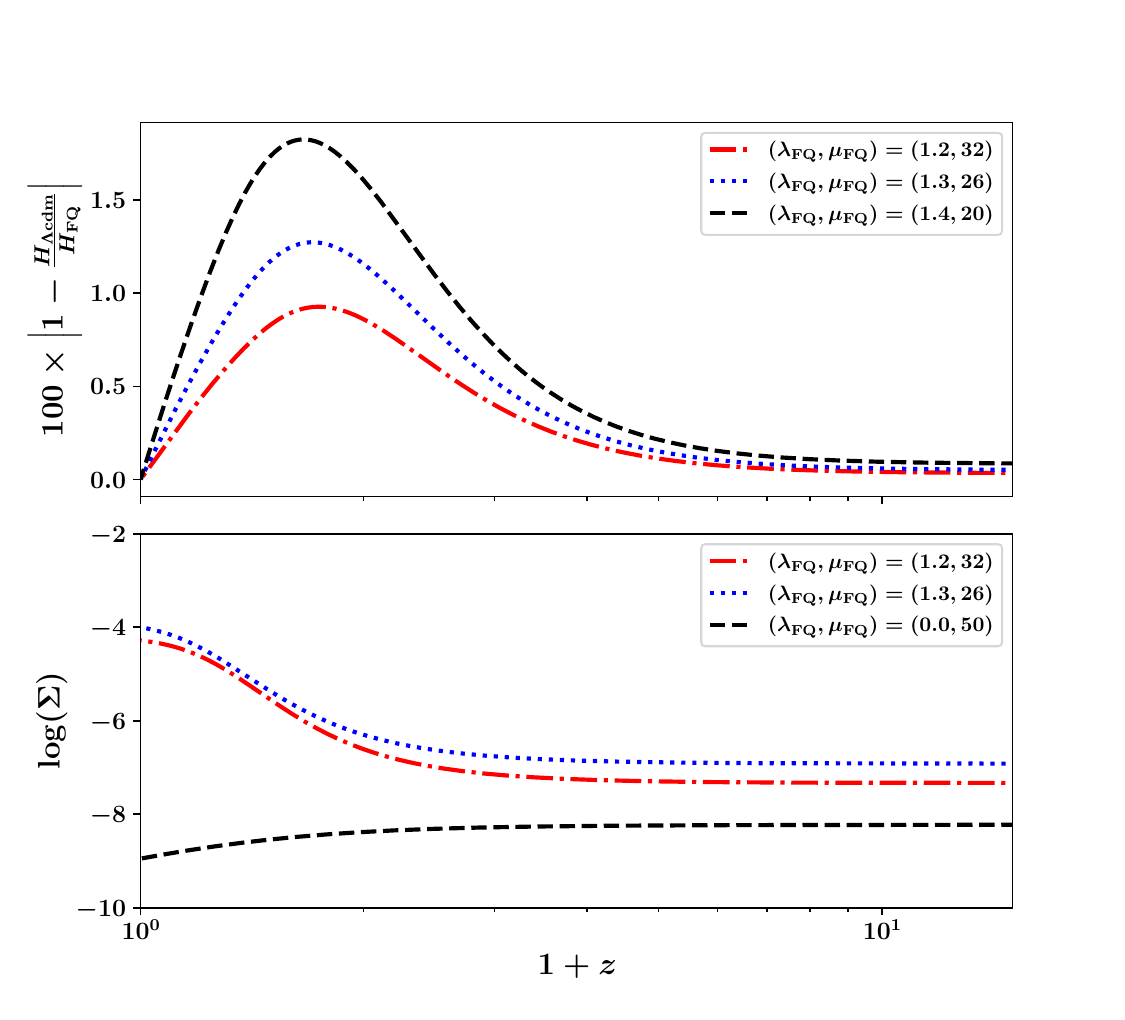}
\caption{\textbf{(Upper)} Percentage difference in $H(z)$ between the FQ model and the $\Lambda$CDM model. When $\lambda_\FQ$ and $\mu_\FQ$ are chosen such that the anisotropic solution is the attractor of the system, the deviation remains small, making it consistent with cosmological observations. \textbf{(Lower)} Evolution of the shear $\Sigma$ for these cases. The shear remains appreciable at present, reaching values of order $10^{-2}$, challenging the conventional assumption that $\Sigma_0$ must be negligible to maintain agreement between theory and observations. The smallest anisotropy is attained at the boundary of the parameter space, specifically at $\lambda_\FQ = 0$ and $\mu_\FQ = 50$.}
\label{Fig: H_2f}
\end{figure}

In the upper panel of Fig.~\ref{Fig: H_2f}, we plot the evolution of the percentage difference in the Hubble parameter with respect to the $\Lambda$CDM model for different values of $\lambda_\FQ$ and $\mu_\FQ$, ensuring that the anisotropic solution is the attractor of the system. For the parameter combinations in this figure, we note that the percentage difference in $H(z)$ for both models does not exceed 2\%, even when $\lambda_\FQ \gtrsim 1$, i.e., when a steep potential is considered.  

The lower panel of this figure shows the evolution of $\Sigma$, revealing that $\Sigma_{0} \sim 10^{-2}$, consistent with the mean value reported in Table \ref{Table: means}. This challenges the common assumption that present-day anisotropy must be nearly negligible. 

\begin{figure}[]
\centering
\includegraphics[width=\linewidth, height=6cm]{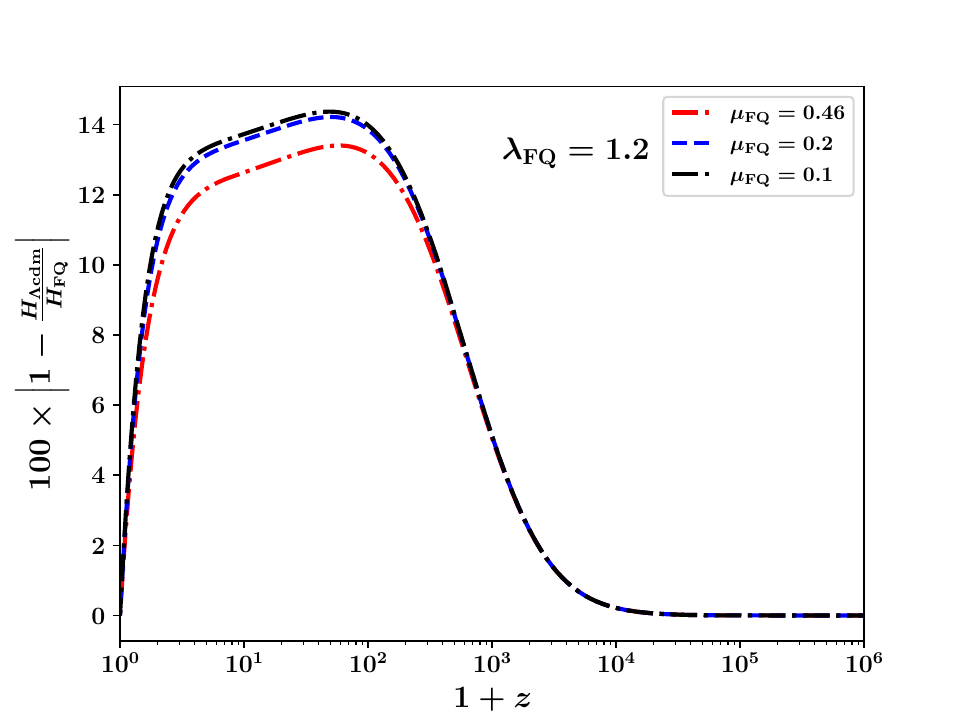}
\caption{Percentage difference in $H(z)$ between the FQ model and the $\Lambda$CDM model. When $\lambda_\FQ$ and $\mu_\FQ$ are chosen such that the isotropic solution is the attractor of the system, the deviation becomes significant, reaching up to 15\% in some cases, making it inconsistent with cosmological observations.}
\label{Fig: H_2f_2}
\end{figure}

On the other hand, the isotropic solution is not favored by observations, as evident in Fig.~\ref{Fig: H_2f_2}. In this figure, we see that for fixed $\lambda_\FQ = 1.2$, small values of $\mu_\FQ$, which ensure the isotropic solution as the attractor, lead to a large percentage difference in $H(z)$. For instance, for $\mu_\FQ = 0.46$ (e.g., the dot-dashed line in Fig.~\ref{Fig: H_2f_2}), the percentage difference in $H$ can reach up to 13\% for a significant portion of the expansion history.

Based on these remarks, the statistical analysis is expected to rule out the isotropic solution as observationally viable. Consequently, the exclusion of zero from the $1\sigma$ contour of $\Sigma_{0}$ is consistent with this scenario. In summary, observations favor an anisotropic dark energy solution as the attractor of the model, leading to the exclusion of $\Sigma_0 = 0$ at the $2\sigma$ confidence level.

\begin{figure}[t!]
\centering
\includegraphics[width=6cm, height=6cm]{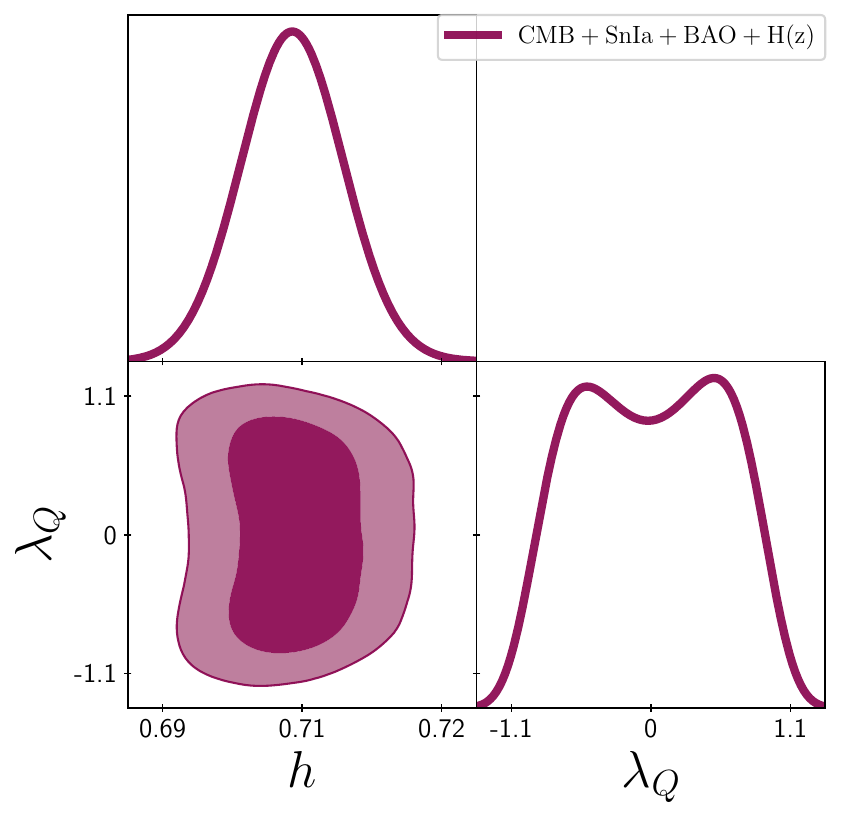}
\caption{Confidence contours for the parameters $\{h, \lambda_Q\}$ of the isotropic quintessence model at the $1\sigma$ and $2\sigma$ confidence levels. The only parameter related to dark energy, $\lambda_Q$, is small, indicating a nearly flat potential that closely mimics the behavior of a cosmological constant.}
\label{Fig: Contours_IsoQ}
\end{figure}

A particularly noteworthy result is that $\lambda_\FQ \gtrsim 1$. String compactifications typically predict a steep potential for the scalar field, which is often coupled to additional fields arising from the compactification process~\cite{Agrawal:2018own,Obied:2018sgi}. In the standard isotropic quintessence (IQ) model~\cite{copeland2006dynamics}, a small potential slope, $\lambda_Q$, is preferred, as illustrated in Fig.~\ref{Fig: Contours_IsoQ} and Table~\ref{Table: means_2}. For $\lambda_Q \sim 0$, the potential flattens, mimicking a cosmological constant. In contrast, the FQ model yields a scenario where the potential slope satisfies $\lambda_\FQ \gtrsim 1$. This result is enabled by the interaction of the scalar field with a 2-form field, which induces a late-time anisotropy in the accelerated expansion. \\

%%%%%%%%%%%%%%%% Anisotropic Solid Dark Energy %%%%%%%%%%%%%%%%
$\bullet$ \textbf{Inhomogeneous scalar fields (SDE):} 
%%%%%%%%%%%%%%%% Anisotropic Solid Dark Energy %%%%%%%%%%%%%%%%
The dynamical system analysis of the SDE model reveals three possible anisotropic dark energy attractors, depending on the values of $n$ and $m$. A remarkable feature of this model is that $\Sigma_0 \neq 0$ despite $w_\text{DE} \sim -1$. This indicates that $w_\text{DE} \approx -1$ is not a sufficient condition for isotropic expansion. However, when $n = m$, both Lagrangian functions, $F^1$ and $F^2$, evolve identically, leading to an isotropic Universe.

Although Fig.~\ref{Fig: Contours_sde} does not impose strong constraints on $n$ and $m$, the MCMC analysis finds that $n \sim m$ at the $1\sigma$ confidence level, suggesting that observations favor isotropic expansion. Additionally, for the present-day shear, we obtain an upper bound of $|\Sigma_{0}| \lesssim 10^{-3}$ within the $1\sigma$ region, allowing for solutions with negligible shear.

\begin{figure}[]
\centering
\includegraphics[width=\linewidth]{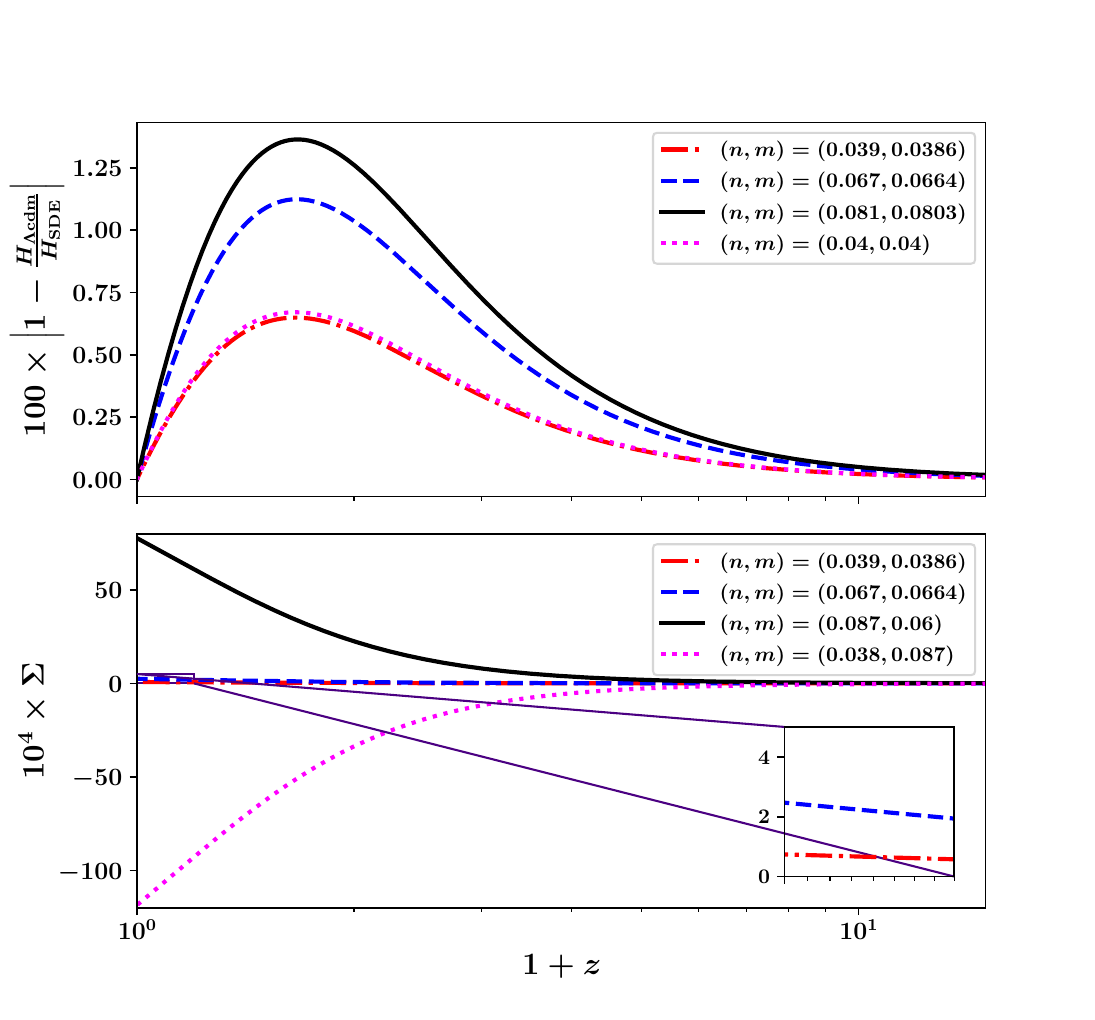}
\caption{\textbf{(Upper)} Percentage difference in $H(z)$ between the SDE model and the $\Lambda$CDM model. When $n \sim m$, this difference remains small, aligning with cosmological observations and favoring an isotropic evolution of the Universe. \textbf{(Lower)} Time evolution of the shear $\Sigma$ for various parameter sets ${n, m}$. The results show that $\Sigma_0 \lesssim 10^{-4}$ throughout the entire expansion history, making its effect negligible.}
\label{Fig: H_sde}
\end{figure}

Figure~\ref{Fig: H_sde} illustrates the time evolution of the percentage difference in the Hubble parameter between the SDE model and $\Lambda$CDM (upper panel), as well as the evolution of $\Sigma$ (lower panel) for various values of $n$ and $m$. The dot-dashed and dashed lines correspond to parameters within the $1\sigma$ and $2\sigma$ confidence regions, respectively, while the solid line represents values outside these regions, and the dotted line denotes the isotropic case. Deviations from $\Lambda$CDM grow as $n$ and $m$ move away from the $1\sigma$ region, reinforcing the conclusion that the dataset favors solutions where the background evolution closely resembles $\Lambda$CDM, which occurs when $n \sim m$. Regarding anisotropy, the numerical solution shows that $\Sigma_{0}$ is typically of order $10^{-4}$, consistent with the MCMC results. 

The 2D contour plot in Figure~\ref{Fig: Contours_sde} shows that certain combinations of the parameters $(n,m)$ are not allowed. For instance, the points $(n,m) = (0.087, 0.06)$ and $(n,m)=(0.03,0.087)$ lie approximately at the boundary of the non-excluded region. Plotting the shear parameter $\Sigma$ for these combinations provides insight into the limiting values that $\Sigma_{0}$ can reach throughout the MCMC analysis. As shown in Figure~\ref{Fig: H_sde}, $\Sigma_{0}$ ranges approximately from $7.7\times 10^{-3}$ to $-1.1\times 10^{-2}$, consistent with the constraints illustrated in Figure~\ref{Fig: Contours_sde}.\\

\begin{figure}[]
\centering
\includegraphics[width=\linewidth]{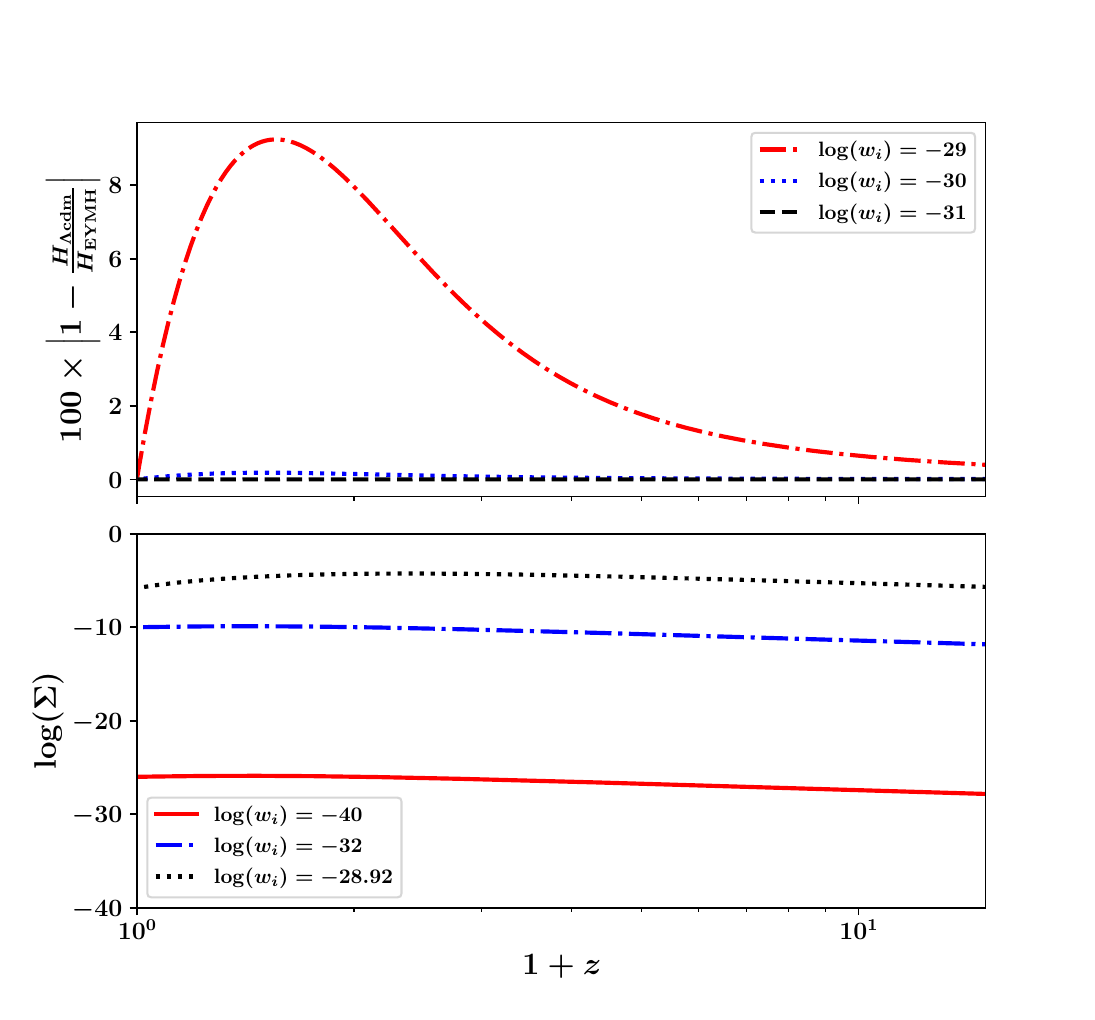}
\caption{\textbf{(Upper)} Percentage difference of $H(z)$ as predicted by the EYMH model relative to the $\Lambda$CDM model. The difference remains small for larger negative values of $\log w_i$, corresponding to smaller values of $w_i$. This parameter quantifies the coupling strength between the Higgs-like field and the gauge vector field, which drives the anisotropy. Consequently, cosmological observations favor an isotropic expansion. \textbf{(Lower)} Time evolution of the shear $\Sigma$ for several values of $\log w_i$. The results show that $\Sigma_0$ is negligible for the values of $\log w_i$ preferred by observations.}
\label{Fig: H_eymh}
\end{figure}

%%%%%%  Anisotropic Einstein Yang Mill Higgs %%%%%%%%%%%%%%%%%%%%
$\bullet$ \textbf{Scalar field with internal symmetry (EYMH):} 
%%%%%%  Anisotropic Einstein Yang Mill Higgs %%%%%%%%%%%%%%%%%%%%
For the EYMH model, the sole free parameter, $w_{i}$, remains unconstrained by the data. The anisotropic stress is of the order $|\Sigma_{0}| \lesssim 10^{-4}$. The lower panel of Figure~\ref{Fig: H_eymh} confirms that $\Sigma_{0}$ can take values around $10^{-4}$ or vanish entirely.

In the upper panel of Figure~\ref{Fig: H_eymh}, we present the percentage difference between the Hubble parameter predicted by the EYMH model and that of the $\Lambda$CDM model, following the same analysis applied to previous models. The results indicate that deviations from $\Lambda$CDM become more pronounced as $\log w_{i}$ approaches the prior limit of $-29$, explaining the small region of $\log w_{i}$ excluded from the $1\sigma$ confidence level in the 2D contours of Figure~\ref{Fig: Contours_eymh}. For values of $\log w_{i}$ within the $1\sigma$ region, the Hubble parameter $H$ closely follows the $\Lambda$CDM evolution, reinforcing the preference for an isotropic expansion in light of the observational data. \\

%%%%%%%%%%%%%%%%%%%%%%%%%%%%%%%%%%%%%%
$\bullet$ \textbf{Vector field (VF):} 
%%%%%%%%%%%%%%%%%%%%%%%%%%%%%%%%%%%%%%
In the scenario where a single vector field governs the anisotropic accelerated expansion, the MCMC analysis reveals that the model parameter $\lambda_\VF$ remains unconstrained. This result is anticipated, as demonstrated in Ref.~\cite{Orjuela-Quintana:2022jrg}. Under the assumption of small anisotropy, the energy density associated with the vector field satisfies:
\begin{equation}
    \rho_A \approx \frac{\dot{\psi}^2e^{4\sigma}}{2a^2} + V_\VF(X_\VF), \quad X_\VF = - \frac{1}{2}\frac{\psi^2 e^{4\sigma}}{a^2}.
\end{equation}
If the vector field is the primary driver of accelerated expansion, the evolution equation for the shear simplifies to:
\begin{equation}
    \ddot{\sigma} + 3H\dot{\sigma} \approx -\lambda_\VF \left( \frac{\psi^2 e^{4\sigma}}{\mpl^2 a^2} \right) \frac{V_\VF}{3 \mpl^2}.
\end{equation}
This equation implies that $\sigma$ tends to acquire negative values. Consequently, the potential $V_\VF$ asymptotically flattens, approaching a constant, thereby replicating the behavior of the $\Lambda$CDM model. Since this outcome is independent of $\lambda_\VF$, the evolution of the Hubble parameter $H$ remains largely unaffected by variations in this parameter. Similarly, as the vector field ceases to evolve, the anisotropic stress loses its source and asymptotically vanishes, a result that aligns with the MCMC analysis. \\

%%%%%%%%%%%%%%%%%%%%%%%%%%%%%%%%%%%%%%%%%%%%%%%%%%%%%%%%
$\bullet$ \textbf{2-form Field coupled to CDM (2F):} 
%%%%%%%%%%%%%%%%%%%%%%%%%%%%%%%%%%%%%%%%%%%%%%%%%%%%%%%%
The analysis presented in Ref.~\cite{Orjuela-Quintana:2022jrg} establishes that the 2F model admits two attractor solutions: an isotropic dark energy solution, realized when $0 \leq \lambda_\twoF < 1/6$, and an anisotropic dark energy solution, occurring when $1/6 \leq \lambda_\twoF \leq 1/2$. The inference analysis reveals that the parameters $\lambda_\twoF$ and $\mu_\twoF$ in the 2F model remain unconstrained, implying that observational data do not rule out either attractor. However, despite the viability of the isotropic attractor, the shear at present, $\Sigma_{0}$, cannot vanish.

\begin{figure}[]
\centering
\includegraphics[width=\linewidth]{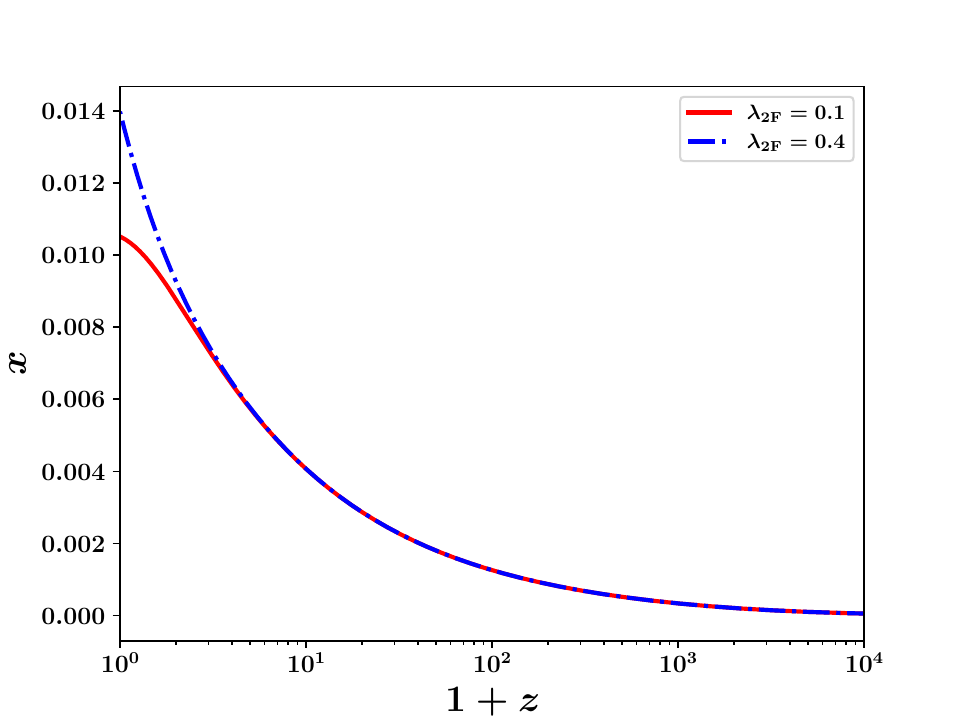}
\caption{Evolution of the dynamical variable \mbox{$x \equiv (\varphi e^{-2\sigma})/(\mpl a^2)$}, which quantifies the magnitude of the 2-form field in the 2F model. The trajectories for $\lambda_\twoF = 0.1$ (isotropic attractor) and $\lambda_\twoF = 0.4$ (anisotropic attractor), exhibit similar behavior, indicating that observational data do not favor one attractor over the other.}
\label{Fig: x_2f_cdm}
\end{figure}

To elucidate this behavior, the evolution of the 2-form field is examined through the dynamical variable $\mpl a^2 x = \varphi e^{-2\sigma}$ in Figure~\ref{Fig: x_2f_cdm}. The results indicate that, for different values of $\lambda_\twoF$, the evolution of $x$ remains nearly identical in both the isotropic and anisotropic scenarios. Moreover, as shown in Figure~2 of Ref.~\cite{Orjuela-Quintana:2022jrg}, the ratio $x/H$ begins to decay after $z = 0$, suggesting that the contribution of the 2-form field becomes negligible at very late times. Additionally, the dynamical variables of the model exhibit minimal differences between the isotropic and anisotropic cases prior to $z = 0$. Given that the shear $\Sigma$ is sourced by the 2-form field, which persists until late times, it follows that $\Sigma_{0}$ remains nonzero. This finding is consistent with the MCMC results, reinforcing the conclusion that $\Sigma_{0}$ cannot vanish.

%%%%%%%%%%%%%%%%%%%%%%%%%% AIC %%%%%%%%%%%%%%%%%%%%%%%%%%%%
\subsection{Information criteria for model comparison}
%%%%%%%%%%%%%%%%%%%%%%%%%% AIC %%%%%%%%%%%%%%%%%%%%%%%%%%%%

A key aspect of Bayesian analysis is determining how well a model fits the data. To compare models, especially when they differ in the number of parameters, various information criteria are used \cite{Liddle:2004nh}. One commonly employed criterion is the Akaike Information Criterion (AIC), defined as~\cite{Trotta:2008qt}:
\begin{equation}
    \mathrm{AIC} = -2\log \mc{L}_{\mathrm{max}} + 2k_{p} + \frac{2k_{p}(k_{p}+1)}{N_{\mathrm{data}}-k_{p}-1}, 
\end{equation}
where $\mc{L}_{\mathrm{max}}$ represents the maximum likelihood, $k_
{p}$ is the number of parameters in the model, and $N_{\mathrm{data}}$ is the number of observational data points.

To apply the AIC, we compute the difference between the AIC value of a given model and that of the best-fitting model within the group under comparison, expressed as $\Delta\mathrm{AIC} = \mathrm{AIC}_{\mathrm{model}} - \mathrm{AIC}_{\mathrm{min}}$. We use Jeffreys' scale to evaluate the relative favorability of the models \cite{Nesseris:2012cq}. According to this scale, a $4<\Delta\mathrm{AIC}<7$ indicates moderate evidence against the model with the higher AIC value, while $\Delta\mathrm{AIC}\geq 10$ suggests strong evidence against it. When $\Delta\mathrm{AIC}\leq 2$, the models are considered statistically equivalent.

In this analysis, the total number of data points is $N_{\mathrm{data}} = 1743$, composed of 4 points from the CMB, 2 from BAO, 36 from $H(z)$, and 1701 from SnIa observations. The maximum likelihood values and the number of parameters for each model are recorded in Table \ref{Table: AIC}.
\begin{table}[t!] 
\setlength{\tabcolsep}{10pt} 
\begin{tabular}{ccccc} \hline \hline 
Model & $-\log \mc{L}_\text{max}$ & $k_p$ &AIC & $\Delta$ AIC \\ \hline \hline 
$\Lambda$CDM & 670.374 & 4 & 1348.77 & 0.0 \\ \hline 
AQ & 670.262 & 6 & 1352.57 & 3.80 \\ \hline 
FQ & 670.184 & 6 & 1352.42 & 3.65 \\ \hline 
2F & 670.380 & 6 & 1352.81 & 4.04 \\ \hline 
SDE & 670.394 & 6 & 1352.84 & 4.07 \\ \hline 
EYMH & 670.373 & 5 & 1350.78 & 2.01 \\ \hline 
VF & 670.380 & 5 & 1350.79 & 2.02 \\ \hline 
IQ & 670.277 & 5 & 1350.59 & 1.82\\ \hline 
\end{tabular} 
\caption{AIC parameters for the $\Lambda$CDM model and the dark energy models. Since $\Delta\text{AIC} \sim 4$ for most of the dark energy models, they are moderately disfavored compared to $\Lambda$CDM.} \label{Table: AIC} 
\end{table}

As shown in Table \ref{Table: AIC}, the additional parameters in the dark energy models, compared to $\Lambda$CDM, generally penalize their favorability. For the AQ and FQ models we have $\Delta\mathrm{AIC}$ values of 3.80 and 3.65, respectively, suggesting they are moderately disfavored relative to $\Lambda$CDM. This implies that introducing anisotropy does not significantly improve the fit to observational data. The EYMH and VF models, with $\Delta\mathrm{AIC}$ values of 2.01 and 2.02, respectively, exhibit slight preference for the standard model. The quintessence model, with $\Delta\mathrm{AIC} = 1.82$, remains relatively competitive but still leans in favor of $\Lambda$CDM. The AIC criterion also indicates moderate evidence against the SDE and 2F models.

For completeness, we also report the Bayesian information criterion (BIC), a widely used metric for assessing cosmological models \cite{Trotta:2008qt}. In the limit of a large sample size, the BIC is defined as
\begin{equation}
\text{BIC} = -2\log \mathcal{L}_{\text{max}} + k_{p} \log N_{\mathrm{data}}.
\end{equation}
To compare two models, we compute the difference $\Delta~\text{BIC}  = \text{BIC}_{\text{model}} - \text{BIC}_{\text{min}}$ where $\text{BIC}_{\text{min}}$ corresponds to the lowest BIC value among the models under consideration. The standard interpretation of $\Delta\text{BIC}$ is as follows:
\begin{itemize}
\item $0 < \Delta\text{BIC} < 2$: the models are statistically equivalent.
\item $2 < \Delta\text{BIC} < 6$: positive evidence against the model with the higher BIC.
\item $6 < \Delta\text{BIC} < 10$: strong evidence against the model with the higher BIC.
\item $\Delta\text{BIC} > 10$: very strong evidence against the model with the higher BIC.
\end{itemize}

Table~\ref{Table: BIC} presents the BIC values for the dark energy models under consideration, compared to the $\Lambda$CDM model. As in the case of the AIC analysis, $\Lambda$CDM remains the most favored model according to the datasets considered. In particular, all dark energy models yield $\Delta\text{BIC} > 7$, which constitutes strong evidence against them relative to $\Lambda$CDM under the standard BIC interpretation.

\begin{table}[t!] 
\setlength{\tabcolsep}{10pt} 
\begin{tabular}{ccc} \hline \hline 
Model & BIC & $\Delta$ BIC \\ \hline \hline 
$\Lambda$CDM & 1370.6 & 0.0 \\ \hline 
AQ & 1385.3 & 14.70 \\ \hline 
FQ & 1385.54 & 14.93 \\ \hline 
2F & 1370.60 & 14.54 \\ \hline 
SDE  & 1385.57 & 14.96 \\ \hline 
EYMH  & 1378.06 & 7.46 \\ \hline 
VF &  1378.08 & 7.47 \\ \hline 
IQ & 1377.87 & 7.26\\ \hline 
\end{tabular} 
\caption{BIC values for the $\Lambda$CDM model and the dark energy models considered. All dark energy models satisfy $\Delta\text{BIC} > 7$, indicating strong evidence against them relative to $\Lambda$CDM.} 
\label{Table: BIC} 
\end{table}

%%%%%%%%%%%%%%%%%%%%%%%%%%
\section{Conclusions} 
\label{Sec: Conclusions}
%%%%%%%%%%%%%%%%%%%%%%%%%%
The assumption of large-scale isotropy is a fundamental pillar of the standard cosmological model. However, persistent observational anomalies and theoretical challenges have prompted investigations into anisotropic extensions. In particular, anisotropic dark energy has been proposed as a potential solution to several outstanding issues in cosmology, including the anomalous alignment of the CMB lowest multipoles, discrepancies between various cosmic dipoles and the CMB dipole, and even the ongoing $H_0$ tension, which some studies suggest could be alleviated by an anisotropic modification of the $\Lambda$CDM model.

The simplest homogeneous yet anisotropic scenario is described by the Bianchi I metric with residual axial symmetry, where the universe expands at two distinct rates. However, Wald's renowned no-hair theorem establishes that a cosmological constant cannot sustain this background anisotropy, encoded in the shear parameter $\Sigma$, causing it to decay rapidly over time. Likewise, a single homogeneous scalar field is insufficient to generate persistent anisotropy. Consequently, sustaining anisotropy requires the presence of a cosmic fluid that contributes to the anisotropic-stress tensor.

Several mechanisms can break isotropy, and in this work, we have explored a selection of them. Our primary objective is to constrain the present-day shear parameter $\Sigma_0$ across a range of representative anisotropic models, assessing whether these mechanisms can sustain anisotropy while remaining consistent with established cosmological parameters, namely $\Omega_{r, 0} \sim 10^{-4}$, $\Omega_{c,0} \sim 0.2$, $\omega_b \sim 0.02$, $h \sim 0.7$, and $\Omega_{\text{DE},0} \sim 0.75$.

The anisotropic models considered in this analysis include a scalar field coupled to a vector field, a scalar field coupled to a 2-form field, a triad of inhomogeneous scalar fields, a scalar field with an internal gauge symmetry, a standalone vector field, and a standalone 2-form field. By incorporating these models into \texttt{CLASS} and conducting an MCMC parameter estimation using \texttt{MontePython}, we constrained $\Sigma_0$ using observational data from CMB, BAO, H(z), and Sn Ia. Given that the CMB shift parameters are derived under the assumption of a $\Lambda$CDM background, we also performed an independent parameter space exploration excluding CMB data. Additionally, we conducted a Bayesian model comparison to assess the relative favorability of these anisotropic scenarios against the standard $\Lambda$CDM model. By considering these diverse sources of anisotropy, our study aims to provide a comprehensive and representative assessment of the expected anisotropic signatures of dark energy in light of current observational constraints.

Our findings indicate that model selection generally favors $\Lambda$CDM due to its minimal parameter set. However, certain anisotropic models remain viable within current observational uncertainties. Specifically, while most models yield constraints consistent with a nearly isotropic universe ($|\Sigma_0| \lesssim 10^{-4}$), one notable exception emerges: the FQ model (a scalar field coupled to a 2-form field). This model permits a nonzero shear at the $2\sigma$ confidence level, providing a rare case in which cosmic anisotropy may persist at late times. Furthermore, excluding the CMB dataset does not qualitatively alter this conclusion.

Beyond the exclusion of isotropy ($\Sigma_0 = 0$) at $2\sigma$ confidence, the FQ model exhibits additional compelling features. Notably, its anisotropic accelerated attractor is observationally preferred over its isotropic counterpart. Our analysis reveals that, despite allowing for an appreciable anisotropy ($\Sigma_0 \sim 0.01$), the FQ model predicts an expansion history similar to $\Lambda$CDM, demonstrating that significant anisotropic effects can exist without drastically altering background cosmology. Moreover, we found that this behavior is driven by a steep potential for the homogeneous scalar field, characterized by $\lambda_\FQ \gtrsim 1$, which is regulated by the coupling parameter $\mu_\FQ$ between the scalar field and the 2-form field. This result is particularly intriguing, as steep potentials commonly arise in string compactifications, suggesting that the FQ model may have a natural embedding within supergravity (SUGRA) frameworks. A deeper investigation into the origin of such couplings in high-energy theories is therefore warranted, which we leave for future work.

Despite these insights, several open questions remain. The present constraints are derived primarily from background evolution and linear perturbations, yet a comprehensive analysis of the impact of late-time anisotropy on cosmological observables, such as CMB temperature anisotropies and the matter power spectrum, is still lacking. In particular, the FQ model’s potential effects on these observables merit further study. Addressing these aspects will be important for gaining a more complete understanding of anisotropic dark energy models.

In summary, our results highlight the value of further exploring anisotropic cosmologies from both theoretical and observational perspectives. While isotropy remains the simplest and most widely supported assumption, even small deviations could offer valuable clues about some of the open questions in modern cosmology. Future studies, incorporating more precise datasets and different theoretical approaches, will help clarify whether cosmic anisotropy has a meaningful role in shaping the universe's evolution. In particular, we plan to compute the linear-order perturbations for the models considered and investigate how key cosmological quantities—such as the anisotropies in the CMB temperature and the matter power spectrum—are modified in a Bianchi I universe.

%%%%%%%%%%%%%%%%%%%%%%%%%%%%%%
\section*{Acknowledgements}
%%%%%%%%%%%%%%%%%%%%%%%%%%%%%%
We are grateful with Prof. Jose Beltr\'an Jimenez for helpful discussions. This work was supported by Vicerrectoría de Investigaciones - Universidad del Valle Grant No. 71373. \\

%%%%%%%%%%%%%%%%%%%%%%%%%%%%%%%%%%%%%%%%%%%%%%%%%%%%%%%%%%%%%%%%%%%%%%%%%%%%%%%%%%%%%%%%%%%%%%
%\textbf{Declarations of generative AI and AI-assisted technologies in the writing process}\\
%%%%%%%%%%%%%%%%%%%%%%%%%%%%%%%%%%%%%%%%%%%%%%%%%%%%%%%%%%%%%%%%%%%%%%%%%%%%%%%%%%%%%%%%%%%%%%

%During the preparation of this work the authors used \texttt{ChatGPT by OpenAI} in order to improve the grammar and clarity of the text. After using this tool/service, the authors reviewed and edited the content as needed and take full responsibility for the content of the publication. 

%%%%%%%%%%%%%%%
\appendix
%%%%%%%%%%%%%%%

%%%%%%%%%%%%%%%%%%%%%%%%%%%%%%%%%%%%%%%%%%%%%%%%%%%%%%%%%%%%%
\section{Sources of Anisotropy: Details of Calculations}
\label{App: Sources}
%%%%%%%%%%%%%%%%%%%%%%%%%%%%%%%%%%%%%%%%%%%%%%%%%%%%%%%%%%%%%

In this appendix, we will provide some details needed to understand the calculations presented in Sec.~\ref{Sec: Sources}. The procedure closely resembles that summarized in Sec.~\ref{Sec: Framework}. The procedure can be summarized as follows: we take the variation of the Lagrangian $\mc{L} = \mc{L}_r + \mc{L}_m + \mc{L}_\text{DE}$ with respect to the metric $g^{\mu\nu}$. This will yield to the field equations, i.e., 
\begin{equation}
    \mpl G_{\mu\nu} = T_{\mu\nu}^{(r)} + T_{\mu\nu}^{(m)} + T_{\mu\nu}^{\text{DE}},
\end{equation}
where $G_{\mu\nu}$ is the Einstein tensor, $T_{\mu\nu}^{(r)}$ and $T_{\mu\nu}^{(m)}$ are the energy-momentum tensors for radiations and pressure-less matter fluids, respectively, and $T_{\mu\nu}^{(\text{DE})}$ denotes the energy-momentum tensor for dark energy. The corresponding evolution equation for the fields driving the dark energy mechanism are obtained through variation of $\mc{L}$ with respect to the corresponding fields. 

Then, by inserting the Bianchi I metric in Eq.~\eqref{Eq: Bianchi I} and the profile of the fields compatible with the symmetries of this metric in the field equations, we obtain the Friedman equations as:
\begin{align}
    3 \mpl^2 H^2 &= \rho_r + \rho_m + \rho_\text{DE} + 3 \mpl^2 \dot{\sigma}^2, \\
    -2 \mpl^2 \dot{H} &= \frac{4}{3}\rho_r + \rho_m + \rho_\text{DE}( 1 + w_\text{DE}) + 6 \mpl^2 \dot{\sigma}^2,
\end{align}
where $\rho_r$, $\rho_m$, and $\rho_\text{DE}$, are the densities of radiation, matter, and dark energy. We have assumed that the pressures of these fluids are: $p_r = \rho_r/3$ for radiation, $p_m = 0$ for matter, and $p_\text{DE} = w_\text{DE} \rho_\text{DE}$, with $w_\text{DE}$ denoting the equation of state of dark energy. The evolution equation for the geometrical shear is computed using Eq.~\eqref{Eq: Evo Geometrical Shear}. Similarly, we get the equations of motion for the corresponding dark energy fields.

In the following, we will apply this procedure to all models in Sec.~\ref{Sec: Sources}.

%%%%%%%%%%%%%%%%%%%%%%%%%%%%%%%%%%%%%%%%%%%%%%%%%%%%%%%%
\subsection{Scalar field coupled to a vector field}
\label{App: AQ Model}
%%%%%%%%%%%%%%%%%%%%%%%%%%%%%%%%%%%%%%%%%%%%%%%%%%%%%%%%

Varying the Lagrangian in Eq.~\eqref{Eq: action_AQ} with respect to the tensor metric $g^{\mu\nu}$, we obtain the following energy-momentum tensor: 
\begin{align}
    T_{\mu\nu}^{(\text{DE})} &= \nabla_{\mu}\phi\nabla_{\nu}\phi - g_{\mu\nu} \left(\frac{1}{2} \nabla_\alpha \phi \nabla^\alpha \phi + V_{\text{AQ}}(\phi) \right) \nonumber \\
    &+ f_{\text{AQ}}^2(\phi) F_{\mu\alpha} F_\nu^\alpha - \frac{1}{4} g_{\mu\nu}f_{\text{AQ}}(\phi)^{2}F_{\alpha\beta}F^{\alpha\beta}.
\end{align}
The equation of motion for the vector field is obtained by varying the action with respect to $A_\mu$, yielding:
\begin{equation}
\label{Eq: field_equation_A_AQ} 
    \nabla_\mu \left(f_{\text{AQ}}^2(\phi) F^{\mu\nu} \right) = 0. 
\end{equation} 
Similarly, by varying $\mc{L}$ with respect to $\phi$, the equation of motion for the scalar field is obtained:
\begin{equation}
\label{Eq: field_equation_phi_AQ}
    \square \phi - V_{\text{AQ},\phi} = \left(2 \frac{f_{\text{AQ},\phi}}{f_{\text{AQ}}}\right)\frac{f_{\text{AQ}}^2}{4}F_{\mu\nu}F^{\mu\nu},
\end{equation}
where $V_{\AQ,\phi} \equiv \dd V_\AQ/\dd\phi$. Now, from the Einstein equation the Friedman equations for the model and the evolution equation of the geometrical are derived as:
\begin{align}
   3 \mpl^2 H^2 &= \rho_{m} + \rho_{r} + \rho_{A}+\rho_{\phi}+3 \mpl^{2}\dot{\sigma}^{2}, \label{Eq: Friedman_1_AQ}  \\
   - 2 \mpl^2 \dot{H} &= \rho_{m} + \frac{4}{3}\rho_{r} + 2X\mc{P} + \frac{4}{3}\rho_A + 6\mpl^{2}\dot{\sigma}^{2}, \\
   \ddot{\sigma} + 3H\dot{\sigma} &= \frac{2\rho_{A}}{3\mpl^2}, \label{Eq: dd sigma_AQ}
\end{align}
where we have defined the contribution of the scalar field and the vector field to the density respectively as:
\begin{equation}
\label{Eq: Densities fields AQ}
    \rho_{\phi} \equiv \frac{1}{2} \dot{\phi}^2 + V_{\text{AQ}}, \quad \rho_{A} \equiv \frac{1}{2}f_{\text{AQ}}^{2}(\phi)\frac{\dot{A}e^{4\sigma}}{a^{2}}.
\end{equation}
Finally, from Eq. \eqref{Eq: field_equation_phi_AQ} the field equations for $\phi$ is expressed as:
\begin{equation}
\ddot{\phi} + 3H\dot{\phi} -\frac{\lambda}{m_{\text{p}}}V_{\text{AQ}} -2\frac{f_{\text{AQ},\phi}}{f_{\text{AQ}}}\rho_{A} = 0.
\end{equation}

%%%%%%%%%%%%%%%%%%%%%%%%%%%%%%%%%%%%%%%%%%%%%%%%%%%%%%
\subsection{Scalar Field Coupled to a 2-form Field}
\label{App: 2F Model}
%%%%%%%%%%%%%%%%%%%%%%%%%%%%%%%%%%%%%%%%%%%%%%%%%%%%%%

Upon varying the Lagrangian of the 2-form model in Eq.~\eqref{Eq: action_2F} with respect to the metric, we obtain the energy-momentum tensor:
\begin{align} 
    T_{\mu\nu}^{(\text{DE})} &= \nabla_{\mu}\phi\nabla_{\nu}\phi - g_{\mu\nu} \left(\frac{1}{2} \nabla_\alpha \phi \nabla^\alpha \phi + V(\phi) \right) \\ 
    &+ \frac{1}{2} f(\phi) H_{\mu\alpha\beta} H_{\nu}^{~\alpha\beta} - \frac{1}{12} f(\phi) H_{\alpha\beta\gamma} H^{\alpha\beta\gamma} g_{\mu\nu}, \nonumber 
\end{align}
Taking the variation of the action with respect to the scalar field, $\phi$, and the 2-form field, $B_{\mu\nu}$, yields the equations of motion:
\begin{align} 
    \nabla_{\mu} \nabla^{\mu} \phi - V_{\text{FQ,}\phi} &= 0, \\
    \nabla_{\alpha} \left(f_{\text{FQ}}(\phi) H^{\mu\nu\alpha} \right) &= 0. 
\end{align}   
The Friedman equations and the evolution equation for the geometrical shear are given by:
\begin{align} 
    3 \mpl^2 H^2 &= \rho_m + \rho_r + \rho_\phi + \rho_{B} + 3 \mpl^2 \dot{\sigma}^2, \label{Eq: Friedman_1_2F}\\
    -2 \mpl^2 \dot{H} &= \rho_m + \frac{4}{3} \rho_r + \dot{\phi}^{2} + \frac{2}{3} \rho_{B} + 6 \mpl^{2} \dot{\sigma}^{2}, \\
    \ddot{\sigma} + 3H \dot{\sigma} &= -\frac{2}{3} \frac{\rho_{B}}{\mpl^2}, \label{Eq: field_equation_sigma_2F}
\end{align}
where  $\rho_{B} = f_{\text{FQ}}(\phi)a^{-4}e^{-4\sigma}\dot{v}_{B}^{2}/2$.

%%%%%%%%%%%%%%%%%%%%%%%%%%%%%%%%%%%%%%%%%%%%%
\subsection{Inhomogeneous scalar fields}
%%%%%%%%%%%%%%%%%%%%%%%%%%%%%%%%%%%%%%%%%%%%%

From the variation of the action in Eq.~\eqref{Eq: action_SDE}, the energy-momentum tensor of this model is given by:
\begin{equation}
\label{Eq: tensor_SDE}
    T_{\mu\nu}^{(\text{DE})} = 2 \sum_{I} F_{X^{I}}\nabla_{\mu}\phi^{I}\nabla_{\nu}\phi^{I} - g_{\mu\nu}\sum_{I} F^{I},
\end{equation}
while the equation of motion for the solid, obtained by varying the action with respect to $\phi^{I}$, is:
\begin{equation}
\nabla_{\mu}\left(F_{X^{I}}\nabla^{\mu}\phi^{I}\right) = 0.
\label{Eq: field_equation_SDE}
\end{equation}
We derive the Friedman equations and the evolution equation for the geometrical shear as:
\begin{align}
    3 \mpl^2 H^{2} &= \rho_{m} + \rho_{r} + F^{1} + 2F^{2} + 3  \mpl^2 \dot{\sigma}^{2}, \label{Eq: Friedman_1_SDE} \\
    -2 \mpl^2 \dot{H} &= \frac{2}{3}X^{1}F_{X^{1}} + \frac{4}{3}X^{2}F_{X^{2}} + 6 \mpl^2 \dot{\sigma}^{2} \nonumber \\
     &+ \rho_{m} + \frac{4}{3}\rho_{r}, \\
    \ddot{\sigma} + 3H\dot{\sigma} &= \frac{2}{3 \mpl^{2}} \left( X^{2}F_{X^{2}} - X^{1}F_{X^{1}} \right).
\label{Eq: field_equation_sigma_SDE}
\end{align}

%%%%%%%%%%%%%%%%%%%%%%%%%%%%%%%%%%%%%%%%%%%%%%%%%%%%%
\subsection{Scalar field with internal symmetry}
%%%%%%%%%%%%%%%%%%%%%%%%%%%%%%%%%%%%%%%%%%%%%%%%%%%%%

The equations of motion for the Higgs field and the gauge vector field are obtained by varying the action with respect to $\mc{H}^{a}$ and $A^a_\mu$, respectively, as:
\begin{align}
    \nabla_{\mu}F_{a}^{~\mu\nu} - g \varepsilon^{b}_{ca}F_{b}^{\mu\nu}A^{c}_{\mu} - g\varepsilon^{b}_{ac}g^{\mu\nu}D_{\mu}\mc{H}_{b}\mc{H}^{c} &= 0,
    \label{Eq: field_equation_A_EYMH} \\
    \nabla_{\mu}D^{\mu}\mc{H}_{a} + 2 g\varepsilon^{c}_{~ba}A^{b}_{\mu}D^{\mu}\mc{H}_{c} - V_{\mc{H}_{a}} &= 0. \label{Eq: field_equation_PHI_EYMH} 
\end{align} 
The corresponding Friedman equations and the evolution equation for the geometrical shear are given by: 
\begin{align}
    3 \mpl^2 H^{2} &= \rho_{m} + \rho_{r} + \rho_{\mathrm{YM}} + \rho_{\mc{H}} + 3 \mpl^2 \dot{\sigma}^{2}, \label{Eq: Friedman_1_EYM} \\
    -2 \mpl^2 \dot{H} &= 3 \mpl^2 H^{2} + \frac{1}{3}\rho_{r} + p_{\mathrm{DE}} + 3 \mpl^2 \dot{\sigma}^2, \\
    \ddot{\sigma} + 3H\dot{\sigma} &= \frac{1}{3 \mpl^2} \Big[(G_{1}\dot{I})^{2} - (G_{2}\dot{J})^{2} + g^{2}(G_{2}J)^{4} \nonumber \\ 
    &- g^{2}(G_{1}I)^{2}(G_{2}J)^{2} + g^{2}\Phi^{2}(G_{2}J)^{2}\Big], \label{Eq: sigma_EYMH}
\end{align} 
where we have defined:
\begin{align}
    \rho_{\mathrm{YM}} &\equiv \frac{1}{2}(G_{1}\dot{I})^{2} + (G_{2}\dot{J})^{2} \nonumber \\
    &+ g^{2}(G_{1}I)^{2}(G_{2}J)^{2} + \frac{1}{2}g^{2}(G_{2}J)^{4}, \\
    \rho_{\mc{H}} &\equiv \frac{1}{2}\dot{\Phi}^{2} + g^{2}\Phi^{2}(G_{2}J)^{2} + V, \\
    p_{\mathrm{DE}} &\equiv \frac{1}{3}\rho_{\mathrm{YM}} + \frac{1}{2}\dot{\Phi}^{2} - \frac{1}{3}g^{2}(G_{2}J)^{2}\Phi^{2} - V.
\end{align} 

%%%%%%%%%%%%%%%%%%%%%%%%%%%%%
\subsection{Vector field}
%%%%%%%%%%%%%%%%%%%%%%%%%%%%%

By varying the action in Eq.~\eqref{Eq: action_avf} with respect to $g^{\mu\nu}$ and $A_\mu$, the energy-momentum tensor and the equation of motion for the vector field for this model read:
\begin{align}
    T_{\mu\nu}^{(\VF)} &= F_{\mu\alpha}F_{\nu}^{~\alpha} -\frac{1}{4}F_{\alpha\beta}F^{\alpha\beta} + 2A_{\mu}A_{\nu}V_{X} \\
    &- g_{\mu\nu}V \nonumber \\
    \nabla_{\mu}F^{\mu\nu} &= A^{\nu}V_{X}. \label{Eq: field_equation_vf}
\end{align}
Substituting the configuration of $A_{\mu}$ into the Einstein field equations, we obtain the Friedman equations and the evolution equation for the geometrical shear:
\begin{align}
    3 \mpl^2 H^2 &= \rho_{m} + \rho_{r} + \frac{\dot{\psi}^{2}e^{4\sigma}}{a^{2}} + V(X) + 3 \mpl^{2} \dot{\sigma}^{2}, \\ 
    -2 \mpl^{2} \dot{H} &= \rho_{m} + \frac{4}{3}\rho_{r} + \frac{2}{3}\frac{\dot{\psi}^{2}e^{4\sigma}}{a^{2}} - \frac{1}{3}V_{X}\frac{\psi^{2}e^{4\sigma}}{a^{2}} \nonumber \\
    &+ 6\mpl^{2}\dot{\sigma}^{2}, \\
    \ddot{\sigma} + 3H\dot{\sigma} &= \frac{\dot{\psi}^{2}e^{4\sigma}}{3\mpl^{2}a^{2}} + \frac{V_{X}\psi^{2}e^{4\sigma}}{3\mpl^{2}a^{2}}. \label{Eq: Evo sigma_vf}
\end{align}

%%%%%%%%%%%%%%%%%%%%%%%%%%%%%
\subsection{2-form field} 
%%%%%%%%%%%%%%%%%%%%%%%%%%%%%

In the case of the 2-form model, the energy momentum tensor is given by varying the action in Eq.~\eqref{Eq: action_2F} with respect to the metric. We obtain:
\begin{align}
    T^{B}_{\mu\nu} &\equiv \frac{1}{2} H_{\mu\alpha\beta}H_{\nu}^{~\alpha\beta} - g_{\mu\nu}\left(\frac{1}{12}H_{\alpha\beta\gamma}H^{\alpha\beta\gamma} + V \right) \nonumber \\ 
    &+ 4\left(V_{X} - f_{X}\mc{L}_{c}\right)B_{\mu\alpha}B_{\nu}^{~\alpha}.
\end{align}
By varying the action with respect to the 2-form field, the equation of motion for $B_{\mu\nu}$ is derived as:
\begin{equation}
    \nabla_{\alpha}H^{\mu\nu\alpha} = 4B^{\mu\nu}\left(V_{X} - f_{X}\mc{L}_{c}\right).
\label{Eq: field_equation_B}
\end{equation}
The Friedman equations and the evolution equation for $\sigma$ are given by
\begin{align}
    3 \mpl^2 H^{2} &= \rho_{r} + \rho_{b} + f(X)\rho_{c} + \frac{1}{2}\frac{\dot{\varphi}^{2}e^{-4\sigma}}{a^{2}} \nonumber\\ 
    &+ V(X) + 3 \mpl^2 \dot{\sigma}^{2}, \\
    - 2 \mpl^2 \dot{H} &= \frac{4}{3}\rho_{r} + \rho_{b} + f(X)\rho_{c} + 3 \mpl^2 \dot{\sigma}^{2} \nonumber \\
    &+ \frac{1}{3}\frac{\dot{\varphi}^{2}e^{-4\sigma}}{a^{4}} + \frac{8}{3} V_{X}\frac{\varphi^{2}e^{-4\sigma}}{a^{4}} \nonumber \\
    &+ \frac{8}{3} f_{X}\rho_{c}\frac{\varphi^{2}e^{-4\sigma}}{a^{4}}. \\
    \ddot{\sigma} + 3 H\dot{\sigma} &= -\frac{\dot{\varphi}^{2}e^{-4\sigma}}{4a^{2}\mpl^2 } + \frac{4}{3} V_{X} \frac{\varphi^{2}e^{-4\sigma}}{a^{2}\mpl^2 } \nonumber \\
    &+ \frac{4}{3} f_{X}\rho_{c} \frac{\varphi^{2}e^{-4\sigma}}{a^{2}\mpl^2 }.
\label{Eq: field_equation_sigma_B}
\end{align}

%%%%%%%%%%%%%%%%%%%%%%%%%%%%%%%%%%%%%%%%%%%%%%%%%%%%
\section{Correction to the Luminosity Distance}
\label{App: Luminosity Distance}
%%%%%%%%%%%%%%%%%%%%%%%%%%%%%%%%%%%%%%%%%%%%%%%%%%%%

\begin{figure}
\centering
\includegraphics[width=\linewidth]{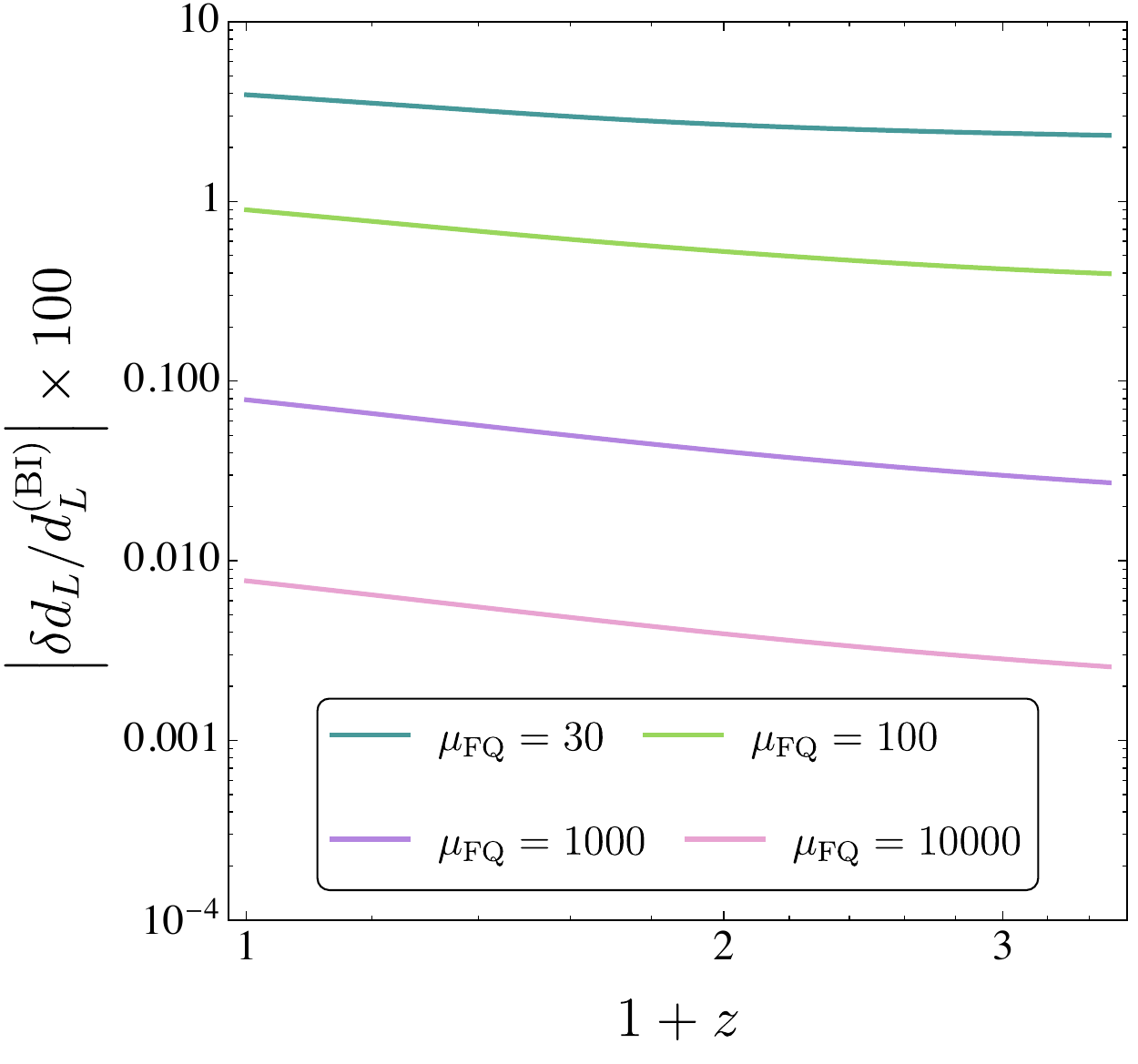}
\caption{Time evolution of the percentage error, $\Delta d_L$, in the determination of the luminosity distance using Eq.\eqref{Eq: dL H_ani} compared to the corrected expression in Eq.\eqref{Eq: Corrected dL}. The redshift range corresponds to the region where SnIa measurements are performed. The error reaches up to $\Delta d_L \lesssim 5\%$ for $\mu_\FQ = 30$, where $\Sigma_0 = -0.021$, while for larger values of $\mu_\FQ$, which correspond to smaller anisotropies, the error remains small at $\Delta d_L \lesssim 0.1\%$.}
\label{Fig: Comparison dL}
\end{figure}

In an anisotropic universe, the expansion rate varies with direction, leading to an anisotropic redshift. As a result, all observables are affected. In our approach, we have assumed an average scale factor $a(t)$, defining an average redshift defined as $a = 1/(1 + z)$ and yielding to an average Hubble parameter $H(z)$. Then, we have computed the luminosity distance as:
\begin{equation}
\label{Eq: dL H_ani}
    d_L^{(\text{ani})} = \int_0^z \frac{\dd z}{H(z)}
\end{equation}
where the anisotropy is introduced through the Hubble parameter. This treatment is equivalent to modeling anisotropy as a fluid with energy density $3\mpl^2 \dot{\sigma}^2$. Our assumption is that this method allows us to apply standard isotropic relations by replacing the isotropic Hubble rate with $H(z)$ in the presence of anisotropy. To assess the accuracy of this approach—which is commonly used in the literature, see for instance Refs.~\cite{Akarsu:2021max, Yadav:2023yyb, Yadav:2024pvr}—we compare it against a more detailed computation of the luminosity distance, $d_L^{(\text{BI})}$ in an axially symmetric Bianchi I background.

%%%%%%%%%%%%%%%%%%%%%%%%%%%%%%%%%%%%%%%%%%%%%%%%%%%%%%%%%%%%%%%%%%%%%%%%%%%%
\subsection{Luminosity Distance in an Axially Symmetric Bianchi I Universe}
%%%%%%%%%%%%%%%%%%%%%%%%%%%%%%%%%%%%%%%%%%%%%%%%%%%%%%%%%%%%%%%%%%%%%%%%%%%%
Recalling the metric in Eq.~\eqref{Eq: Bianchi I}, the line element for a photon trajectory ($\dd s^2 = 0$) is given by:
\begin{equation}
    \dd t^2 = a^2(t)\left\{ e^{-4\sigma(t)} \dd x^2 + e^{2\sigma(t)}\left( \dd y^2 + \dd z^2 \right)\right\}.
\end{equation}
Exploiting the axial symmetry of the metric, we express the spatial coordinates in spherical form for a radially emitted photon:
\begin{align}
    \dd x &= \cos\theta \, \dd r, \\ 
    \dd y &= \sin\theta \sin\varphi \, \dd r, \\ 
    \dd z &= \sin\theta \cos\varphi \, \dd r,
\end{align}
where $r$ is the radial comoving distance, and $(\theta, \varphi)$ describe the angular position of the photon source. Substituting these coordinates and assuming small anisotropy, the expression simplifies to:
\begin{equation}
    \dd t^2 \approx a^2(t)\left\{1 - \sigma(t)[1 + 3\cos(2\theta)]\right\} \, \dd r^2.
\end{equation}
The maximum anisotropic effect occurs when the photon propagates parallel to the fastest expanding direction ($\theta = 0$), yielding:
\begin{equation}
    \dd t \approx a(t)\left\{1 - 2\sigma(t)\right\} \, \dd r.
\end{equation}
Since the comoving distance is defined as the integral of $\dd r$, we obtain:
\begin{equation}
    \chi(z) \equiv \int_0^r \dd r = \int_{t_i}^{t_0} \frac{\dd t}{a(t)} + 2\int_{t_i}^{t_0} \dd t \frac{\sigma(t)}{a(t)}.
\end{equation}
The first term corresponds to the comoving distance in Eq.~\eqref{Eq: dL H_ani}, while the second accounts for the correction due to anisotropy. Consequently, the luminosity distance in terms of the average redshift is given by:
\begin{equation}
\label{Eq: Corrected dL}
    d_L^{(\text{BI})} = d_L^\text{(ani)} + \delta d_L,
\end{equation}
where the correction term is:
\begin{equation}
\label{Eq: Correction to dL}
    \delta d_L = 2(1+z)\int_0^z \dd z \, \frac{\sigma(z)}{H(z)}.
\end{equation}

%%%%%%%%%%%%%%%%%%%%%%%%%%%%%%%%%%%%%%%%%%%%%
\subsection{Comparison of Approximations}
%%%%%%%%%%%%%%%%%%%%%%%%%%%%%%%%%%%%%%%%%%%%%

In Fig.\ref{Fig: Comparison dL}, we compare our approximation using Eq.\eqref{Eq: dL H_ani} with the more precise estimation from Eq.~\eqref{Eq: Corrected dL}, for a fixed $\lambda_\FQ = 2$ and varying $\mu_\FQ$. For $\mu_\FQ = 30$, the present-day shear is $\Sigma_0 = -0.021$, while for $\mu_\FQ = 10^4$, it decreases to $\Sigma_0 = -7.4 \times 10^{-5}$, showing that anisotropy grows for smaller values of $\mu_\FQ$. This was expected since $\Sigma \propto \lambda_\FQ/\mu_\FQ$ in its attractor, as mentioned in Ref.~\cite{BeltranAlmeida:2019fou}.

To quantify the accuracy of our approach, we define the percentage error in the luminosity distance as:
\begin{equation}
    \Delta d_L = \left\vert \frac{\delta d_L}{d_L^{(\text{BI})}} \right\vert \times 100.
\end{equation}
From Eq.~\eqref{Eq: Correction to dL}, we expect that a larger anisotropy leads to a larger correction. This is confirmed in Fig.~\ref{Fig: Comparison dL}, where we plot the evolution of the percentage error over the redshift range relevant for Sn Ia measurements. As expected, $\Delta d_L$ is smaller for larger $\mu_\FQ$ (corresponding to low anisotropy) and increases as $\mu_\FQ$ decreases (corresponding to high anisotropy). For $\mu_\FQ = 30$, the percentage error can reach up to $\Delta d_L \lesssim 5\%$. 

While this error remains relatively small, it is notable given that $\Sigma_0 = - 0.021$ represents a significant anisotropy by standard expectations. Given the current precision of Sn Ia measurements, we emphasize the need for caution when applying simplified approaches. The method we have used—Eq.~\eqref{Eq: dL H_ani}—is standard in the literature (see, e.g., Refs.~\cite{Akarsu:2021max, Yadav:2024pvr}). In those studies, the anisotropy remains below 0.001 throughout cosmic history, ensuring that the correction term remains negligible. However, for larger anisotropy, a more refined approach may be necessary as that pointed out in Refs.~\cite{Campanelli:2010zx, Verma:2024lex}.

%%%%%%%%%%%%%%%%%%%%%%%%%%%%%%%%%%%%%%%%%%%%%%%%%%%%%%
\section{Full 2D Probability Posteriors}
\label{App: Full Posteriors}
%%%%%%%%%%%%%%%%%%%%%%%%%%%%%%%%%%%%%%%%%%%%%%%%%%%%%%

Here, we present the complete posterior probability contours for the $\Lambda$CDM model and the six alternative cosmological models considered in this work. For each case, we report the number of accepted steps in the MCMC chains, the final acceptance rate, and the Gelman-Rubin convergence diagnostic for all parameters involved in the analysis. \\

%%%%%%%%%%%%%%%%%%%%%%%%%%%%%%%%%%%%%%
$\bullet$ \textbf{Scalar field coupled to a vector field (AQ):} 
%%%%%%%%%%%%%%%%%%%%%%%%%%%%%%%%%%%%%%

For this model, the number of accepted MCMC steps was 859~225, with an acceptance rate of 24\%. The Gelman–Rubin convergence diagnostic for the set of parameters is:

\begin{align}
&\Omega_{c,0} \rightarrow 0.06, &\omega_{b}& \rightarrow 0.01,  &h\rightarrow 0.03, \nonumber  \\
&N_\eff \rightarrow 0.1, &\lambda_\AQ& \rightarrow 0.002, &\mu_\AQ \rightarrow 0.01, \nonumber \\
&M \rightarrow 0.02, &\Omega_{\mathrm{DE},0}& \rightarrow 0.06, &\log|\Sigma_{0}| \rightarrow 0.02 .  \nonumber 
\end{align}

Values of $R-1 < 0.1$ typically indicate acceptable convergence of the chains; $R-1 < 0.05$ implies good convergence, while $R-1 < 0.01$ corresponds to strong convergence \cite{10.1214/ss/1177011136}. Although some parameters exhibit $R-1$ values above the ideal threshold, the large number of accepted steps suggests that further sampling would not significantly alter the results.

\begin{figure*}[t!]
\centering
\begin{minipage}[b]{0.85\textwidth}
\includegraphics[width=\textwidth]{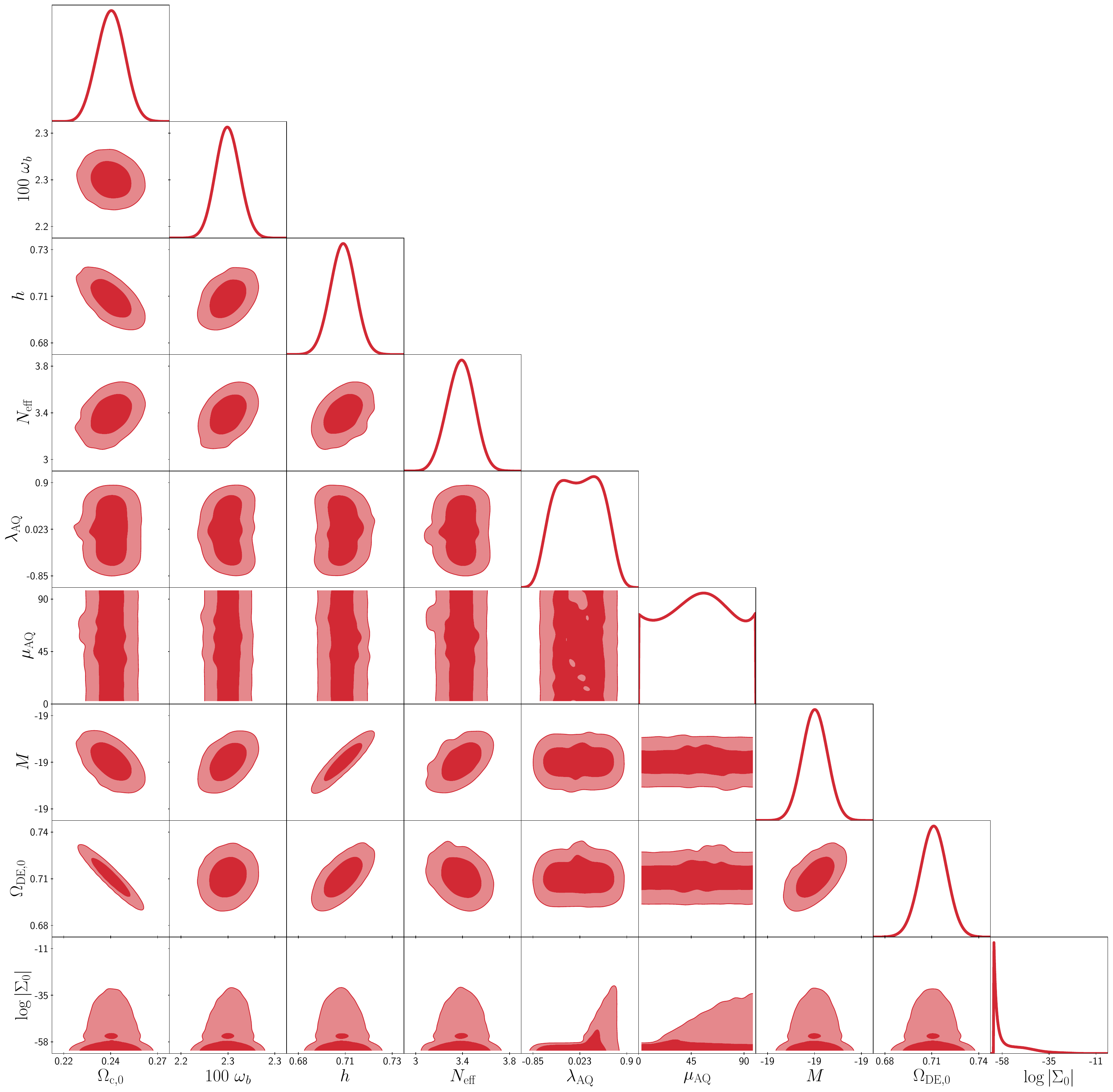}
\end{minipage}
\caption{Full posterior probability contours for the scalar field coupled to a vector field model, using the combined CMB, BAO, $H(z)$, and SnIa datasets. The parameters under consideration include the standard cosmological parameters $\{\Omega_{c,0}, \omega_b, h, N_\mathrm{eff}\}$, the model-specific parameters $\{\lambda_{\AQ}, \mu_{\AQ}\}$, the shear parameter $\Sigma_0$ (shown as $\log|\Sigma_0|$), and the dark energy density today $\Omega_{\mathrm{DE},0}$ (as a derived parameter). The absolute magnitude of Sn Ia, $M$, is included as a nuisance parameter.}
\label{Fig: Full_Contours_AQ}
\end{figure*}

Figure~\ref{Fig: Full_Contours_AQ} displays the full posterior probability contours for the model parameters of the scalar field coupled to a vector field. The plot includes both the cosmological parameters and the nuisance parameter $M$, representing the absolute magnitude of Type Ia supernovae. \\

%%%%%%%%%%%%%%%%%%%%%%%%%%%%%%%%%%%%%%
$\bullet$ \textbf{Scalar field coupled to a 2-Form field (FQ):} 
%%%%%%%%%%%%%%%%%%%%%%%%%%%%%%%%%%%%%%
For this model, the number of accepted MCMC steps was 835~997, with an acceptance rate of 25\%. The Gelman–Rubin convergence diagnostic, $R-1$, for each parameter is:
\begin{align}
\Omega_{c,0} &\rightarrow 0.0004, &\omega_{b}& \rightarrow 0.0006,  &h &\rightarrow 0.003, \nonumber  \\
N_\eff &\rightarrow 0.001, &\lambda_\FQ& \rightarrow 0.0008, &\mu_\FQ &\rightarrow 0.004, \nonumber \\
M &\rightarrow 0.004, &\Omega_{\mathrm{DE},0}& \rightarrow 0.0006, &\log|\Sigma_{0}| &\rightarrow 0.0001 .  \nonumber 
\end{align}
For all parameters, the $R-1$ values satisfy the criterion for good convergence.

\begin{figure*}[t!]
\centering
\begin{minipage}[b]{0.95\textwidth}
\includegraphics[width=\textwidth]{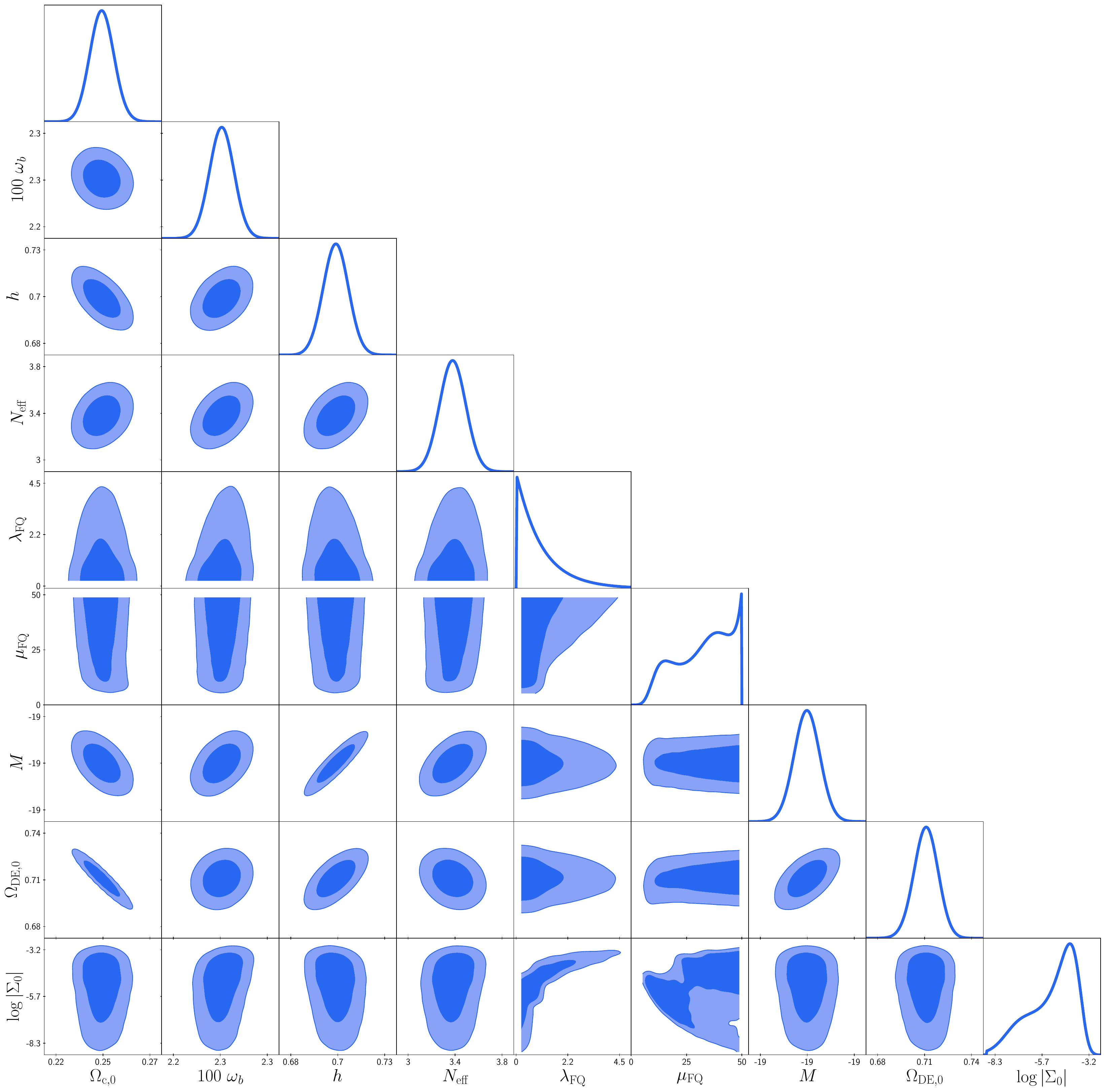}
\end{minipage}
\caption{Full posterior probability contours for the scalar field coupled to a 2-form field model, using the combined CMB, BAO, $H(z)$, and Sn Ia datasets. The parameters shown include the standard cosmological parameters $\{\Omega_{c,0}, \omega_b, h, N_\mathrm{eff}\}$, along with the model-specific parameters $\lambda_\FQ$ and $\mu_\FQ$, the derived parameter $\Omega_{\mathrm{DE},0}$, the shear $\Sigma_0$ (shown as $\log|\Sigma_0|$), and the nuisance parameter $M$.}
\label{Fig: Full_Contours_FQ}
\end{figure*}

Figure~\ref{Fig: Full_Contours_FQ} shows the full set of 2D posterior probability contours obtained using the combined CMB, BAO, $H(z)$, and Sn Ia datasets. \\

%%%%%%%%%%%%%%%%%%%%%%%%%%%%%%%%%%%%%%
$\bullet$ \textbf{Inhomogeneous scalar fields (SDE):} 
%%%%%%%%%%%%%%%%%%%%%%%%%%%%%%%%%%%%%%

For the SDE model, the number of accepted MCMC steps was 590~892, with an acceptance rate of 26\%. The Gelman–Rubin convergence diagnostic, $R-1$, for each parameter is:
\begin{align}
\Omega_{c,0} &\rightarrow 0.0006, &\omega_{b}& \rightarrow 0.0005,  &h &\rightarrow 0.0002, \nonumber  \\
N_\eff &\rightarrow  0.001, &n& \rightarrow 0.002, &m &\rightarrow 0.002384, \nonumber \\
M &\rightarrow 0.0003, &\Omega_{\mathrm{DE},0}& \rightarrow 0.0006, &10^{4}\Sigma_{0} &\rightarrow 0.0002.  \nonumber 
\end{align}
All parameters satisfy the convergence criterion, indicating good convergence of the chains.

\begin{figure*}[t!]
\centering
\begin{minipage}[b]{0.85\textwidth}
\includegraphics[width=\textwidth]{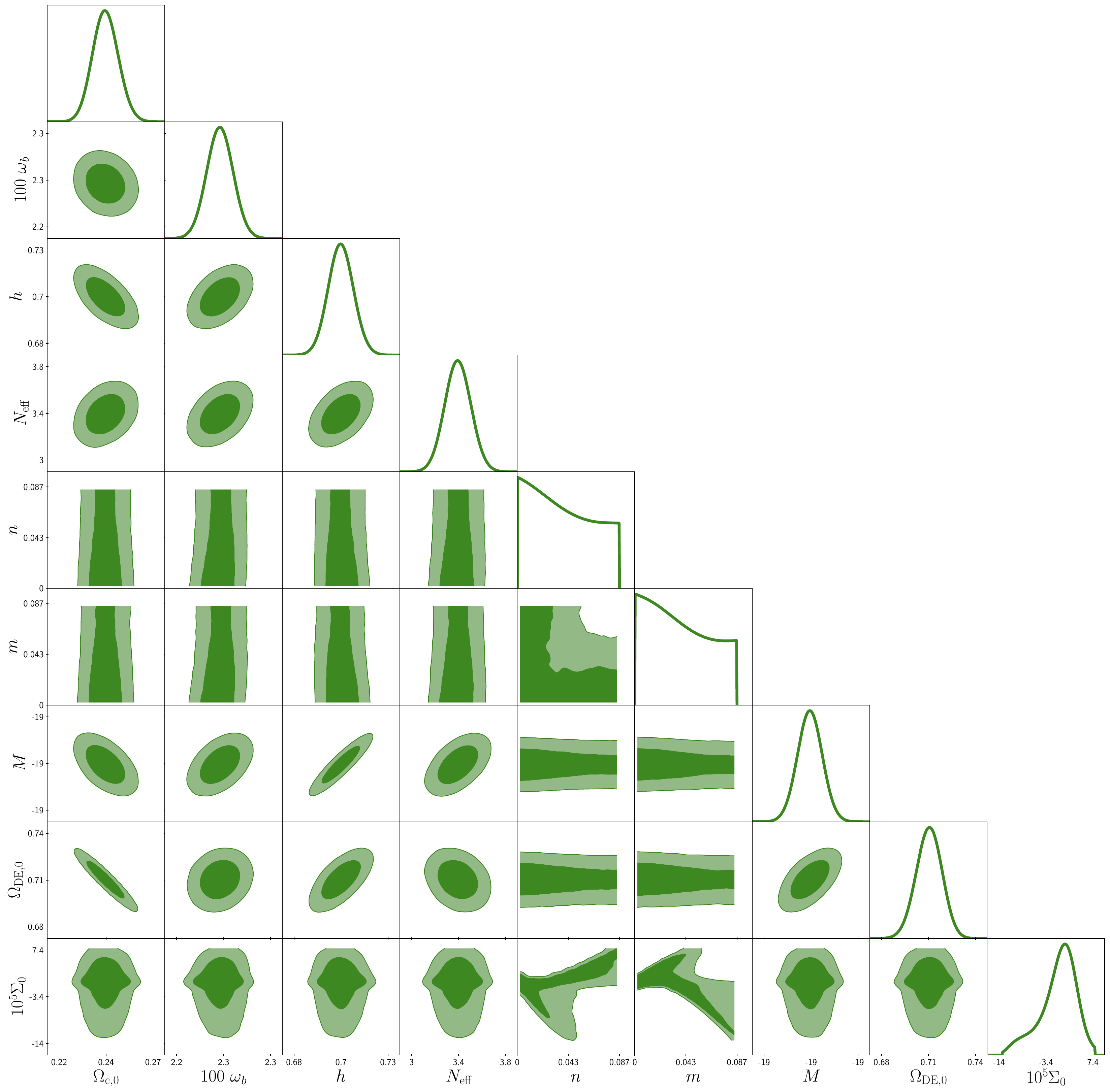}
\end{minipage}
\caption{Full posterior probability contours for the inhomogeneous scalar field model (SDE), obtained using the combined CMB, BAO, $H(z)$, and Sn Ia datasets. The parameters shown include the standard cosmological parameters $\{\Omega_{c,0}, \omega_b, h, N_\mathrm{eff}\}$, the model-specific parameters $n$ and $m$, the derived dark energy density $\Omega_{\mathrm{DE},0}$, the shear $\Sigma_0$ (rescaled by $10^5$ for readability), and the nuisance parameter $M$.}
\label{Fig: Full_Contours_SDE}
\end{figure*}

Figure~\ref{Fig: Full_Contours_SDE} displays the full set of 2D posterior probability contours for the combined CMB, BAO, $H(z)$, and Sn Ia datasets. \\

%%%%%%%%%%%%%%%%%%%%%%%%%%%%%%%%%%%%%%
$\bullet$ \textbf{Scalar fields with internal symmetry (EYMH):} 
%%%%%%%%%%%%%%%%%%%%%%%%%%%%%%%%%%%%%%

For the EYMH model, the number of accepted MCMC steps was 865~196, with an acceptance rate of 26\%. The Gelman–Rubin convergence diagnostic, $R - 1$, for each parameter is:
\begin{align}
\Omega_{c,0} &\rightarrow 0.00008, &\omega_{b}& \rightarrow 0.00006,  &h &\rightarrow 0.0001, \nonumber  \\
N_\eff &\rightarrow 0.00003, &w_i& \rightarrow 0.0001, &M &\rightarrow 0.00009, \nonumber \\
\Omega_{\mathrm{DE},0} &\rightarrow 0.00008, &\log|\Sigma_{0}|& \rightarrow 0.0001. &  &   \nonumber 
\end{align}

\begin{figure*}[t!]
\centering
\begin{minipage}[b]{0.85\textwidth}
\includegraphics[width=\textwidth]{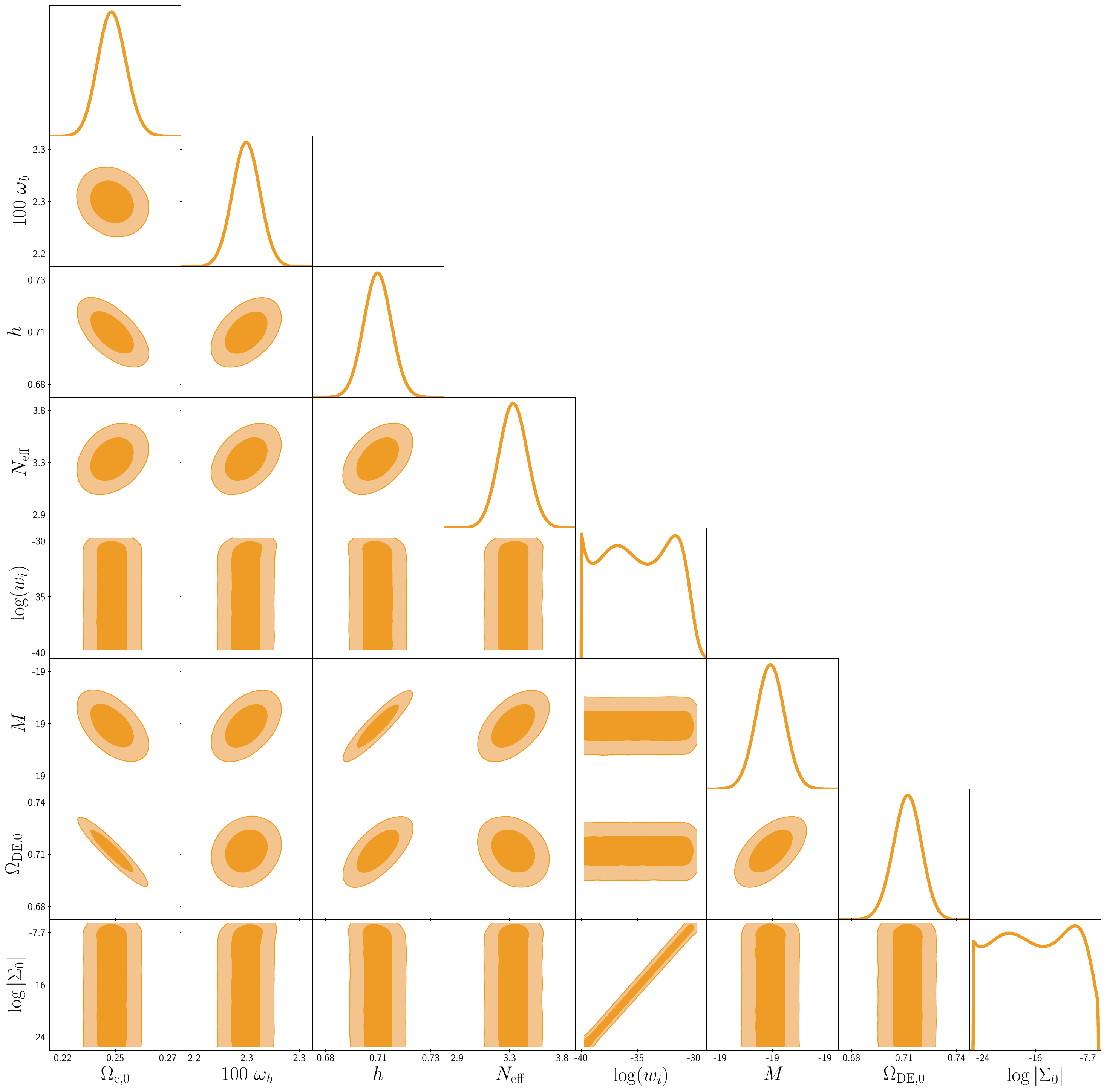}
\end{minipage}
\caption{Full posterior probability contours for the scalar field model with internal symmetry (EYMH), obtained using the combined CMB, BAO, $H(z)$, and Sn Ia datasets. The parameters shown include the standard cosmological parameters $\{\Omega_{c,0}, \omega_b, h, N_\mathrm{eff}\}$, the model-specific parameter $w_i$, the derived dark energy density $\Omega_{\mathrm{DE},0}$, the shear $\Sigma_0$ (shown as $\log|\Sigma_0|$), and the nuisance parameter $M$.}
\label{Fig: Full_Contours_EYMH}
\end{figure*}

Figure~\ref{Fig: Full_Contours_EYMH} presents the full set of 2D posterior probability contours for the combined CMB, BAO, $H(z)$, and Sn Ia datasets. \\

%%%%%%%%%%%%%%%%%%%%%%%%%%%%%%%%%%%%%%
$\bullet$ \textbf{2-Form coupled to CDM (2F):} 
%%%%%%%%%%%%%%%%%%%%%%%%%%%%%%%%%%%%%%

For this model, the number of accepted MCMC steps is 753~674, with an acceptance rate of 26\%. The Gelman–Rubin convergence diagnostic, $R - 1$, for each parameter is:
\begin{align}
\Omega_{c,0} &\rightarrow 0.003, &\omega_{b}& \rightarrow  0.001,  &h &\rightarrow 0.0009, \nonumber  \\
N_\eff &\rightarrow 0.007, &\lambda_\twoF& \rightarrow 0.001, &100\mu_\twoF &\rightarrow 0.0004, \nonumber \\
M &\rightarrow 0.001, &\Omega_{\mathrm{DE},0}& \rightarrow 0.003, &10^{4}\Sigma_{0} &\rightarrow 0.0007.  \nonumber 
\end{align}

\begin{figure*}[t!]
\centering
\begin{minipage}[b]{0.85\textwidth}
\includegraphics[width=\textwidth]{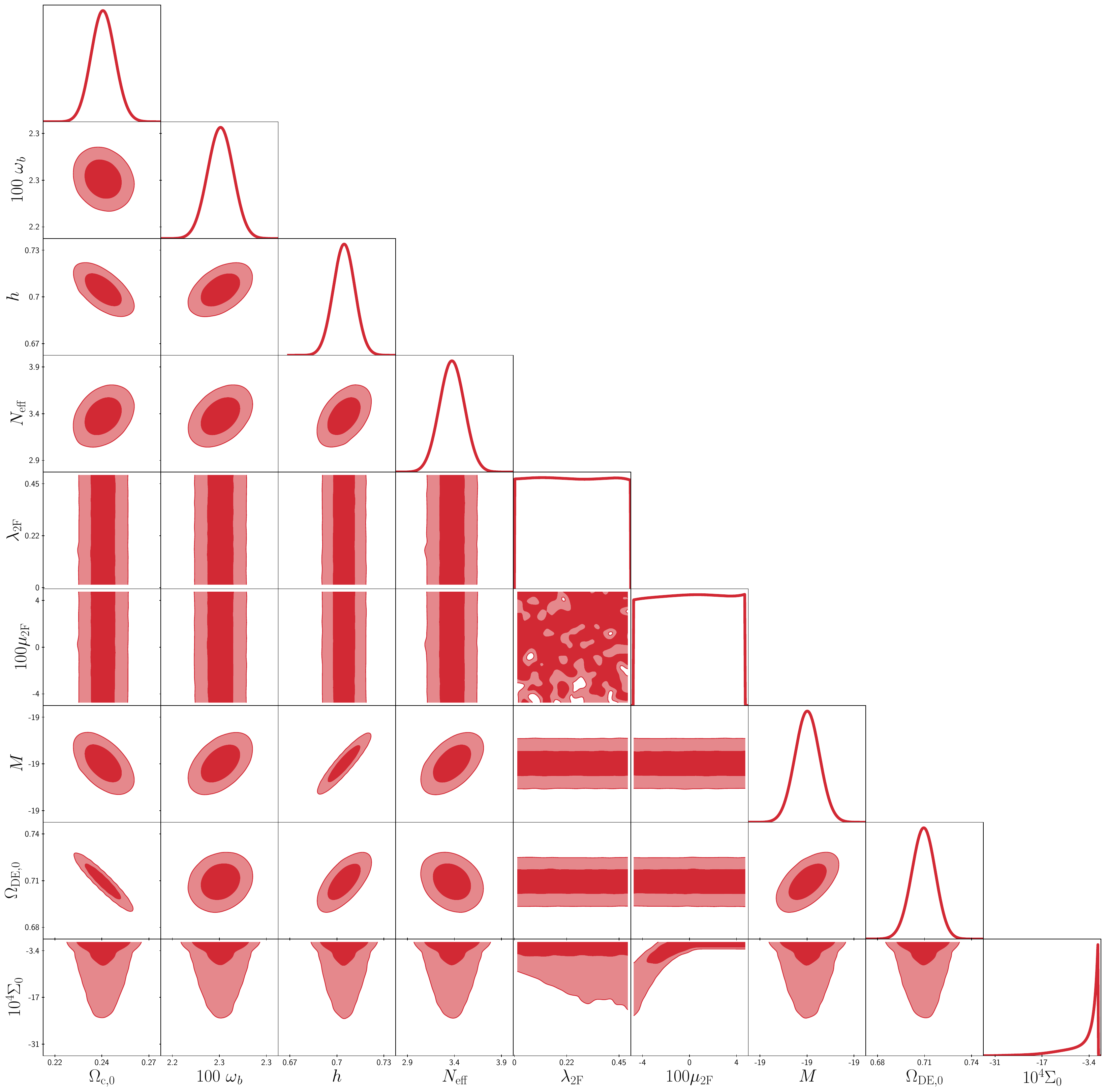}
\end{minipage}
\caption{Full posterior probability contours for the 2-form field model coupled to cold dark matter (2F), obtained using the combined CMB, BAO, $H(z)$, and Sn Ia datasets. The parameters shown include the standard cosmological parameters $\{\Omega_{c,0}, \omega_b, h, N_\mathrm{eff}\}$, the model-specific parameters $\lambda_{\twoF}$ and $\mu_{\twoF}$, the derived dark energy density $\Omega_{\mathrm{DE},0}$, the shear $\Sigma_0$ (rescaled as $10^4\Sigma_0$ for clarity), and the nuisance parameter $M$.}
\label{Fig: Full_Contours_TF}
\end{figure*}

Figure~\ref{Fig: Full_Contours_TF} shows the full 2D posterior probability contours for the combined CMB, BAO, $H(z)$, and Sn Ia datasets. \\

%%%%%%%%%%%%%%%%%%%%%%%%%%%%%%%%%%%%%%
$\bullet$ \textbf{Vector field (VF):} 
%%%%%%%%%%%%%%%%%%%%%%%%%%%%%%%%%%%%%%
For this model, the number of accepted MCMC steps is 622~736, with an acceptance rate of 19\%. The Gelman–Rubin convergence diagnostic, $R - 1$, for each parameter is:
\begin{align}
\Omega_{c,0} &\rightarrow 0.0001, &\omega_{b}& \rightarrow 0.0001,  &h &\rightarrow 0.0001, \nonumber  \\
N_\eff &\rightarrow 0.00005, &\lambda_\VF& \rightarrow 0.0003, &M &\rightarrow 0.0001, \nonumber \\
\Omega_{\mathrm{DE},0} &\rightarrow 0.0001, &\log|\Sigma_{0}|& \rightarrow 0.0002. &  &   \nonumber 
\end{align}

\begin{figure*}[t!]
\centering
\begin{minipage}[b]{0.85\textwidth}
\includegraphics[width=\textwidth]{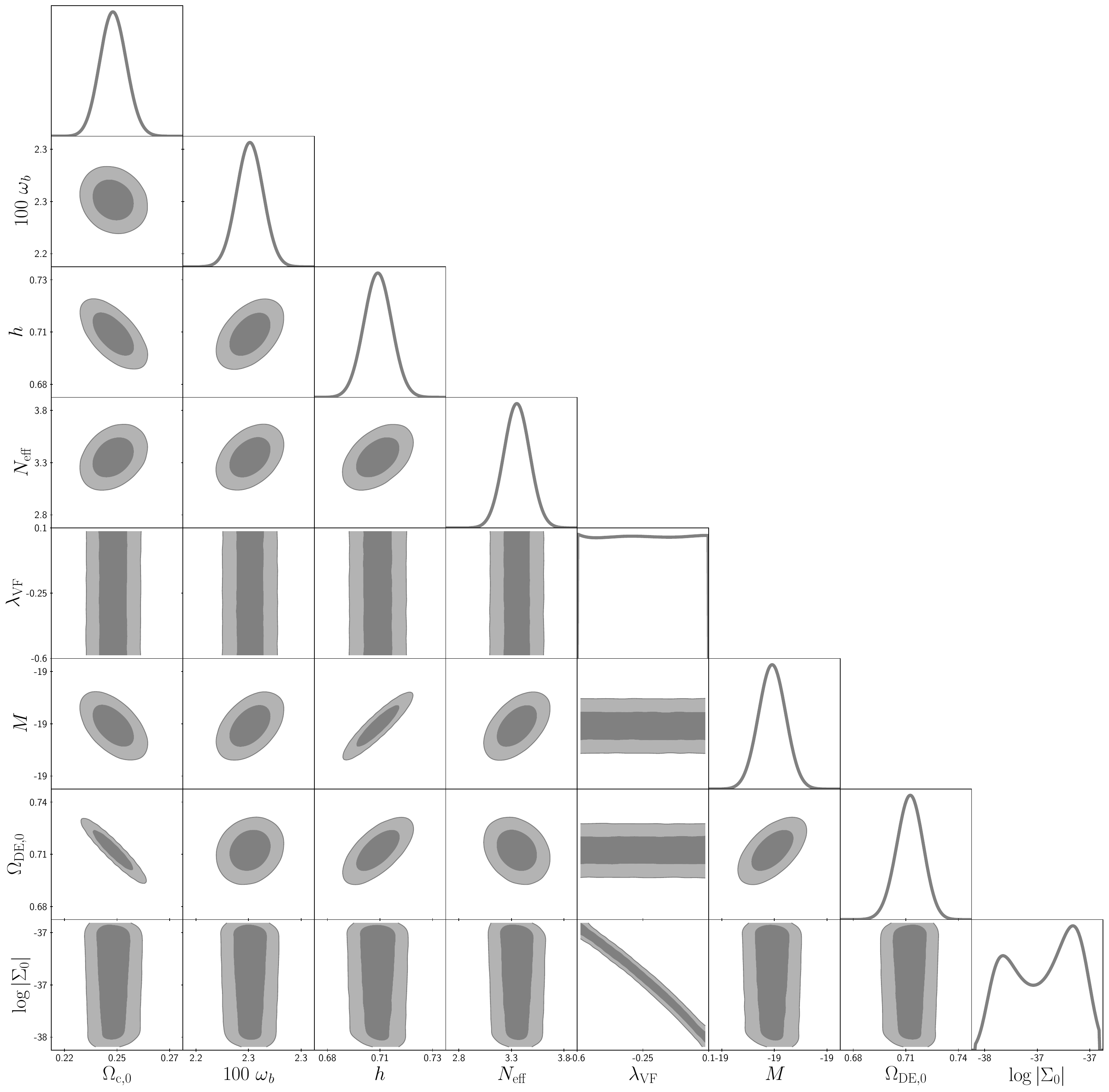}
\end{minipage}
\caption{Full posterior probability contours for the anisotropic vector field (VF) model, obtained using the combined CMB, BAO, $H(z)$, and Sn Ia datasets. The parameters shown include the standard cosmological parameters $\{\Omega_{c,0}, \omega_b, h, N_\mathrm{eff}\}$, the model-specific parameter $\lambda_\VF$, the derived dark energy density $\Omega_{\mathrm{DE},0}$, the shear $\Sigma_0$ (represented as $\log|\Sigma_0|$), and the nuisance parameter $M$.}
\label{Fig: Full_Contours_VF}
\end{figure*}

Figure~\ref{Fig: Full_Contours_VF} displays the full 2D posterior probability contours obtained from the combined CMB, BAO, $H(z)$, and Sn Ia datasets. \\

%%%%%%%%%%%%%%%%%%%%%%%%%%%%%%%%%%%%%%
$\bullet$ \textbf{$\Lambda$CDM:}
%%%%%%%%%%%%%%%%%%%%%%%%%%%%%%%%%%%%%%
As a reference, we also present the results for the $\Lambda$CDM model. The number of accepted MCMC steps is 821~756, with an acceptance rate of 32\%. The Gelman–Rubin convergence diagnostic, $R - 1$, for each parameter is:
\begin{align}
\Omega_{c,0} &\rightarrow 0.00009, &\omega_{b}& \rightarrow 0.0001,  &h &\rightarrow 0.00009, \nonumber  \\
N_\eff &\rightarrow 0.0001, &M& \rightarrow 0.00007, &\Omega_{\mathrm{DE},0} &\rightarrow  0.00009. \nonumber
\end{align}

\begin{figure*}[t!]
\centering
\begin{minipage}[b]{0.85\textwidth}
\includegraphics[width=\textwidth]{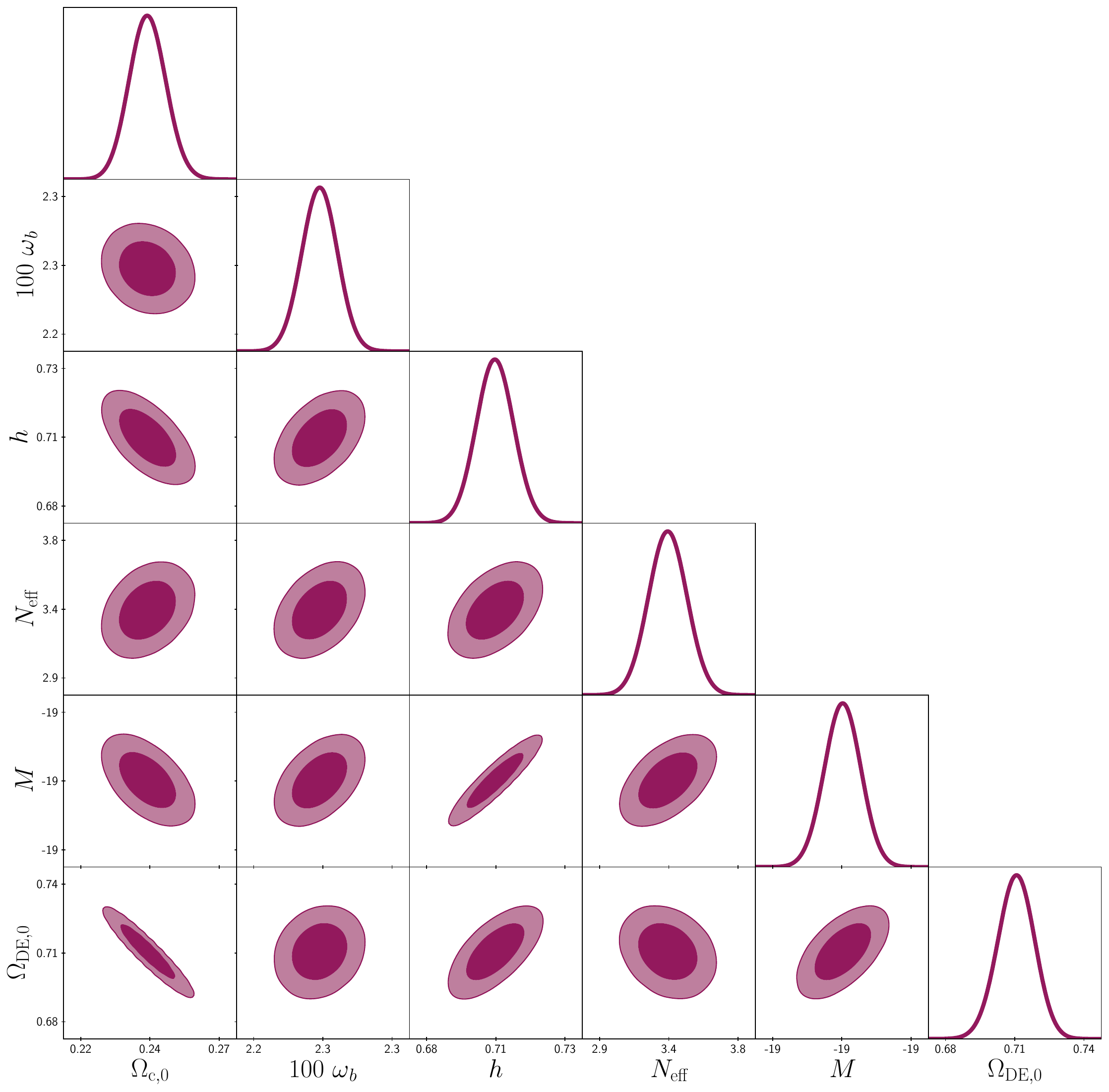}
\end{minipage}
\caption{Full posterior probability contours for the standard $\Lambda$CDM model, obtained using the combined CMB, BAO, $H(z)$, and Sn Ia datasets. The parameters shown include $\{\Omega_{c,0}, \omega_b, h, N_\mathrm{eff}\}$, the derived dark energy density $\Omega_{\mathrm{DE},0}$, and the nuisance parameter $M$.}
\label{Fig: Full_Contours_Lcdm}
\end{figure*}

Finally, Figure~\ref{Fig: Full_Contours_Lcdm} shows the full 2D posterior probability contours for the $\Lambda$CDM model.

\clearpage
\FloatBarrier
\bibliography{Bibli.bib} 

\end{document}